\documentclass[prx,twocolumn,nofootinbib,amsmath,amssymb,aps,longbibliography]{revtex4-2}
\usepackage{graphicx}% Include figure files
\usepackage{dcolumn}% Align table columns on decimal point
\usepackage{bm}% bold math
\usepackage{physics}
\usepackage{amsmath}
\usepackage{dsfont}
\usepackage{xcolor}
\usepackage{lmodern}
\usepackage{appendix}
\usepackage{amsthm}
\usepackage{ulem}

\usepackage{hyperref}
\hypersetup{colorlinks=true,linkcolor=blue,citecolor=blue,urlcolor=blue}

\usepackage{titlesec}
\titleformat{\paragraph}
{\normalfont\normalsize\bfseries}{\theparagraph}{1em}{}
\titlespacing*{\paragraph}
{0pt}{3.25ex plus 1ex minus .2ex}{1.5ex plus .2ex}

\usepackage{hyperref}
\usepackage[bottom]{footmisc}
\usepackage{scrextend}
\deffootnote[1em]{1.5em}{1em}{\textsuperscript{\thefootnotemark}}

\newtheorem{remark}{Remark}

\definecolor{aliceblue}{rgb}{0.94, 0.97, 1.0}

\usepackage{mwe}
\usepackage[framemethod=tikz]{mdframed}
\usetikzlibrary{shadows}
\usepackage{tikz}
    \usetikzlibrary{shadows}
\usepackage{tcolorbox}
    \tcbuselibrary{skins}

\mdfdefinestyle{mystyle}{%
    innerlinewidth = 0.0pt,
    outerlinewidth = 1pt,
    linecolor = blue,
    tikzsetting = {draw = black, line width = 0.5pt},
    roundcorner=10pt,
    linecolor=blue,
    linewidth=1pt,
    topline=true,
    frametitleaboveskip=1.5\dimexpr-\ht\strutbox\relax,
}

\def\Xint#1{\mathchoice
   {\XXint\displaystyle\textstyle{#1}}%
   {\XXint\textstyle\scriptstyle{#1}}%
   {\XXint\scriptstyle\scriptscriptstyle{#1}}%
   {\XXint\scriptscriptstyle\scriptscriptstyle{#1}}%
   \!\int}
\def\XXint#1#2#3{{\setbox0=\hbox{$#1{#2#3}{\int}$}
     \vcenter{\hbox{$#2#3$}}\kern-.5\wd0}}

\def\dashint{\Xint-}

\begin{document}
	
	%\preprint{APS/123-QED}
	
	\title{Renormalization in the Theory of Open Quantum Systems\\ via the Self-Consistency Condition}% Force line breaks with \\
	
\author{Marek Winczewski${}^{(1,2)}$}\email{Marek.Winczewski@ug.edu.pl}
\author{Robert Alicki${}^{(1)}$}
\affiliation{${}^{(1)}$International Centre for Theory of Quantum Technologies, University of Gda\'nsk, Wita Stwosza 63, 80-308 Gda\'nsk, Poland}
\affiliation{${}^{(2)}$Institute of Theoretical Physics and Astrophysics and National Quantum Information Centre in Gda\'nsk, University of Gda\'nsk, 80--952 Gda\'nsk, Poland}
%\affiliation{${}^{(3)}$National Quantum Information Centre in Gda\'{n}sk,
%Faculty of Mathematics, Physics and Informatics, University of Gda\'{n}sk, 80–952 Gda\'{n}sk, Poland}

	\date{\today}
	
\begin{abstract}
    We investigate the topic of renormalization in the theory of weakly interacting open quantum systems. Our starting point is an open quantum system interacting with a single heat bath. For a given setup, we discuss that the stationary state of the Davies-GKSL equation is thermodynamically inconsistent with the presence of the Lamb-Stark shift term. For this reason, we postulate the self-consistency condition for the dynamical equations. The condition fixes the renormalization procedure and recovers the thermodynamical consistency. In this way, we rederive the cumulant equation to illustrate how the self-consistency condition enters the derivation of the dynamical equations. The physical interpretation of the renormalization procedure is discussed in terms of the Born approximation. Furthermore, we compare the Lamb-Stark shift term (dynamical correction) with the second-order (static) correction to the so-called mean-force (Gibbs state) Hamiltonian. The discrepancy between the static and the dynamical correction questions the physical meaning of the dynamical one.  Finally, we formulate a simplified renormalization scheme that can be directly applied to Davies-GKSL or Bloch-Redfield equations.
\end{abstract}
	
\maketitle

%\tableofcontents

\section{Introduction}

Well-controlled and subjected to precise measurements ``small" quantum system $\mathcal{S}$  weakly interacting with a stationary ``large" reservoir $\mathcal{R}$   lies at the heart of quantum information processing, quantum thermodynamics, and, more generally,  quantum technology. The ab initio computations involving the solution of the Schr{\"o}dinger equation for the total system $\mathcal{S+R}$ are usually not feasible because of the unknown detailed structure and parametrization of the total Hamiltonian and computational challenges related to a large number of reservoir's degrees of freedom. Therefore, it is necessary to develop semi-phenomenological approximation schemes for the time-dependent reduced density matrix of the open system $\mathcal{S}$  involving only a small number of parameters determined partially by the experimental data and partially by computations based on certain model Hamiltonians. 

\begin{figure}[h!]
    \centering
    \includegraphics[width=0.75 \columnwidth]{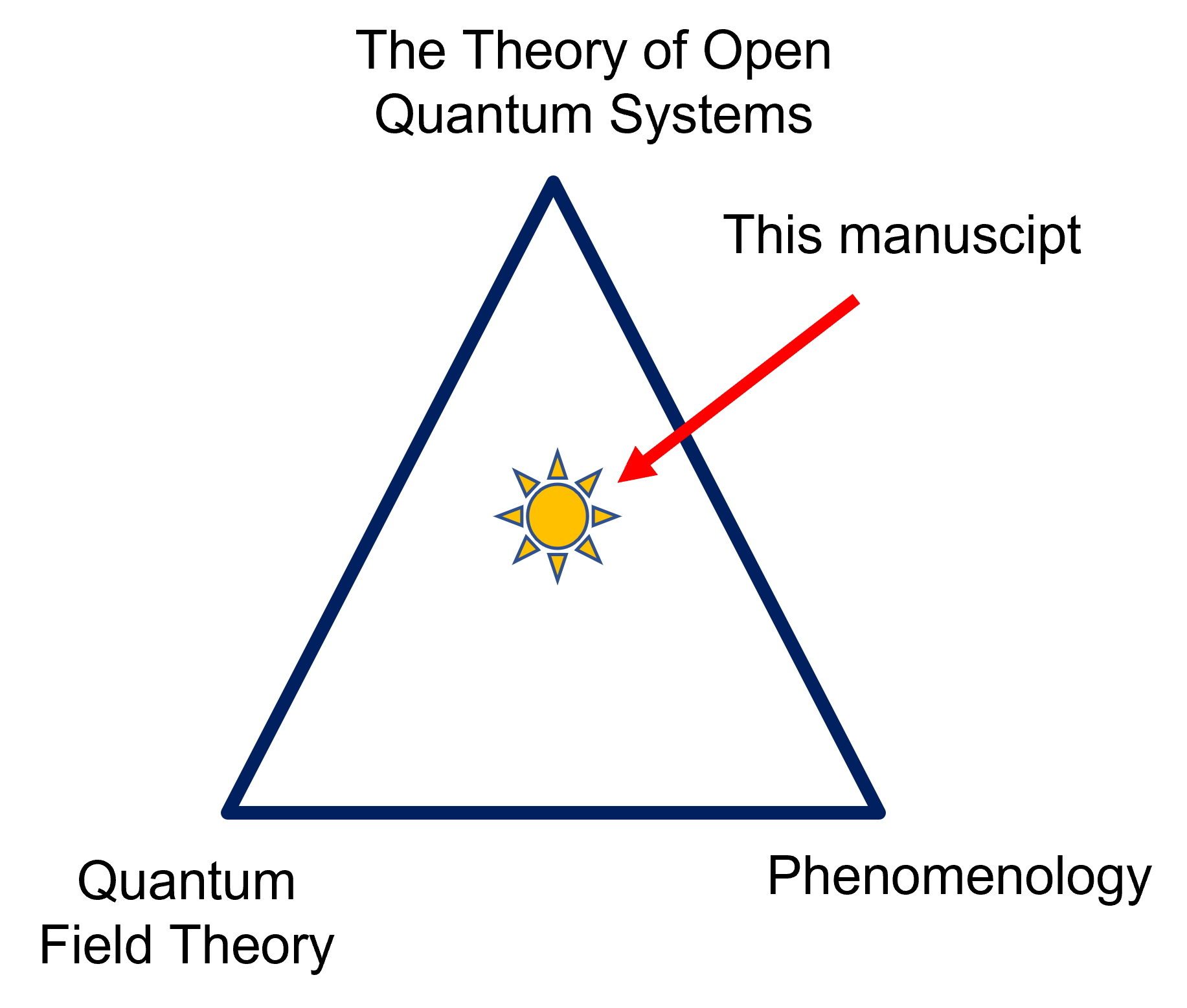}
    \caption{ Picture illustrating the approach we assimilate in this manuscript. }
    \label{fig:Approach}
\end{figure}

The challenges posed by the rapidly developing quantum technologies appear to exceed the applicability of theoretical methods that were successful in the past. In this way, the (purely) phenomenological description of interacting quantum systems seems too heuristic and insufficient to capture exclusively quantum phenomena of contemporary central interest. On the other hand, the great machinery of quantum field theory (QFT) fails to describe situations in which the system's time evolution is a relevant figure of merit. For example, the description of quantum gates, elementary units of a quantum computer, in terms of IN ($t=-\infty$) and OUT ($t=+\infty$) states that are inherent for QFT treatment would be highly unsatisfactory as for the long times ($t=+\infty$) the state of the system thermalizes. It is, therefore, crucial to develop a possibly accurate scheme in the language of the theory of open quantum systems. In this manuscript, we propose an approach to the reduced dynamics of open quantum systems that facilities the qualities of both phenomenological and field-theoretic description (see Figure~\ref{fig:Approach}). Namely, the dynamics is determined with just a few experimentally determined parameters and encompasses ideas from the QFT, such as renormalization and the dressing of systems.

The requirement for the weak interaction usually referred to as the Born approximation, is the heart of the dynamical equations of the theory of open quantum systems~\cite{Breuer+2006}. In a nutshell, the Born approximation consists of two steps: (i) the perturbative expansion of dynamics is truncated to the second-order with respect to coupling constant, (ii) the initial state of the total system is of the product form between the open system and the reservoir (cf.~\cite{Alipour_2020,Merkli_2021,Trushechkin_2021}). In particular, the famous Davies-GKSL\footnote{Davies-Gorini–Kossakowski–Sudarshan–Lindblad.} Markovian master equation is exceedingly successful in describing the evolution of open quantum systems in the weak coupling regime~\cite{davies1974markovian,  gorini1976completely, lindblad1976generators}.  Still, due to explicit separation of time scales concerning the internal evolution of the system and the reservoir relaxation rates, the Davis-GKSL equation in its different versions\footnote{Here, we mean the ''global" and the ''local" approaches \cite{Cattaneo_2021_comment}} can not simultaneously describe all time regimes of the evolution of the quantum system\footnote{The initial (super-short) times $t< \frac{1}{\omega_c}$ are also (currently) out of reach for all (universal) dynamical equations that attempt to describe the dynamics using just a few phenomenological parameters. This limitation is because the initial dynamics depends on the cut-off frequency that is usually not well characterized and depends on the specific setup. On the other hand, we have to assume (at least) the pure existence of some cut-off frequency~$\omega_c$ for the system Hamiltonians to be well defined.}. On the other hand, the equally acclaimed Bloch-Redfield (master) equation does not distinguish any particular time scale and paves the way for the correct description in the whole time regime~\cite{bloch1957generalized, redfield1957theory}. Unfortunately, the Bloch-Redfield equation violates positivity of the density operator as it is not of the Gorini–Kossakowski–Sudarshan–Lindblad (GKSL) form. The drawback of the dynamics not being CPTP is of fundamental importance, and for this reason, it has been addressed in literature many times. As a result, a plethora of different dynamical equations for the time evolution of the reduced density matrix of an open system has been constructed~\cite{schaller2008preservation,Majenz_2013,Davidovic2020,trushechkin2021unified}.

In general, all the standard approaches mentioned above contain the so-called Lamb-Stark shift term \cite{Breuer+2006}. The presence of the aforementioned term in the dynamical equation leads to a (alleged) shift in the energy levels of the open system cause by the interaction with the environment. However, the dissipation still proceeds with respect to energy levels that are not shifted (renormalized). This leads to a thermodynamical inconsistency of the described approaches. On the one hand, the energy levels of the open system are renormalized via the Lamb-Stark shift term. On the other hand, the evolution leads to the stationary state with populations identical to Gibbs state with respect to a Hamiltonian in which energy levels are not shifted. Indeed, the Davies-GKSL equation predicts the stationary state to be a Gibbs state with respect to the not renormalized (bare) Hamiltonian, whereas the Bloch-Redfield equation predicts the same populations as the Davies-GKSL equation but exhibits steady-state coherence (see Ref. \cite{Lobejko_2022} for detailed discussion). 
%This inconsistency leads to a discrepancy in predicting the populations of the final state of evolution.
As the drawback of this inconsistency, the standard approaches always fail to correctly describe the stationary state and dissipation rates simultaneously. As we will show, this issue can be overcome with a proposed here renormalization procedure.
\color{black}

The first equation that yields CPTP dynamics and is relevant for all time scales (except initial times) was derived in Ref.~\cite{Alicki1989}, and has been recently independently discovered by A.~Rivas ~\cite{Rivas_2017}, where the name refined weak coupling limit is used. This equation, for which we adapt the name the cumulant equation is one of the central objects in our considerations (see also reference~\cite{Winczewski_2021}). The first derivation in Ref.~\cite{Alicki1989} is based on a deeper insight into the meaning of the Born approximation that employs the idea of quantum noise. Namely, in its first part, i.e. (i), the Born approximation refers to the description of the interaction between the open system and the reservoir in the spirit of the quantum analogs of the central limit theorem (CLT)~\cite{Goderis_1989,Goderis_1990,Accardi_1990,Accardi_2002}. In this manner, the interaction of the open system with the bath is expressed as a sum of numerous (ideally infinite) weak and independent perturbations. In principle, this form of interaction accounts for the open system to experience a (quantum) Gaussian noise due to the reservoir. On the other hand, this kind of characterization of the interaction has to be confronted with the Hamiltonian dynamics of the total system. Therefore, it is essential to decompose the quantum noise into two parts. The first component (drift) corresponds to the coherent evolution of the system. It contains Hamiltonian-like terms associated with the (bare) free Hamiltonian of the open system combined with possible (renormalization) corrections. The second part accounts for pure dissipation. At this place, the idea of renormalization enters the derivation of the cumulant equation. Namely, it is postulated that in the interaction picture with respect to the physical, renormalized Hamiltonian of the system, the dynamics should be purely dissipative, i.e., it should not contain any Hamiltonian-like terms. \\
~~\\
{\it In the interaction picture with respect to the physical (renormalized) Hamiltonian $H_\mathcal{S}$, the reduced dynamics of the system $\mathcal{S}$ should be purely dissipative, i.e., it should not include any Hamiltonian-like terms.}\\
~~\\
In this manuscript, we refer to the above postulate as to the self-consistency condition. Indeed, the Hamiltonian-like terms (Lamb-Stark shift term) can be compensated with the aforementioned corrections to the bare Hamiltonian of the open system. As we discuss in this manuscript, the presence of the corrections is indispensably connected to the assumption of the initial state to be in the product form, i.e., the second part, (ii) of the Born approximation. However, the renormalization procedure that removes all Hamiltonian-like terms yields different dynamical equations than the standard derivation of Davies-GKSL~\cite{Breuer+2006} or Bloch-Redfield equation, in which the Lamb-Stark shift term (Hamiltonian-like) is explicitly present\footnote{Still, negligence of this term is a common practice.}. As we show, the renormalized equations do not exhibit the aforementioned conflict in simultaneous prediction of the populations of the stationary state and the dissipation rates. Additionally, we take a chance to exhaustively explain the first derivation of the cumulant equation. Still, we do not hesitate to provide some minor generalizations and novel contributions.  

In order to prove the relevance of the self-consistency condition, we firstly formulate an intuition about the return to the equilibrium in the open quantum system setting. The intuition prompts us that a joint quantum system consisting of the small system $\mathcal{S}$, and a large reservoir $\mathcal{R}$ in the thermal state (heat bath) should {\it return to the equilibrium state} whenever an interaction between the units is present. Moreover, because of the discrepancy in ``sizes" of systems $\mathcal{S}$ and $\mathcal{R}$, and the requirement for weak interaction, the equilibrium state should be the global thermal state with the same temperature as the initial temperature of the reservoir $\mathcal{R}$. In fact, this intuition has its confirmation in mathematically rigorous results of the quantum theory of infinite systems~\cite{Araki_1963,Araki_1973,Bratteli_1987,Bratteli_1997,Dereziski_2007}. Therefore, the intuition we present allows us to ask an important question about the reduced state of the system $\mathcal{S}$ after the equilibrium is reached \cite{Hilt_2011_MeanForce,Trushechkin_2021_MeanForce}.
~~\\
~~\\
{\it If the reduced state of systems $\mathcal{S}$ is truly a Gibbs (thermal) state with inverse temperature $\beta$ then with respect to which Hamiltonian and what is its relation to the bare one?}\\
~~\\

The Hamiltonian in the above question is the so-called mean-force (Gibbs state) Hamiltonian \cite{Hilt_2011_MeanForce,Trushechkin_2021_MeanForce,Lobejko_2022,Timofeev_2022}. Informally speaking, the mean-force Hamiltonian describes the ''effective" force exerted on the open system by the reservoir. In this manner, the relevance of the renormalization procedure for the dynamical equations is guaranteed via confronting it with the ``static" results from the quantum theory of infinite systems~\cite{Araki_1963,Araki_1973,Bratteli_1987,Bratteli_1997,Dereziski_2007}. Moreover, the mean-force Hamiltonian approach provides a platform at which the interpretation of the Lamb-Stark shift term can be questioned.
\color{black}

The manuscript is organized as follows. In Section \ref{sec:intuition} we formulate the intuition about the relaxation to equilibrium for the total system. We expect that if the initial state of the reservoir is the thermal (KMS) state with the inverse temperature $\beta$, then the total system will equilibrate to the global thermal (KMS) state with the same inverse temperature. In Section \ref{sec:setup} we describe the specific system under the study in terms of Hamiltonians and initial states. Section \ref{sec:MME} is devoted to the analysis of the standard derivation of the Davies-GKSL equation~\cite{Breuer+2006}. In particular, we show how the first-order renormalization is built-in in the standard derivation of the Davies-GKSL equation. Furthermore, subsection \ref{sec:sub:LSterm} takes a deeper insight into problems concerning Lamb-Stark shift term from the phenomenological perspective. We discuss the problems concerning the presence of Lamb-Stark shift term and the condition for its negligence. The problem of inconsistency with thermodynamics in the standard approaches is remarked. In Section \ref{sec:PartialTrace} we calculate the partial trace with respect to the reservoir of the total system $\mathcal{S}+\mathcal{R}$ thermal (KMS) state, and in this way we rederive the so-called mean-force Hamiltonian \cite{Lobejko_2022,Timofeev_2022}. This important technical result is obtained using {\it Explicit description of the Zassenhaus formula} obtained by T. Kimura~\cite{Kimura_2017}. The partial trace provides a ``static" form of renormalization and the answer to the question stated before in Section \ref{sec:intuition}. In the next Section \ref{sec:cumulant} we sketch the derivation of the cumulant equation equation (the complete derivation is present in Section \ref{sec:app:derivationCumulantEquation} of the Appendix). We discuss the relation between the cumulant equation and the Bloch-Redfield master equation. In particular, we transform the cumulant equation into a differential equation and show that the Bloch-Redfield equation emerges as its approximation. Furthermore, we calculate the long time-limit for the cumulant equation (cf. \cite{Rivas_2017}), and discuss its stationary state \cite{MW_preparation}. We show how the renormalization procedure motivated by the previously postulated self-consistency condition yield equations that reproduce the correct stationary state populations, i.e., populations of the Gibbs state with respect to the renormalized (physical) Hamiltonian. Section \ref{sec:discussion} is devoted to the discussion about the physical meaning of the renormalization procedure, interpretation of the renormalized Hamiltonian, and the counterterms (Lamb-Stark shift term). Moreover, we provide a simplified scheme for the renormalization that can be applied to Davies-GKSL or Bloch-Redfield equations. The last Section \ref{sec:conclusion} is left for the concluding remarks.

\section{Intuition}\label{sec:intuition}

Our starting point is the intuition wandering around the zeroth law of thermodynamics. Still, the intuitive considerations present in this Section have their mathematically rigorous justification in the theory of infinite systems~\cite{Araki_1963,Araki_1973,Bratteli_1987,Bratteli_1997,Dereziski_2007}. Consider small quantum system $\mathcal{S}$, and thermal reservoir (heat bath) $\mathcal{R}$ at the inverse temperature $\beta$. The heat bath is a much larger system than $\mathcal{S}$ as it possesses an infinite number of degrees of freedom contrary to small system $\mathcal{S}$. The total, closed system $\mathcal{S}+\mathcal{R}$ is prepared initially (at $t=0$) in a product state $\rho_\mathcal{S} \otimes \rho_{\mathcal{R},\beta}$. Where $\rho_{\mathcal{R},\beta}$ is a thermal (KMS) state of the reservoir, and $\rho_\mathcal{S}$ is unspecified, i.e.,  any possible state of~$\mathcal{S}$.
\begin{figure}[h!]
    \centering
    \includegraphics[width=0.80 \columnwidth]{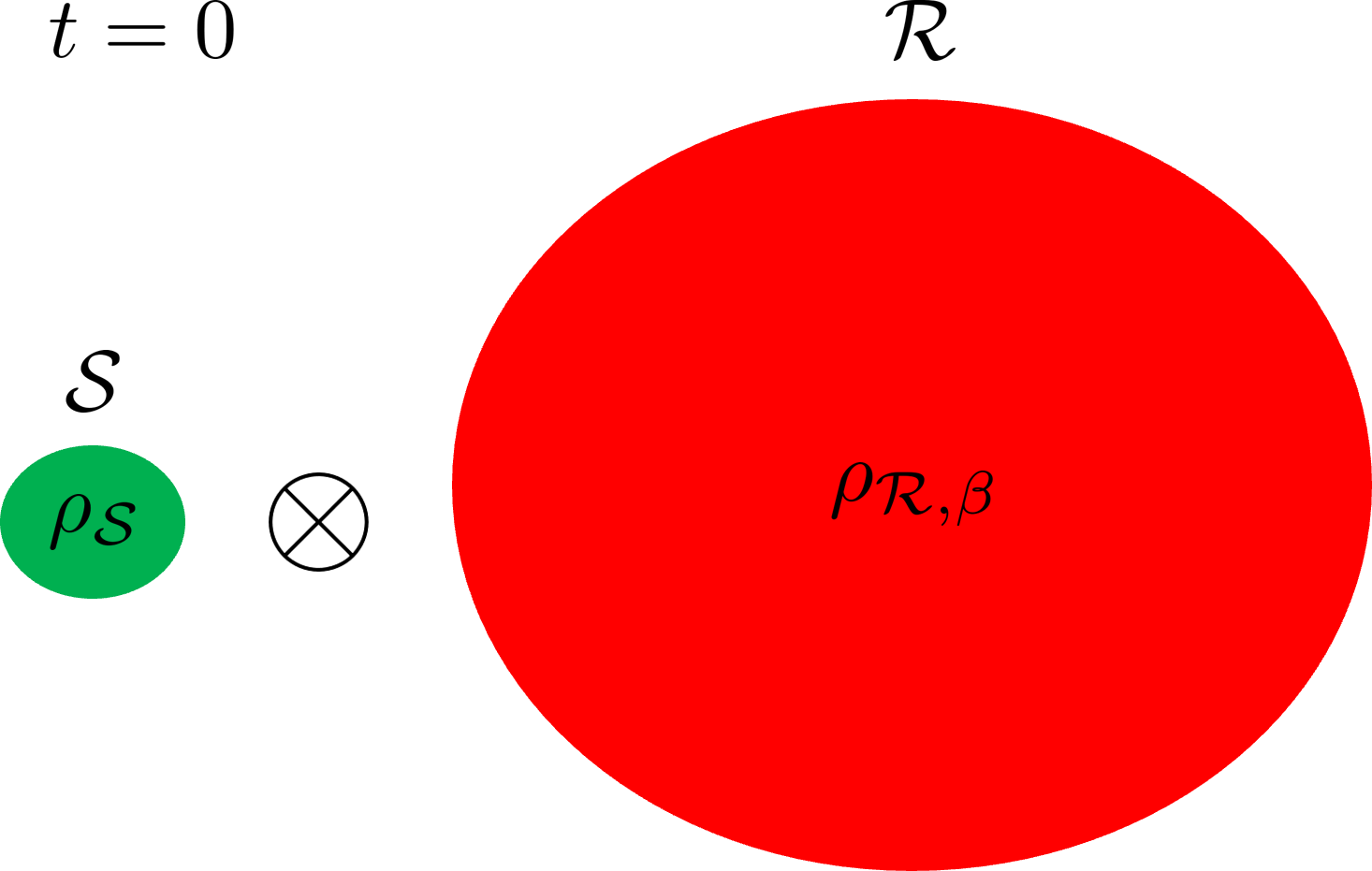}
    \caption{ Pictographic description on the initial state of the system. In the Born approximation the inital state is of the product form.}
    \label{fig:initial}
\end{figure}

\begin{figure}[h!]
    \centering
    \includegraphics[width=0.60 \columnwidth]{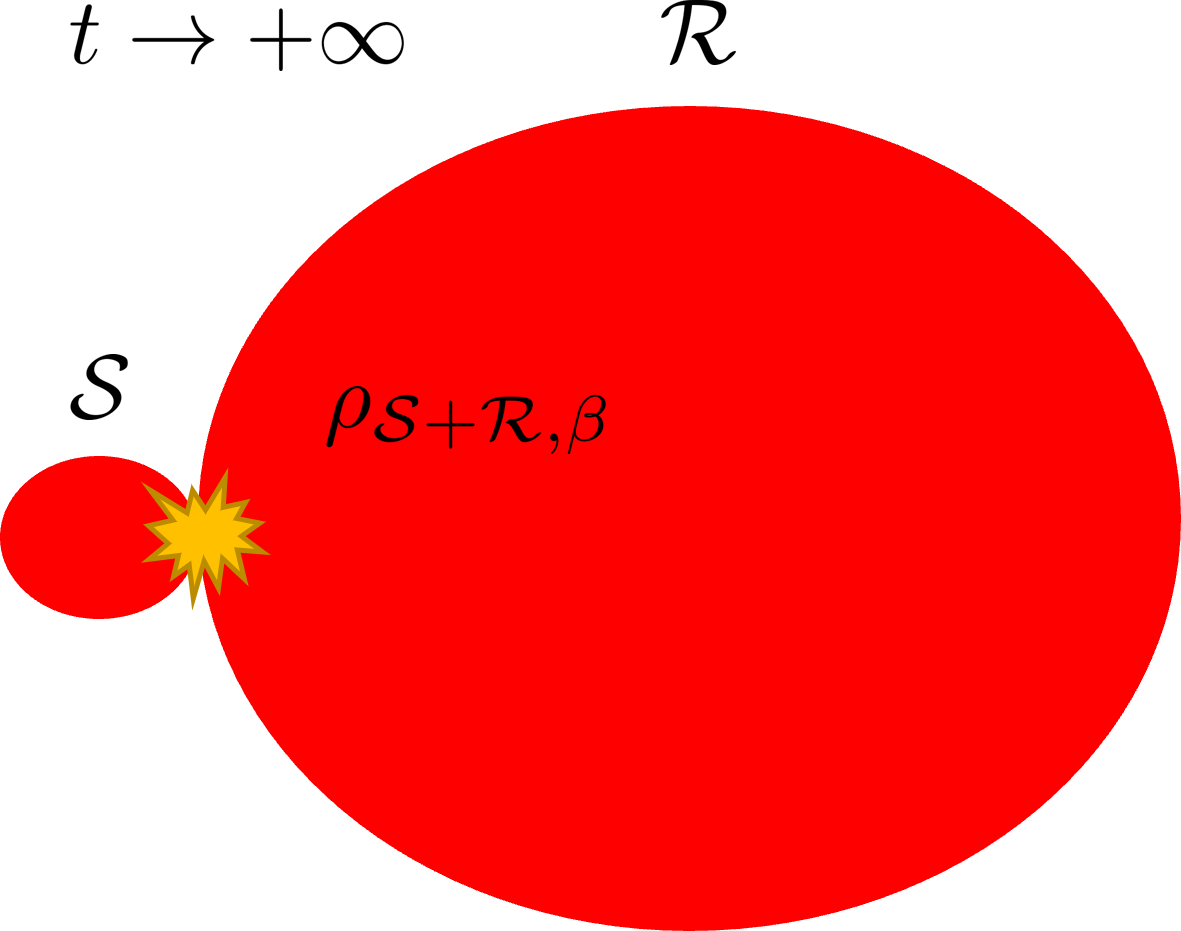}
    \caption{ Pictographic description on the initial state of the stationary state of the system established after long time since the interaction started.
      }
    \label{fig:final}
\end{figure}

Suppose now that at time $t=0$ the system $\mathcal{S}$ and the heat bath $\mathcal{R}$ start to interact (see Figure \ref{fig:initial}). We assume here that the coupling is reasonably weak. Then, after a judiciously long time, a new equilibrium\footnote{Precisely speaking, the state of the total system asymptotically approaches the equilibrium in the sense of locally measurable quantities.} state is established, the form of which is dependent on the interaction. However, because of its size, the reservoir is scarcely affected by the interaction. Therefore, the new equilibrium state is perfectly approximated with KMS (thermal) state $\rho_{\mathcal{S}+\mathcal{R},\beta}$, with the same inverse temperature $\beta$ (see Figure \ref{fig:final}). This approximation is the better, the larger is the heat bath, and the smaller is the system~$\mathcal{S}$. Alternatively, in this situation, one can think of system $\mathcal{S}$ as a perfect thermometer and $\mathcal{R}$ as an examined object.

One can ask now {\it ``what is the reduced state of the system $\mathcal{S}$?"}. The most obvious, and apparently almost correct, answer to this question is  {\it ``thermal state with inverse temperature $\beta$, i.e., $\rho_{\mathcal{S},\beta}$"}, however, this is not enough. Because the system $S$ is small, the interaction with a much larger system $\mathcal{R}$ can alter its properties, e.g., energy levels. In this way, a more elaborate question arises on {\it if the reduced state of systems $\mathcal{S}$ is truly a Gibbs (thermal) state with inverse temperature $\beta$ then with respect to which Hamiltonian and what is its relation to the bare one?}. Finding the adequate Hamiltonian is, therefore, a matter of proper renormalization procedure. This task is a goal of the forthcoming Sections.

\section{The setup}\label{sec:setup}

In order to show and justify the idea of the renormalization procedure proposed by us, we consider the most basic, newbie training, yet nontrivial setup. In this way, we consider small, open quantum system $\mathcal{S}$ interacting with large quantum reservoir $\mathcal{R}$ in a KMS state that we call the heat bath interchangeably. At this place, we would like to specify that by ``small," we mean a system with a discrete spectrum, e.g., a qubit, an atom or a molecule, and by ``large," we mean a reservoir with an infinite number of degrees of freedom that can be modeled by a large, i.e., numerous (ideally infinite), collection of non-interacting quantum harmonic oscillators (modes), e.g. electromagnetic, phonon or fermionic field.

We denote the states of any system $\mathcal{A} \in \left\{\mathcal{S},\mathcal{R},\mathcal{S}+\mathcal{R}\right\}$, with $\rho_\mathcal{A} \in \mathcal{B}(\mathcal{H}_\mathcal{A})$. Additionally, subscript $\beta$ in $\rho_{\mathcal{A},\beta}$, referring to inverse temperature $\beta$, stands for a system $\mathcal{A}$ being in the KMS (thermal) state defined with:
\begin{align}
    &\rho_{\mathcal{A},\beta} := Z_{\mathcal{A}}^{-1} e^{-\beta H_\mathcal{A}}, \label{eqn:def_thermal1}\\
    &Z_{\mathcal{A}} : = \tr \left[e^{-\beta H_\mathcal{A}}\right],\label{eqn:def_thermal2}
\end{align}
where $Z_{\mathcal{A}}^{-1}$ is called the partition function.

Let, $H_\mathcal{S}^{(0)}$ be the free (also called bare) Hamiltonian of the system $\mathcal{S}$, and $H_\mathcal{R}$ be the Hamiltonian of the reservoir $\mathcal{R}$. Further, we chose the interaction Hamiltonian $H^{(0)}_\mathcal{I}$ to have the following standard form:
\begin{align}
    H^{(0)}_\mathcal{I}= \sum_i S_i \otimes R^{(0)}_i,
\end{align}
with $S_i$, and $R^{(0)}_i$ being self-adjoint operators. For simplicity we assume that $S_i$, and $R^{(0)}_i$ are time independent (this assumption will be relaxed later in Section \ref{sec:cumulant}). Moreover we assume that the diagonal elements of interaction operators $S_i$ are zero (See Remark \ref{rem:Polaron}). We also assume that the interaction does not allow for any dark (meta-stable) states that are decoupled, so that unique equilibrium state can be reached. In fact, this is the most physically relevant case as the higher-order contributions to the interaction finally reproduce the ergodicity. Additionally, we make a technical assumption on the spectral density of the reservoir. Namely, we assume that the spectral density of the reservoir does not grow exponentially for large frequencies.

Finally, we can write down the Hamiltonian of the total, closed system $\mathcal{S}+\mathcal{R}$:
\begin{align}\label{eqn:Htot}
    H_{\mathcal{S}+\mathcal{R}}=H^{(0)}_{\mathcal{S}}+H_{\mathcal{R}}+ \lambda H^{(0)}_\mathcal{I},
\end{align}
where $\lambda=1$ is an auxiliary object, that will prove its usefulness in enumerating powers of the coupling constant in perturbative expansions.

\begin{remark}\label{rem:centering}
Through the manuscript we do not follow (unless specified otherwise) the usual assumption of $R^{(0)}_i$ being centered with respect to the state of the reservoir, i.e.,  $\left<R^{(0)}_i\right>_{\rho_{\mathcal{R},\beta}}=0$\footnote{We follow a notation in which $\left<A\right>_\rho=\tr \left[A \rho \right].$}. As we will show centering of $R^{(0)}_i$ operators is the first step of the renormalization.
\end{remark}

\begin{remark}\label{rem:Polaron}
For the sake of simplicity, we assume that the diagonal elements of $S_i$ vanish (in the measurable energy eigenbasis). This ensures us that close to equilibrium the ``mean force" that the system $\mathcal{S}$ exerts on the reservoir $\mathcal{R}$ vanishes. This is the mirrored situation to the ``centering" of reservoir operators described in Remark \ref{rem:centering}. An opposite case in which diagonal elements of $S_i$ do not vanish should be treated, for example, with the polaron transform.  
\end{remark}

\section{Case study: Davies-GKSL equation}\label{sec:MME}

In this Section, we consider a situation in which one is interested in the time evolution of the open system $\mathcal{S}$ in the setup described in Section \ref{sec:setup}. However, our focus is concentrated much more on the stationary state (if it exists) obtained at the end of the evolution, that is, when~$t \to +\infty$.   

One of the possibilities is to make use of the Davies-GKSL equation (Markovian master equation in secular approximation) that is known to be extremely successful in describing the dynamics of open quantum systems in weak coupling regime~\cite{davies1974markovian, gorini1976completely, lindblad1976generators}.
Due to the secular approximation performed in the microscopic derivation of the Davies-GKSL equation, the validity of its solution is limited to the time scales that are not too short. Nonetheless, the stationary state, obtained for very long times, should be devoid of this limitation and therefore correctly determined.

\subsection{The stationary state - the first step of renormalization}\label{sec:subsec:1stOrderRenormalizationMME}

In the Davies-GKSL equation that is derived under Born approximation the initial state is assumed to be a product $\rho_\mathcal{S} \otimes \rho_{\mathcal{S},\beta}$. Then, provided $H^{(0)}_\mathcal{I}$ is such that quantum dynamical semigroup generated by Davies-GKSL equation is ergodic~\cite{Spohn_1977,Fagnola_2001}, and operators $R^{(0)}_i$ are centered, i.e.,  $\left<R^{(0)}_i\right>_{\rho_{\mathcal{R},\beta}}=0$, there exists the unique stationary state~\cite{Breuer+2006}. The latter condition, is however not satisfied a priory in our case (see Remark \ref{rem:centering}).  

Still, one can recover the standard approach to the Davies-GKSL equation if prior to the derivation we reshuffle terms in equation (\ref{eqn:Htot}) so that the total Hamiltonian $H_{\mathcal{S}+\mathcal{R}}$ stays unchanged:
\begin{align}
    &H^{(0)}_\mathcal{S} \longrightarrow H^{(1)}_\mathcal{S}=H_\mathcal{S}^{(0)}+\sum_i S_i  \left<R^{(0)}_i\right>_{\rho_{\mathcal{R},\beta}},\label{eqn:Davies1stOrderA}\\
    &H^{(0)}_\mathcal{I} \longrightarrow H^{(1)}_\mathcal{I} =H_\mathcal{I}^{(0)}-\sum_i S_i \left<R^{(0)}_i\right>_{\rho_{\mathcal{R},\beta}}.\label{eqn:Davies1stOrderB}
\end{align} 
In this way the above transformation defines the centered operators $R^{(1)}_i$, and centered interaction Hamiltonian
\begin{align}
    &R^{(1)}_i=R^{(0)}_i-\left<R^{(0)}_i\right>_{\rho_{\mathcal{R},\beta}},\\
    &H^{(1)}_\mathcal{I}= \sum_i S_i \otimes R^{(1)}_i,
\end{align}
Additionally $H^{(1)}_\mathcal{S}$ is now the new (first-order renormalized) Hamiltonian of the system $\mathcal{S}$, that defines the stationary state 
\begin{align}
    &\rho^{(1)}_{\mathcal{S},\beta} = {Z^{(1)}_\mathcal{S}}^{-1} e^{-\beta H^{(1)}_\mathcal{S}}, ~~Z^{(1)}_\mathcal{S} = \tr \left[e^{-\beta H^{(1)}_\mathcal{S}}\right],\label{eqn:MMEstationary}
\end{align}
of the Davies-GKSL equation that following the standard miscroscopic derivation~\cite{Breuer+2006} has the form (in Schr{\"o}dinger picture):
\begin{align}
    &\frac{d}{dt} {\rho}_\mathcal{S}(t)=L {\rho}_\mathcal{S}(t) \\
    &=-i \left[H_\mathcal{S}^{(1)}+\lambda^2 H_{LS},{\rho}_\mathcal{S}(t)\right]+\lambda^2 \mathcal{D}({\rho}_\mathcal{S}(t)), \label{eqn:MME}
\end{align}
where $L$ is the generator of the quantum dynamical semigroup of the Davies-GKSL equation and $\mathcal{D}({\rho}_\mathcal{S}(t))$ is a component of the Davies-GKSL equation so-called the dissipator defined with:
\begin{align}
    &\mathcal{D}({\rho}_\mathcal{S}(t))=
    \sum_{\omega_1} \sum_{ij} \gamma_{ij}(\omega_1) \nonumber\\
    &\times\left(S_j(\omega_1){\rho}_\mathcal{S}(t)S_i^\dagger(\omega_1) -\frac{1}{2}\left\{S_i^\dagger(\omega_1)S_j(\omega_1),{\rho}_\mathcal{S}(t)\right\}\right). \label{eqn:Davies_dissipator}
\end{align}
In the above sum, $\{\omega_1\}$'s are the Bohr frequencies of the system $\mathcal{S}$ calculated with respect to the Hamiltonian $H^{(1)}_\mathcal{S}$\footnote{In the manuscript, we assimilate a notation in which Bohr frequencies of Hamiltonians $H_\mathcal{S}^{(k)}$ are denoted $\omega_k$, for $k=0,1,2$.}. Furthermore, 
\begin{align}
    &H_{LS}= \sum_{\omega_1} \sum_{ij} \mathcal{S}_{ij}(\omega_1) S_i^\dagger(\omega_1)S_j(\omega_1), \label{eqn:app:MME:HLS}
\end{align}
is the Lamb-Stark shift Hamiltonian, and
\begin{align}
\gamma_{ij}(\omega) = \Gamma_{ij}(\omega)+\Gamma^*_{ji}(\omega), \label{eqn:MarkovianRelaxationRate}
\end{align}
are the Markovian relaxation rates. The above objects can be calculated with
\begin{align}
    &\mathcal{S}_{ij}(\omega) = \frac{\Gamma_{ij}(\omega)-\Gamma^*_{ji}(\omega)}{2i}, \label{eqn:def:coe1}\\
    &\Gamma_{ij}(\omega)=\int_0^{+\infty}ds~e^{i\omega s} \left<\tilde{R}^{(1)}_i(s)R^{(1)}_j\right>_{\rho_{\mathcal{R},\beta}},\label{eqn:def:coe2}\\
    % &R^{(1)}_i=R^{(0)}_i-\left<R^{(0)}_i\right>_{\rho_{\mathcal{R},\beta}},\\
    &S_i(\omega_1) = \sum_{\epsilon^{\prime}_1-\epsilon_1=\omega_1} \Pi(\epsilon_1) S_i\Pi(\epsilon^\prime_1),\label{eqn:def:coe3}
\end{align}
where $\Pi(\epsilon_1)$ are projections into eigenspaces of $H^{(1)}_\mathcal{S}$ associated with eigenvalues $\{\epsilon_1\}$ (see Section \ref{sec:app:Properties} for more details).

The above result, despite its triviality, can be seen as the first step of renormalization. Indeed, the relaxed and recovered later on assumption (see Remark \ref{rem:centering}) is very often satisfied from the very beginning. On the other hand, the interaction changes the properties of the system $\mathcal{S}$, in such a way that the new, first-order renormalized Hamiltonian $H^{(1)}_\mathcal{S}$ must be considered in order to describe its dynamics and stationary state of Davies-GKSL equation correctly.

Suppose now that the Davies-GKSL equation correctly predicts the reduced state of the system $\mathcal{S}$ after the equilibrium is reached. Then the above considerations, on the contrary to their triviality, show that the question stated in Section \ref{sec:intuition}, about Hamiltonian with respect to which the reduced state of the system $\mathcal{S}$ is a Gibbs (thermal) state is nontrivial itself. This nontriviality is because one can wrongly associate the reduced state of the system $\mathcal{S}$ with $H^{(0)}_\mathcal{S}$ instead of $H^{(1)}_\mathcal{S}$, if the analysis of the form of interaction is skipped. This is because the interaction Hamiltonian $H^{(1)}_\mathcal{I}$ determines the reduced state of the system $\mathcal{S}$ after the equilibrium is reached. In the next Section, we discuss if another steps of the renormalization are required for the dynamical equation to satisfy thermodynamical consistency.

\subsection{Lamb-Stark shift term - the next renormalization step}\label{sec:sub:LSterm}

Let us analyze in this place one more element of the Davies-GKSL equation. The term including $H_{LS}$ is usually referred to as a term including Lamb (vacuum) and Stark (reservoir's state-dependent) shifts that lead to the renormalization of eigenstates of $H^{(1)}_\mathcal{S}$. For the sake of the clarity of the presentation, let us assume in this Section that the quantities in equations \eqref{eqn:MarkovianRelaxationRate}-\eqref{eqn:def:coe1} can be written in the following form\footnote{The state of the reservoir is still the KMS state, that is in this case a quasi-free gauge-invariant state completely characterized by the occupation numbers.}:
\begin{align}
    &\gamma_{ij}(\omega) = J_{ij}(\omega)\left(N_\beta(\omega)+1\right), \label{eqn:MarkovianRelaxationRateFourier}\\
    &\mathcal{S}_{ij}(\omega)= \dashint_{0}^{\omega_c} d\Omega~J_{ij}(\Omega) \left(\frac{N_\beta(\omega)+1}{\omega-\Omega}+\frac{N_\beta(\omega)}{\omega+\Omega}\right),\label{eqn:def:coe1Fourier}
\end{align}
where $J_{ij}$ is the so-called spectral density function, $N_\beta$ is the Bose–Einstein distribution distribution, $\omega_c$ is the cut-off frequency and $\dashint$ denotes the Cauchy's principal value of the integral. Nonetheless, the above assumption is valid for the majority of the situations considered in the literature.

However, the presence of the Lamb-Stark shift term in equation \eqref{eqn:MME} leads to the inconsistency because dissipation governs by other terms (the dissipator) in equation (\ref{eqn:app:MME:HLS}) is assumed to happen with respect to energy levels of the open system $\mathcal{S}$ that are not renormalized (not shifted). Moreover, only for the case of negligible small Lamb-Stark shift associated with $H_{LS}$ the equation \eqref{eqn:MME}  is consistent with thermodynamics. Namely, for the thermal bath  at the inverse temperature $\beta$,  due to KMS condition for the correlation functions $\left<{R}^{(1)}_i(s)R^{(1)}_j\right>_{\rho_{\mathcal{R},\beta}}$, the additional relation holds:
\begin{align}
    \gamma_{ij}(-\omega) = e^{-\beta\omega}\gamma_{ji}(\omega).
\end{align}
Therefore, the  dynamics in equation \eqref{eqn:MME} drives 
the open quantum system $\mathcal{S}$ to the Gibbs state defined by the bare Hamiltonian $H_\mathcal{S}^{(1)}$ and not, as expected on the thermodynamic ground, the (alleged\footnote{As we show later, the interpretation of the Lamb-Stark shift term is (at least) questionable.}) physical one containing the Lamb-Stark shift $H_\mathcal{S}^{(1)}+H_{LS}$.

To discuss the problem of  Lamb-Stark shift computation and its relevance for the real system, we consider a bath consisting of many harmonic oscillators such that  
\begin{align}
&H_\mathcal{I}^{(1)}= \lambda~S \otimes \sum_{k} \left( \bar{g}(k) b_k + g(k) b^{\dagger}_k \right), \\
&\quad  H_\mathcal{R} =  \sum_{k} \omega(k) b^{\dagger}_k b_k .
\label{oscillator_bath}
\end{align}
While for a finite number of bath's oscillators, the total  Hamiltonian  $H_{\mathcal{S}+\mathcal{R}} =  H_\mathcal{S} + H_\mathcal{R} + H_\mathcal{I}$ is well-defined, the models with their infinite number, in particular, the infinite volume baths need special attention (cf. reference \cite{Rivas_2010}). Namely, it is well-known  that in those physically important cases, the following additional conditions (written in the continuous modes notation)
\begin{equation}
\int  |g(k)|^2  \omega(k)^{-2} \, dk < \infty
\label{ham_existence}
\end{equation}
must be fulfilled to yield a self-adjoint and possess a ground state  Hamiltonian $H_{\mathcal{S}+\mathcal{R}}$. For realistic models with physically justified dispersion $\omega(k)$, this means a suitable behavior of the coupling magnitude $|g(k)|$  at low-frequency modes (avoiding ``infrared catastrophe")\footnote{For example, a very popular Ohmic coupling is excluded.} and its strong enough suppression for high frequencies usually characterized by the ``ultraviolet cut-off" frequency~$\omega_c$.

Importantly, as long as $\omega < \omega_c$, the values of the relaxation rates $\gamma_{ij}(\omega)$ do not depend
on the cut-off, while the Lamb shift is strongly dependent on it and typically diverges with $\omega_c\to \infty$ (see equations \eqref{eqn:MarkovianRelaxationRateFourier}- \eqref{eqn:def:coe1Fourier}).

In principle, the Lamb-Stark shift effect could be ignored if (i) it is proportional to $\lambda^n$, with $n>2$ (Born approximation) or (ii) it is just small (cf. reference \cite{Rivas_2017}). Nonetheless, (i) does not hold because the term with $H_{LS}$ is proportional to $\lambda^2$ exactly the same as the dissipator, and (ii) does not hold in general (on the contrary to fully relativistic field-theoretic calculations). Moreover, if the spectral density of the reservoir $\mathcal{R}$ does not contain a natural cut-off $\omega_c$, just as in an extremely important case of electromagnetic field, then $H_{LS}$ contains integrals with irremovable divergences. This divergence makes the whole Davies-GKSL is ill-defined\footnote{If any auxiliary cut-off that regularizes the dynamical equation is introduced, then the resulting dynamics shall not depend on the value of its parameters.}. In this case, one of the solutions encountered in the literature is not to include the Lamb-Stark shift term, which is (as we show) equivalent to stating that $H^{(1)}_\mathcal{S}$ is already renormalized, physically relevant, and measurable Hamiltonian of the system $\mathcal{S}$. In this way, self-consistency is recovered since the correct (renormalized) Bohr frequencies enter the dissipator. On the other, this drawback can be fixed using the derivation in the interaction picture with respect to the already renormalized physical Hamiltonian. The justification for the equivalence between the two renormalization methods will be presented in the rest of this manuscript.

To illustrate the physical meaning of $\omega_c$, we consider two examples of qubits  (fermionic oscillators) interacting with bosonic fields and present the relevant frequency scales.

\begin{enumerate}
    \item Atom  interacting with electromagnetic field,  $\omega\sim 10^{15}Hz$,  $\abs{\gamma_{ii}(\omega)}\sim 10^9 s^{-1}$ ; $\omega_c$ is not precisely defined, sometimes identified as $\omega_c \sim c/a_0 \simeq 10^{19}Hz$, ($a_0$ - the Bohr radius, $c$ - the speed of light in vacuum).
    \item Superconducting qubit interacting with phonons,  $\omega\sim 5 \cdot 10^{9}Hz$,  $\abs{\gamma_{ii}(\omega)} \sim 10^4 s^{-1}$ ; $\omega_c$ is identified with the Debye frequency $\omega_D\sim 10^{13} Hz $.
\end{enumerate}

In both cases, the separation of relevant frequency (or time) scales,  $\abs{\gamma_{ij}(\omega)} << \omega << \omega_c$,  is by the four to six orders of magnitude what implies the validity of the Markovian approximation in the weak coupling limit approach (cf. reference \cite{Rivas_2017}).  The cut-off frequency inserted into the definition of interaction Hamiltonian makes the corresponding effective theory mathematically consistent and applicable to physical phenomena characterized by time scales much longer than $1/\omega_c$. However, in both representative cases above, the Lamb-Stark shift computed using the renormalized equation is physically meaningless. In the case of an atom interacting with the electromagnetic field, we must refer to the higher-level theory - the quantum electrodynamics, while for the superconducting qubit, we must use phenomenological values because the linear (in $b_k, b^{\dagger}_k$) interaction Hamiltonian is only a low energy approximation, insufficient to account for corrections to bare Hamiltonian generated by higher-order terms. 

Summarizing, the consistent approach to the quantum master equations obtained in the weak coupling limit approach (like the Davies-GKSL equation \eqref{eqn:MME}) involves two steps: 1) determination of the physical Hamiltonian $H_\mathcal{S}$ of the open system $\mathcal{S}$ using either its measured Bohr frequencies, or referring to the renormalization procedure within a higher level, more fundamental theory;  2) determination of the relaxation parameters which can be measured, or computed using the cut-off independent formulas, applied to the low energy effective Hamiltonian $H_\mathcal{S}$.

\begin{remark}
The renormalization procedure for the Lamb -Stark corrections is important not only for a single open system interacting with a reservoir, as it also describes the effects of interaction between many subsystems mediated by a shared reservoir. For example, for an ensemble of two-level ``atoms" localized at different sites and interacting with an electromagnetic field, the total Lamb-Stark correction contains not only (cut-off dependent) individual energy shifts but also two-body (cut-off independent) interactions interpreted as Van der Waals dipole-dipole coupling \cite{Gross_1982,AlickiLendi1987}.
\end{remark}

\section{Partial trace of the total $\mathcal{S}+\mathcal{R}$ system  KMS state}\label{sec:PartialTrace}

We note that the results presented in this section were initially obtained in reference \cite{Lobejko_2022} (see also Ref. \cite{Timofeev_2022}). Still, for the sake of completeness of the paper and due to an alternative derivation technique we do not hesitate to display them. Here, we would like to get back to the idea presented in Section \ref{sec:intuition}. In this place, without any reference to dynamical equations governing the time evolution of the reduced state of the system $\mathcal{S}$ we shall find the reduced state of the system $\mathcal{S}$ provided the state of the total system $\mathcal{S}+\mathcal{R}$ is a KMS state
\begin{align}
    \rho_\mathcal{S} = \tr_\mathcal{R} \left[\rho_{\mathcal{S}+\mathcal{R},\beta}\right] = \tr_\mathcal{R} \left[\frac{e^{-\beta \left(H_\mathcal{S}+H_\mathcal{R}+\lambda H_\mathcal{I}\right)}}{Z_{\mathcal{S}+\mathcal{R},\beta}}\right].
\end{align}
In this way, we formulate the ``static" argument for the justification of the self-consistency condition that we postulate.

In order to achieve our goal, without the loss of generality, we assume that the reduced state of the system $\mathcal{S}$ is of the following form: 
\begin{align}
    &\rho_{\mathcal{S},\beta}^{(mf)} =\frac{e^{-\beta H^{(mf)}_\mathcal{S}}}{Z_\mathcal{S,\beta}}, \label{eqn:partialTr1}\\
    &H^{(mf)}_\mathcal{S} = H_\mathcal{S}^{(0)} + \sum_{k=1}^{\infty}\lambda^k H_{mf,\mathcal{C}}^{(k)}.
\end{align}
Indeed, the reduced state of the system $\mathcal{S}$ should only depend on the bare Hamiltonian $H_\mathcal{S}^{(0)}$, the interaction Hamiltonian $H_\mathcal{I}^{(0)}$, and the state of the reservoir $\rho_{\mathcal{R},\beta}$. With the form given in equation \eqref{eqn:partialTr1} in the case of no interaction the Gibbs (thermal) state of the system $\mathcal{S}$ with respect to $H_\mathcal{S}^{(0)}$ is correctly recovered.

In the above formula operators, $H_{mf,\mathcal{C}}^{(k)}$ are different orders corrections to the bare Hamiltonian of the system $\mathcal{S}$ that are about to be determined. In order to achieve this task\footnote{Cf. for parallel works \cite{Lobejko_2022,Timofeev_2022} we different techniques are applied.}, we make use of an explicit description of the Zassenhaus formula proposed by T. Kimura in reference~\cite{Kimura_2017} (cf. reference \cite{Goold_2020,Takashi_2008,Hilt_2011_MeanForce,Trushechkin_2021_MeanForce}). Despite the fact that the technique employed by us allows calculating the corrections of any order, we limit ourselves to the first two corrections $H_{mf,\mathcal{C}}^{(1)}$, $H_{mf,\mathcal{C}}^{(2)}$.  This simplification is because, due to the Born approximation, the dynamical equations considered in this manuscript are limited to the second-order in $\lambda$ order as well. Therefore, a renormalization of $H_\mathcal{S}^{(0)}$ up to the second-order is sufficient for our purposes. 

We find the first two corrections to the bare Hamiltonian $H_\mathcal{S}^{(0)}$ of the system $\mathcal{S}$ to be expressed in terms of interaction operators (see equations (\ref{eqn:def:coe1}-\ref{eqn:def:coe3}))
\begin{align}
    &H_{mf,\mathcal{C}}^{(1)} = \sum_i S_i
    \left<R_i^{(0)}\right>_{ \rho_{\mathcal{R},\beta}}, \label{eqn:PT:cor1}\\
    &H_{mf,\mathcal{C}}^{(2)} = \sum_{{\omega_1},{\omega_1}^\prime} \sum_{ij}  \Upsilon^{(2)}_{ij}({\omega_1},{\omega_1}^\prime) S_i^\dagger({\omega_1})S_j({\omega_1}^\prime),\label{eqn:PT:cor2}\\
    & \Upsilon^{(2)}_{ij}({\omega_1},{\omega_1}^\prime)=\frac{1}{e^{\beta ({\omega_1}^\prime-{\omega_1})}-1}   \left(e^{\beta(\omega_1^\prime-\omega_1)}  \mathcal{S}_{ij}(\omega_1) -\mathcal{S}_{ij}(\omega_1^\prime) 
    \right.\nonumber\\
    &\left.-e^{\beta \omega_1^\prime}\left(\mathcal{S}_{ji}(-\omega_1^\prime)-\mathcal{S}_{ji}(-\omega_1)\right)\right) ,
\end{align}
where the terms for which $\omega^\prime=\omega$ are determined via a limiting procedure $\omega^\prime \to \omega$ 
\begin{align}
     &\Upsilon^{(2)}_{ij}({\omega_1},{\omega_1})=\mathcal{S}_{ij}(\omega_1)\nonumber\\
     & - \frac{1}{\beta} \left(\frac{\partial}{\partial \omega_1} \mathcal{S}_{ij}(\omega_1)+e^{\beta \omega_1}\frac{\partial}{\partial \omega_1} \mathcal{S}_{ji}(-\omega_1)  \right).
\end{align}
It is therefore evident that the mean-force Hamiltonian correction leads to a different prediction for the renormalization of the energy levels than the Lamb-Shift terms present in the Davies-GKSL (see equation \eqref{eqn:MME}) or the Bloch-Redfield equation \cite{cattaneo2019local}. 
%\sout{As we show in Appendix~\ref{sec:app:PT}, for temperatures large enough (small $\beta$) the shift of the energy levels is negligible \cite{Timofeev_2022,Gelzinis_2020}. This agrees with the predictions of the QFT suggesting that at energy scales that are large enough the particles exist as the bare ones. }
\color{black}

The formulas \eqref{eqn:PT:cor1} and \eqref{eqn:PT:cor2} above define the second-order approximation for the Hamiltonian $H^{(mf)}_\mathcal{S}$, and the state after the partial trace (we skip $\lambda=1$ lastly):
\begin{align}
    &H_\mathcal{S}^{(mf)} \approx H_\mathcal{S}^{(mf,2)}=H_\mathcal{S}^{(0)}+H_{mf,\mathcal{C}}^{(1)}+H_{mf,\mathcal{C}}^{(2)}, \\
    & \rho^{(mf)}_{\mathcal{S},\beta}\approx\rho_{\mathcal{S},\beta}^{(mf,2)}=  \frac{e^{-\beta H_\mathcal{S}^{(mf,2)}}}{Z_{\mathcal{S},\beta}^{(mf,2)}} .
\end{align} 
The details of the calculations of the partial trace are present in Section \ref{sec:app:PT} of the Appendix.

In the first correction $H_{mf,\mathcal{C}}^{(1)}$, in the equation \eqref{eqn:PT:cor1}, we recognize the correction that was also incorporated in the standard microscopic derivation of the Davies-GKSL equation (see equations \eqref{eqn:Davies1stOrderA}--\eqref{eqn:Davies1stOrderB} in Section \ref{sec:MME}), namely the centering of the interaction correction. This correspondence certifies that the requirement for the operators $R_i^{(1)}$ to be centered with respect to the state of the heat bath is essential for reproducing the renormalized, physical Hamiltonian $H_\mathcal{S}$.

On the other hand, the presence of the second-order correction $H_{mf,\mathcal{C}}^{(2)} \neq 0$ that we obtained shows that incorporating only the first-order correction to the bare Hamiltonian $H^{(0)}_\mathcal{S}$ is not enough for the stationary state of the Davies-GKSL in equation \eqref{eqn:MMEstationary} to be consistent with the second-order Born approximation and thermodynamics. Additionally, we note that $\left[H_\mathcal{S}^{(1)},H_{mf\mathcal{C}}^{(2)}\right]\neq0$, what means that the interaction indeed changes the properties of the system $\mathcal{S}$. This change has a physical interpretation in terms of the dressing of the bare system $\mathcal{S}$ with the excitations of the reservoir $\mathcal{R}$ that emerge as a result of the interaction.

The incompatibility between the Lamb-Stark shift Hamiltonian $H_{LS}$ and the second-order correction $H_{mf,\mathcal{C}}^{(2)}$ emphasizes the importance of the correct renormalization procedure for the treatment of open quantum systems. This is because the discrepancy mentioned above questions the interpretation of the Lamb-Stark shift term. Indeed, the comparison between $H_{LS}$ and $H_{mf,\mathcal{C}}^{(2)}$ indicates that the Lamb-Stark shift term $H_{LS}$ does not describe accurately the shift of energy levels caused by the interaction with the reservoir. As we show in the next Section \ref{sec:cumulant}  the self-consistency condition fixes the renormalization procedure and removes the controversy described above.  

\begin{remark}
In fact, the Lamb-Stark shift term $H_{LS}$ and the mean-force correction $H_{mf,\mathcal{C}}^{(2)}$ emerge from different perspectives, i.e, from the static and dynamical pictures \cite{Lobejko_2022}. As we suggest, this is the Lamb-Stark shift term $H_{LS}$ that is misinterpreted and does not accurately describe any physical process (except for the coherent/drift part of quantum noise). However, an experiment would be necessary for the final verification of which correction accurately describes the shift of energy levels (see an experimental proposal in ref. \cite{Gramich_2011}).  
\end{remark}
\color{black}

\section{The cumulant equation}\label{sec:cumulant}

In this Section we present the dynamical equation governing the time evolution of an open quantum system, in which the renormalization of the open system Hamiltonian is inalienable. This type of dynamical equations was firstly derived in reference~\cite{Alicki1989}, and has been recently independently discovered by A. Rivas in reference~\cite{Rivas_2017}, where the name {\it refined weak coupling limit} was used (see also~\cite{Majenz_2013}). Still, we use the nomenclature recently introduced in reference~\cite{Winczewski_2021}. However, in the derivation in the reference~\cite{Rivas_2017}, due to different derivation techniques, the need for renormalization is not discussed. Therefore we base on the first derivation in the reference~\cite{Alicki1989}. 

\subsection{The derivation}

The derivation of the cumulant equation in reference~\cite{Alicki1989} is based on different ideas than the derivation of the Davies-GKSL equation, but on the contrary to the latter, the cumulant equation is exact up to the Born approximation that is incorporated in both derivations. For the sake of generality, in the derivation, we relax the assumption on the reservoir state to be fixed to the KMS state. Instead, we assume that its state is stationary, i.e.,  $\left[H_\mathcal{R},\rho_\mathcal{R}\right]=0$. Moreover, we assume that the total Hamiltonian of the open system $\mathcal{S}$, and the reservoir $\mathcal{R}$ before the renormalization is of the following, time-dependent form:
\begin{align}
    &H_{\mathcal{S}+\mathcal{R}}=H_\mathcal{S}^{(0)}(t)+H_\mathcal{R}+\lambda H_\mathcal{I}^{(0)}(t),\\
    &H_\mathcal{I}^{(0)}(t)=  \sum_i S_i \otimes R^{(0)}_i(t).
\end{align}
The relaxation of an assumption on the time-independence for the bare system and interaction Hamiltonians allows for understanding the behavior of the total system $\mathcal{S}+\mathcal{R}$ in the spirit of the quantum field theory (QFT). This kind of treatment is the most adequate in realistic situations in which effects of state preparation and finite time of initiating the interaction have to be considered. The proposed relaxation helps us to bypass the aforementioned issues. Undoubtedly, to correctly understand the renormalization procedure, we have to consider the presence of the dressing processes that can be explained solely by field theoretical (QFT) considerations. At the same time, we assume that the total Hamiltonian of the system $\mathcal{S}+\mathcal{R}$, is approximately time-independent in a manner explained in Section \ref{sec:app:derivationCumulantEquation} of the Appendix, where the details of the derivation are presented.

% \begin{remark}
%     This situation of the cumulant equation being exact is similar to the case of the Bloch-Redfield equation in its most crude version~\cite{Breuer+2006}. Here, by ``crude," we mean the absence of the long time regularization ($t\to+\infty$) for the integrals containing two-point correlation functions.  However, the Bloch-Redfield equation contains one more hidden assumption. Namely, it assumes that the dynamical map governing the time evolution of the open system is associated with a first-order differential equation. This assumption is not present in the derivation of the cumulant equation, which on contrary to the Bloch-Redfield equation, yields completely positive and trace-preserving (CPTP) dynamics for all initial states of the open system~$\mathcal{S}$. 
% \end{remark}

The renormalization procedure enters the derivation of the cumulant equation in a similar manner as it was in the case of the centering of interaction procedure in the derivation of Davies-GKSL equation (see Section \ref{sec:subsec:1stOrderRenormalizationMME}). Namely, we postulate the existence of the following corrections to the bare Hamiltonian $H_\mathcal{S}^{(0)}$ of the system $\mathcal{S}$, and the interaction Hamiltonian $H_\mathcal{I}^{(0)}$, that give rise to the renormalized, physical Hamiltonian of the open system $H_\mathcal{S}$, and the renormalized interaction Hamiltonian $H_\mathcal{I}$. Indeed, we treat here Hamiltonians $H_\mathcal{S}^{(0)}$, and $H_\mathcal{I}^{(0)}$ as auxiliary objects, whereas only the renormalized species, e.g., $H_\mathcal{S}$, have physical relevance and can be empirically determined. This assignment reflects field-theoretical (QFT) that we (almost) never observe the bare particles. Instead, what we observe in the laboratory are the dressed particles evolution of which is governed by effective Hamiltonians.
\begin{align}
    H_\mathcal{S}&=H_\mathcal{S}^{(0)}(t)+H_\mathcal{C}(t), \label{eqn:RenormalizedHamiltonianS}\\
    H_\mathcal{I}&=H_\mathcal{I}^{(0)}(t)-H_\mathcal{C}(t), \label{eqn:RenormalizedInteraction}\\
    H_\mathcal{C}(t)&=\sum_{k=1}^{\infty}\lambda^k H_\mathcal{C}^{(k)}(t),
\end{align}
where for the sake of simplicity we assume now that the renormalized, physical Hamiltonian $H_\mathcal{S}$, is time-independent, yet the generalization is straightforward\footnote{The time-dependent renormalized Hamiltonian $H_\mathcal{S}(t)$ is relevant for example in the case of the system $\mathcal{S}$ interacting with external classical fields.}. However, the above procedure has two major differences to the procedure in Section \ref{sec:subsec:1stOrderRenormalizationMME}, namely here $(i)$ we are interested in more correction than only the first one $H_\mathcal{C}^{(1)}(t)$, $(ii)$ the corrections are not a priory determined with some conditions on the interaction Hamiltonian $H_\mathcal{I}$, and are in fact variables that should be determined in the derivation. At this place we would like to address the reader, not to identify $H_\mathcal{C}^{(1)}(t)$, and $H_\mathcal{C}^{(2)}(t)$ with the corrections in equations \eqref{eqn:PT:cor1}, and \eqref{eqn:PT:cor2}.
%for a while, as the former one is at the moment still an unknown that can be shown to be equal to the later one with the requirement for the self-consistency described below. 

The detailed derivation of the cumulant equation is the content of Section \ref{sec:app:derivationCumulantEquation}, here for the sake of conciseness we discuss the most important steps within the derivation. Firstly, all relevant operators are transformed into interaction picture with respect to the physical, measurable Hamiltonian $H_\mathcal{S}$, that is not determined yet, and the corresponding interaction picture Von Neumann equation for the total system $\mathcal{S}+\mathcal{R}$ is formally solved. Then the reduced density matrix of the open system $\mathcal{S}$ at time $t$ is obtained with performing a partial trace over the reservoir $\mathcal{R}$ system. In this way a family of one-parameter (interaction picture) dynamical maps $\tilde{W}_t$ is obtained
\begin{align}
    \tilde{\rho}_\mathcal{S} (t)= \tilde{W}_t\tilde{\rho}_\mathcal{S} (0) = \tr_\mathcal{R} \left[e^{-it\left[H_{\mathcal{S}+\mathcal{R}},\cdot\right]}\tilde{\rho}_\mathcal{S} (0) \otimes \tilde{\rho}_\mathcal{R} \right],
\end{align}
where the decoration $\tilde{\cdot}$ denotes the interaction picture with respect to the renormalized Hamiltonian $H_\mathcal{S}$.

On the other hand, due to quantum analogs of Central Limit Theorem (CLT), we know that the interaction between system $\mathcal{S}$ and the reservoir $\mathcal{R}$, being a ``large" collection of harmonic oscillators, can be considered a Gaussian (quantum) stochastic process~\cite{Kubo1963,Goderis_1989,Goderis_1990,Accardi_1990,Accardi_2002}. Similarly, as in reference~\cite{Alicki1989} we expand the dynamical map $\tilde{W}_t$ in terms of generalized cumulants~\cite{Kubo1962,Fox1976,Bianucci2020}. We then perform the second-order approximation for $\tilde{W}_t$ (consistent with Born approximation) by truncating the cumulant generating function (series) to only two terms in order to obtain:
\begin{align}
    \tilde{W}_t \approx \tilde{W}_t^{(2)} = \exp \left(\sum_{n=1}^{2} \lambda^n \tilde{K}^{(n)}(t)\right).
\end{align}
% In fact, considering only two first cumulants $\tilde{K}^{(1)}(t)$, and $\tilde{K}^{(2)}(t)$ corresponds to the evolution of the open system $\mathcal{S}$ being guided by the classical Gaussian noise\footnote{For the classical Gaussian noise all higher-order cumulants vanish.}.
In fact, by assimilating the above approximation, we base on examples of open quantum systems being driven by classical Gaussian noise, for which the higher-order cumulants ($n>2$) vanish for the arbitrary magnitude of the coupling constant. Therefore, the form of the dynamics achieved via truncating the cumulant generating function is not exact, still consistent with the Born approximation. Nevertheless, the dynamics is CPTP as the well-known mathematical structure of the generator in the GKSL form is recovered. Because, we are interested in the dynamics correct solely up to the second-order approximation, from now we identify $\tilde{W}_t^{(2)}$ with $\tilde{W}_t$, and $H_\mathcal{S}^{(2)}$ with $H_\mathcal{S}$ for the ease of notation.

In order to find the cumulant equation, we formulate and apply the following requirement for the self-consistency of the reduced dynamics of an open quantum system.\\
~~\\
{\it In the interaction picture with respect to the physical (renormalized) Hamiltonian $H_\mathcal{S}$, the reduced dynamics of the system $\mathcal{S}$ should be purely dissipative, i.e., it should not include any Hamiltonian-like terms.}\\
~~\\
The above stated condition allows to determine both unknown corrections, $\tilde{H}_\mathcal{C}^{(1)}(t)$, $\tilde{H}_\mathcal{C}^{(2)}(t)$ and consequently $H_\mathcal{S}^{(2)}$, by using them as counterterms to any Hamiltonian-like terms appearing at each order. In the first-order (with respect to $\lambda$), we have:
\begin{align}
    &\tilde{K}^{(1)}(t)=0,\\
    &{H}_\mathcal{C}^{(1)}(t) = \sum_i {S}_i \left<\tilde{R}_i^{(0)}(t)\right>_{{\rho}_\mathcal{R}},
\end{align}
where we recognize the first correction ${H}_\mathcal{C}^{(1)}(t)$, to be the centering of interaction correction. Indeed, if we consider the bare interaction to be time-independent and the reservoir to be in KMS state, then  ${H}_\mathcal{C}^{(1)}(t)$ has identical form as the corrections in Section~\ref{sec:MME}, and Section~\ref{sec:PartialTrace}. Then, in the second-order, we obtain:
\begin{align}
    &\tilde{K}^{(2)}(t) \tilde{\rho}_\mathcal{S} (0) = \sum_{i,j} \sum_{\omega_2, \omega_2^\prime} \gamma_{ij}(\omega_2,\omega_2^\prime,t) \nonumber\\
    &\times \left(S_i (\omega_2) \tilde{\rho}_\mathcal{S} (0) S_j^\dagger (\omega_2^\prime) - \frac{1}{2} \left\{S_j^\dagger (\omega_2^\prime) S_i (\omega_2), \tilde{\rho}_\mathcal{S} (0) \right\} \right), \label{eqn:TheSecondCumulant}\\
    &{H}_\mathcal{C}^{(2)} (t)= \sum_{ij}\sum_{\omega_2,\omega_2^\prime}   \frac{\Gamma_{ji}^{(t)}(\omega_2)-{\Gamma_{ij}^{(t)}}^*(\omega_2^\prime)}{2i}  S_j^\dagger (\omega_2^\prime) S_i(\omega_2) ,   \label{eqn:Cumulant2ndCorrection}
\end{align}
where $\omega_2$ are the Bohr frequencies of the renormalized Hamiltonian $H_\mathcal{S}^{(2)}$, so that the dissipation proceed with respect to renormalized energy levels (of the dressed system), and
\begin{align}
    &\gamma_{ij} (\omega,\omega^\prime,t) = \int_0^t ds \int_0^t dw~ e^{i (\omega^\prime s - \omega w)} \left< \tilde{R}^{(1)}_j (s) \tilde{R}^{(1)}_i (w) \right>_{\tilde{\rho}_\mathcal{R}},\label{eqn:Cumulant_Matrix}\\
    &\Gamma_{ji}^{(t)}(\omega)= \int_0^t du~  e^{i\omega u} \left<\tilde{R}^{(1)}_j (u) {R}^{(1)}_i \right>_{\tilde{\rho}_{\mathcal{R}}}.
\end{align}
We note here, that the superoperator $\tilde{K}^{(2)}(t)$ is of GKSL form, as $\gamma_{ij} (\omega,\omega^\prime,t)$ are elements of a positive semidefinite matrix. Therefore, the dynamics obtained with the cumulant equation is CPTP. %Furthermore, it is evident that the renormalization procedure is not equivalent to any trivial symmetries of a generator in GKSL form \cite{Breuer+2006}.

The cumulant superoperator $\tilde{K}^{(2)}(t)$ possesses solely the dissipative part that incorporates the Bohr frequencies of the (second-order) renormalized Hamiltonian $H^{(2)}_\mathcal{S}$. As the dissipation proceeds, starting from $t=0$, with respect to the renormalized Bohr frequencies, the renormalized equations reflect a situation in which all dressing processes took place in the distant past. Consequently, the initial conditions for the state in the product form can be associated with the partition between the dressed system and the reservoir.

Finally, we can write the cumulant equation in the interaction picture:
\begin{align}
    &\tilde{\rho}_\mathcal{S} (t)=e^{ \tilde{K}^{(2)}(t)}\tilde{\rho}_\mathcal{S} (0),
\end{align}
where transformation to the Schr{\"o}dinger picture is non-trivial (in contrast to the case of Davis-GKSL equation), because superoperators $-it\left[H_\mathcal{S},\cdot\right]$ and $\tilde{K}^{(2)}(t)$ do not commute. The (approximate) form of the cumulant equation in the Schr{\"o}dinger picture is discussed in Section \ref{sec:app:subsec:SchroedingerPicture} of the Appendix. 

The second correction $\tilde{H}_\mathcal{C}^{(2)}(t)$ to the Hamiltonian of the open system $S$ in equation \eqref{eqn:Cumulant2ndCorrection}, is time-dependent, and differs from the correction in equation \eqref{eqn:PT:cor2} (see Section \ref{sec:PartialTrace}). Indeed, the second correction in the cumulant equation is dynamical as it is a counterterm for the (i) dressing processes (some part) caused by the initial state not being compatible with interaction and (ii) finite time of switching on the interaction. This incompatibility is caused by the Born approximation, which imposes the initial state to be in the form of a product\footnote{See, Section \ref{sec:app:derivationCumulantEquation} for the discussion how the renormalization (informally) shifts the dressing of the system to the distant past.}. In order to compare $\tilde{H}_\mathcal{C}^{(2)}(t)$ in equation \eqref{eqn:Cumulant2ndCorrection} with the correction $\tilde{H}_{mf,\mathcal{C}}^{(2)}(t)$ in equation \eqref{eqn:PT:cor2}, we firstly transform the former to the Schr{\"o}dinger picture, and then perform the long time limit.
\begin{align}
      {H}_\mathcal{C}^{(2)}&=\lim_{t \to +\infty} {H}_\mathcal{C}^{(2)} (t) \\
      &= \sum_{ij}\sum_{\omega_2,\omega_2^\prime}  \frac{\Gamma_{ij} (\omega_2^\prime)-\Gamma_{ji}^*(\omega_2)}{2i}   S_i^\dagger (\omega_2) S_j(\omega_2^\prime). \label{eqn:Cumulant2ndCorrectionAsymptotic} 
\end{align}
If we now consider the reservoir to be in the KMS state with the inverse temperature $\beta$, and 
\begin{align}
    \lim_{t \to +\infty} H_\mathcal{S}^{(0)}(t)=H_\mathcal{S}^{(0)}
\end{align}
coincide with the bare Hamiltonian in equation \eqref{eqn:Htot}, then in formula \eqref{eqn:Cumulant2ndCorrectionAsymptotic} we obtain an agreement with the form of Lamb-Stark shift in reference~\cite{cattaneo2019local} (see Section \ref{sec:app:subsec:detailed}). 

Importantly, the second order ''dynamical" correction $\tilde{H}_\mathcal{C}^{(2)}$ obtained in the derivation of the cumulant equation does not match with the second-order correction derived for the ''static" mean-force Hamiltonian $\tilde{H}_{mf,\mathcal{C}}^{(2)}$. In particular, 
\begin{align}
    &\mathrm{diag}_1\left(H_{mf,\mathcal{C}}^{(2)}-H_\mathcal{C}^{(2)}\right) \stackrel{\mathcal{O}(\lambda^4)}{\approx} \sum_{\omega_{1} } \sum_{ij} S_i^\dagger (\omega_{1}) S_j(\omega_{1})\nonumber\\ 
    & \times-\frac{1}{\beta} \left(\frac{\partial}{\partial \omega} \mathcal{S}_{ij}(\omega_1)+e^{\beta \omega_1}\frac{\partial}{\partial \omega}  \mathcal{S}_{ji}(-\omega_1)  \right) ,
\end{align}
where $\mathrm{diag}_1(\cdot)$ represents the projection on the diagonal part with respect to $H_\mathcal{S}^{(1)}$ eigenbasis (see Section \ref{sec:app:subsec:detailed} for details). As we observe the correction $H_{mf,\mathcal{C}}^{(2)}$ and $H_\mathcal{C}^{(2)}$ are different from each other. The $H_{mf,\mathcal{C}}^{(2)}$ possesses an additional term that explicitly depends on the temperature of the reservoir. The discrepancy described above strengthens the argument for the need of the renormalization. This is because, provided the partial trace approach in Section \ref{sec:app:PT} correctly predicts the connection between the bare $H_\mathcal{S}^{(0)}$ and physical Hamiltonian $H_\mathcal{S}$, the Lamb-Stark shift terms appears to be an artificial object, not describing accurately any physical process, as for the description of dressing an additional term seems to be required.  

% Furthermore, in the zeroth-order\footnote{Here, by zeroth-order approximation we mean that there might exist a second-order correction $\delta H(t)$ to $H_\mathcal{S}$ that describes the long-time dynamics of the cumulant equation more accurately.} approximation the cumulant equation possesses a stationary state that is the Gibbs (thermal) state with respect to the renormalized Hamiltonian $H_\mathcal{S}^{(2)}$:
Furthermore, it can be predicted that the cumulant equation possesses a stationary state that is the Gibbs (thermal) state with respect to the (second-order) renormalized  Hamiltonian $H_\mathcal{S}^{(2)}$
\begin{align}
    &\rho^{(2,ss)}_{\mathcal{S}} = \frac{e^{-\beta H_\mathcal{S}^{(2)}}}{Z^{(2)}_{\mathcal{S},\beta}}, 
    \\ &H_\mathcal{S}^{(2)}=H_\mathcal{S}^{(0)}+H_\mathcal{C}^{(1)}+H_\mathcal{C}^{(2)}.
\end{align}
The above claim on the stationary state of the cumulant equation can be explained most easily, with observing that for long times the cumulant $\tilde{K}^{(2)}(t)$, resembles the generator of a (interaction picture) quantum dynamical semigroup $\tilde{L}$ in the following sense (see relation in equation \eqref{eqn:gammaAsymptotic} in Section \ref{sec:app:derivationCumulantEquation} of the Appendix)
\begin{align}
    \lim_{t \to \infty} \frac{1}{t} \tilde{K}^{(2)}(t)= \tilde{L}.
    \label{eqn:cumulant_superoperator_Davies_limit}
\end{align}
Therefore, in the long-times limit $\tilde{K}^{(2)}(t)$, approximates the Davies-GKSL equation~\cite{AlickiLendi1987}, with the correct renormalization (as a consequence of the interaction picture with respect to $H_\mathcal{S}$)
\begin{align}
    &\tilde{\rho}_\mathcal{S}(t)=e^{t\tilde{L}}\tilde{\rho}_\mathcal{S}(0),\\
    &\tilde{L}\tilde{\rho}_\mathcal{S}(0)=  \sum_{i,j} \sum_{\omega_2} \gamma_{ij}(\omega_2) \nonumber\\
    &\times \left(S_i (\omega_2) \tilde{\rho}_\mathcal{S}(0) S_j^\dagger (\omega_2) - \frac{1}{2} \left\{S_j^\dagger (\omega_2 S_i (\omega_2), \tilde{\rho}_\mathcal{S} (0) \right\} \right).
\end{align}
The stationary state of the above equation is a Gibbs state with respect to $H_\mathcal{S}^{(2)}$, that coincides with the predicted stationary state of the cumulant equation.

The above argumentation reproduces the results described in reference \cite{Rivas_2017} for the long-time limit of the dynamics of the cumulant equation (up to renormalization). Still, the procedure described above is not consistent (and fully justified) with the second-order Bohr approximation. This is because the limiting procedure in equation \eqref{eqn:cumulant_superoperator_Davies_limit} suppresses any constant or oscillatory terms (contributing to second-order dynamics) in the cumulant superoperator $\tilde{K}^{(2)}(t)$.

Surprisingly, it can be shown that $\rho^{(2,ss)}_{\mathcal{S}}$ is indeed the stationary state of the cumulant equation \cite{MW_preparation}. Moreover, the pace of convergence to the stationary state is not faster than $t^{-1}$. The formal proof for the mentioned property is, however, out of the scope of this paper.

%\sout{It is out of the scope of this paper to determine whether the stationary state of the cumulant equation exists in an arbitrary order. In the next Section, we show that the second-order approximation of long-time state of the dynamics of the cumulant equation can be determined. This is done via the relation between the cumulant equation and the Bloch-Redfield equation as demonstrated in Section \ref{sec:app:subsec:Relatioon_BR} of the Appendix.}

\color{black}

We finish this Section with the following Remark on the advantage of the cumulant equation over other approaches mentioned in this manuscript.
\begin{remark}
The cumulant equation has advantages over both Davies-GKSL and Bloch-Redfield equations. Indeed, dynamics obtained with the cumulant equation is non-Markovian and CPTP. Moreover, it is valid for all time regimes, in the sense of being exact up to the Born approximation. This is because, on the contrary to the Davies-GKSL equation the derivation of the cumulant equation does not include any separation of time scales. We refer the interested reader to reference~\cite{Winczewski_2021} where certain extensions of the cumulant equation are proposed.
\end{remark}

\subsection{Relation to the Bloch--Redfield equation}

Interestingly, the second cumulant superoperator $\tilde{K}^{(2)}(t)$ has a simple relation with the Liouvillian of the  (renormalized\footnote{In which the Lamb-Stark shift term is skipped. See the discussion in the next Section.}) Bloch-Redfield \cite{Maniscalco_2004,Blum_2012} equation $\tilde{\mathcal{L}}_{BR}(t)$, namely
\begin{align}
\frac{d}{dt} \tilde{K}^{(2)}(t)=\tilde{\mathcal{L}}_{BR}(t).\label{eqn:BR_relation}
\end{align}
This means, that the cumulant equation is equivalent to the formally integrated Bloch-Redfield equation when the time-ordering is skipped (see Appendix \ref{sec:app:subsec:Relatioon_BR} for the details of the comparison). Surprisingly, this is enough to recover the complete positivity of the dynamical equation as the cumulant equation guarantees the completely-positive dynamics. 

Using simple Lie-algebraic techniques \cite{Wulf_2008}, the cumulant equation can be transformed into a differential equation of the following form
\begin{align}
\frac{d}{dt} \tilde{\rho}_S(t)= \left(\frac{e^{[\tilde{K}^{(2)}(t),\cdot]}-1}{[\tilde{K}^{(2)}(t),\cdot]}\frac{d \tilde{K}^{(2)}(t)}{dt}\right)\tilde{\rho}_S(t).\label{eqn:cumulant_differential}
\end{align}
The r.h.s. of the formula above has a superoperator that contains all even orders in the coupling constant (recall $\tilde{K}^{(2)}(t)$ is second-order in the coupling constant). If we now truncate the r.h.s. of equation \eqref{eqn:cumulant_differential} up to the second-order in the coupling constant, and use relation in equation \eqref{eqn:BR_relation} we obtain the interaction picture (renormalized) Bloch-Redfield equation
\begin{align}
\frac{d}{dt} \tilde{\rho}_S(t)=\tilde{\mathcal{L}}_{BR}(t) \tilde{\rho}_S(t).\label{eqn:BR_equation}
\end{align}

The Bloch-Redfield equation (of the form in equation \eqref{eqn:BR_equation}) has a stationary state $\rho^{(2,ss)}_{\mathcal{S},BR}$ in the Schr{\"o}dinger picture: 
\begin{align}
    \rho^{(2,ss)}_{\mathcal{S},BR} \sim e^{\beta(H_\mathcal{S}^{(2)}+\delta H)}. \label{eqn:BR_SS}
\end{align}
%\sout{that is consistent up to the second-order with the long-time limit of the dynamics of the cumulant equation.}
Here, $\delta H$ is the second-order correction to the renormalized Hamiltonian $H_\mathcal{S}$ for which the diagonal elements vanish, and consequently, no additional shift of energy levels is observed. In reference, \cite{Lobejko_2022} it was shown that the off-diagonal elements of the Hamiltonian for the (Gibbs form) stationary state of the not renormalized Bloch-Redfield equation correspond to off-diagonal elements of the mean-force Hamiltonian $H_\mathcal{S}^{(mf)}$. However, in our approach, the choice of $H_\mathcal{S}$ to be $H_\mathcal{S}^{(mf)}$ that reproduces the diagonal elements of the stationary state correctly (with respect $H_\mathcal{S}^{(mf)}$ Hamiltonian Gibbs state) still predicts a correction to the off-diagonal part. In this manner, the stationary state of the Bloch-Redfield equation is always incompatible with the predictions of the mean-force Hamiltonian approach.
%the form of the correction $\delta H$ that we derive is defined via the renormalized Bohr frequencies $\{\omega_2\}$ and does not allow for a straightforward comparison with $H_{mf,\mathcal{C}}^{(2)}$.
%This proves that the renormalization procedure saves the thermodynamic consistency of both the Bloch-Redfield and the cumulant equation (up to the second-order). Still, the stationary state coherence are present despite the renormalization procedure.
The details of the derivation of the above stationary-state and explicit formula for $\delta H$ are given is Section \ref{sec:app:subsec:Relatioon_BR} of the Appendix. 

%\sout{On contrary, finding the stationary state at arbitrary order (if exists) of the cumulant equation is a formidable task. However, the state in equation \eqref{eqn:BR_SS} constitutes the second-order approximation for the long-time state of the cumulant equation.}
The above result shows that the renormalized Bloch-Redfield equation %\sout{and the cumulant equation (up to approximation)}
thermalizes to a Gibbs state with respect to a Hamiltonian that has the same diagonal part as the renormalized $H_\mathcal{S}$ in the eigenbasis of the latter. This proves that the renormalization procedure prevents any additional shift of the energy levels of $H_\mathcal{S}$ and therefore saves the thermodynamical consistency. Sill, we observe the presence of (eternal) coherence is the stationary state~$\rho^{(2,ss)}_{\mathcal{S},BR}$.

\subsection{The long time limit of the cumulant equation}

% It can be easily verified that the following relation holds
% \begin{align}
%     \lim_{t \to \infty} \frac{1}{t} \tilde{K}^{(2)}(t)= \mathcal{L},
%     \label{eqn:cumulant_limit_wrong}
% \end{align}
% where $\mathcal{L}$ is the generator of the quantum dyna Davies-GKSL equation (see equation \eqref{eqn:Davies_dissipator}). Still, the relation above does not imply that
% \begin{align}
%     \gamma_{ij} (\omega,\omega^\prime,t) \stackrel{t\approx+\infty}{\approx} t \gamma_{ji}(\omega) \delta_{\omega,\omega^\prime},  \label{eqn:cumulant_limit_wrong2}
% \end{align}
% since the form of the limit in the l.h.s. of equation \eqref{eqn:cumulant_limit_wrong} suppresses any constant or oscillatory terms. 

As it was indicated in equation \eqref{eqn:cumulant_superoperator_Davies_limit} for long times the cumulant superoperator $\tilde{K}^{(2)}(t)$ is approximated 
%in the zeroth-order 
by the Liouvillian of the Davies-GKSL equation $\tilde{L}$. In this Section we take an effort to find the exact long-time limit of the cumulant superoperator $\tilde{K}^{(2)}(t)$.  
In order, to accomplish the task described above we express the long time limit for the matrix $\gamma_{ij}(\omega,\omega^\prime,t)$ (see equation \eqref{eqn:Cumulant_Matrix}) in terms of quantities known form Davies-GKSL equation (see equations \eqref{eqn:def:coe1}, \eqref{eqn:def:coe2} and \eqref{eqn:def:coe3})
\begin{align}
    &\gamma_{ij}(\omega,\omega^\prime,t) \nonumber \\
    &\stackrel{t \approx +\infty}{\approx}\frac{i}{2}\frac{1}{\omega-\omega^\prime} \left(e^{-it(\omega-\omega^\prime)}-1\right) \left(\gamma_{ji}(\omega)+\gamma_{ji}(\omega^\prime)\right) \nonumber\\
    &-\frac{1}{\omega-\omega^\prime} \left(e^{-it(\omega-\omega^\prime)}+1\right) \left(\mathcal{S}_{ji}(\omega)-\mathcal{S}_{ji}(\omega^\prime)\right), \label{eqn:cumulant_limit}
\end{align}
(with a typical rate of convergence $\mathcal{O}(t^{-2})$) where the terms for which $\omega^\prime=\omega$ are defined via a limiting procedure
\begin{align}
    \gamma_{ij}(\omega,\omega,t) =\lim_{\omega^\prime \to \omega} \gamma_{ij}(\omega,\omega^\prime,t).
\end{align}
The limit described above yields:
\begin{align}
    \gamma_{ij}(\omega,\omega,t)&\stackrel{t\approx +\infty}{\approx}
    t \gamma_{ji}(\omega)
    -2 \frac{\partial}{\partial \omega} \mathcal{S}_{ji}(\omega). \label{eqn:cumulant_limit2}
\end{align}
Formulas in equations \eqref{eqn:cumulant_limit} and \eqref{eqn:cumulant_limit2} correct\footnote{Formula \eqref{eqn:cumulant_limit2} was also compared numerically with formula \eqref{eqn:Cumulant_Matrix} to confirm its validity.}  the relation from equation \eqref{eqn:cumulant_superoperator_Davies_limit} as a direct calculation of the integral does not suppress oscillatory or constant terms. In this way, the above result corrects the solution for the long-time limit of the cumulant superoperator proposed in ref. \cite{Rivas_2017}, and constitutes a basis for the proof on the existence of the stationary state \cite{MW_preparation}. 

%\sout{We comment here on the structure of the long-time limit of the cumulant superoperator $\tilde{K}^{(2)}(t)$ in the context of the eternal coherence. Indeed, as equation \eqref{eqn:cumulant_limit} reveals the matrix $\gamma_{ij}(\omega,\omega^\prime,t)$ is not diagonal (in $\omega$, $\omega^\prime$) for arbitrary long-times. Therefore, the cumulant equation predicts the eternal coherence in the basis of $H_\mathcal{S}$.}

% Furthermore, the above formula can be used to determine the long-time state predicted by the cumulant equation. We look for the long-time state determined up to the second-order in the coupling constant. Additionally, we work in the Schr{\"o}dinfer picture, so that we can compare our result with the prediction of the Bloch-Redfield equation in equation~\eqref{eqn:BR_SS}. In order, to achieve out goal we consider a specific initial state, i.e., $\rho_{\mathcal{S},\beta}$ (see equations~\eqref{eqn:def_thermal1} and~\eqref{eqn:def_thermal2}).

\color{black}

\section{Discussion - reconciliation of approaches}\label{sec:discussion}

% \sout{In the previous Section \ref{sec:cumulant} we have shown how the self-consistency condition makes the stationary state (in the Schr{\"o}dinger picture) of the renormalized Bloch-Redfield equation and the cumulant equation \sout{(up to the second-order)} match \sout{with the reduced state of the system $\mathcal{S}$ obtained from the partial trace in Section~\ref{sec:PartialTrace}} with the diagonal part of Gibbs state with respect to the renormalized Hamiltonian $H_\mathcal{S}$.}

In the previous Section we have shown how the renormalization procedure motivated by the self-consistency condition saves the thermodynamic consistency of the dynamical equation. In the renormalized equations the dissipation proceeds with respect to renormalized energy levels, and the dynamics (thermal reservoir) approaches the stationary state in Gibbs form with respect to renormalized Hamiltonian\footnote{Bloch-Redfield equation exhibits additional steady state coherence}.

On the other hand, the ''standard" Davies-GKSL equation studied in Section \ref{sec:MME} fails to reproduce the stationary state consistent with dissipation rates. Moreover, the alleged shift of energy levels (Lamb-Stark shift term) does not correspond to the predictions of the ''mean-force" Hamiltonian approach (see Section Section~\ref{sec:PartialTrace}). Still, it is possible to apply (at the level of derivation) the self-consistency condition to Davies-GKSL or any other master equation to obtain the equations that reproduce the correct populations of the stationary state, i.e., Gibbs (thermal) state with respect to $H_\mathcal{S}$\footnote{As explained earlier we identify $H_\mathcal{S}$ with $H_\mathcal{S}^{(2)}$.}. Alternatively, one can follow the standard derivation of Davies-GKSL or Bloch-Redfield equations and then proceed with three steps
\begin{enumerate}
    \item Skip the Lamb-Stark shift term.
    \item Identify $H_\mathcal{S}^{(1)}$ with the physical Hamiltonian of the system $H_S$
    \item Incorporate Bohr frequencies calculated with respect to $H_\mathcal{S}^{(1)}$ in the dissipator.
\end{enumerate}
The above procedure, that without formal proofs for its relevance was known before, and results in the same dynamical equations as those obtained by applying the self-consistency condition in the derivations (see Section \ref{sec:app:derivationCumulantEquation}).

The question on the choice of the $H_\mathcal{S}$ still remains open. Ideally, $H_\mathcal{S}$ should be the physical Hamiltonian, i.e., measured for the system that already has thermalized. If the former is not accessible, then the next choice can be the mean-force Hamiltonian $H_\mathcal{S}^{(mf)}$. We expect that the latter approach would be satisfactory, provided that $H_\mathcal{S}^{(0)}$ is known and one possesses an accurate model of the interaction.  
\color{black}

In some situations, the second corrections $\tilde{H}_{mf,\mathcal{C}}^{(2)}(t)$ and $\tilde{H}_\mathcal{C}^{(2)}(t)$ to the bare Hamiltonian  $H_\mathcal{S}^{(0)}$ can contain divergent integrals, what leads to infinite renormalization of energy levels (in exactly the same way as mention before in Section \ref{sec:sub:LSterm} for $H_{LS}$). This situation is encountered in an extremely important case of the reservoir being an electromagnetic field in free space\footnote{With the interaction of transition dipole moment type.} (e.g. quantum optics). Moreover, consequently either $H_\mathcal{S}^{(0)}$ or $H_\mathcal{S}$ is ill-defined as well. Let us notice now that (in principle) $H_\mathcal{S}$ can always be measured\footnote{For this reason we anticipated and called $H_\mathcal{S}$ the physical Hamiltonian.}, and $H_\mathcal{S}^{(0)}$ can be determined only when the system $\mathcal{S}$ is separated from the reservoir and non-interacting with it. Because the aforementioned condition can not be always satisfied $H_\mathcal{S}^{(0)}$ should be treated as an auxiliary object, with unclear interpretation. Moreover, in some physically relevant situations, we can not prepare the state of the total system $\mathcal{S}+\mathcal{R}$ in a product form; for example, we never observe bare electrons in nature but a species dressed with excitations of electromagnetic and Dirac fields\footnote{High energy physics that unveils bare electrons is not within  the weak coupling regime.}. Additionally, the usual assumption on the coupling that can be instantaneously switched on is physically not relevant. The above considerations explain how the renormalization procedure that we propose counteracts to Born approximation (assumption on the initial to be in a product form) that can be irrelevant for some systems, especially quantum optical ones. Notice that the dissipators in the renormalized equations are defined via the renormalized Bohr frequencies. %\sout{In particular, the correction $\tilde{H}_\mathcal{C}^{(2)}(t)$ can be understood as a counterterm to the dressing processes that are induced by the Born approximation.} 
Finally, thanks to the presence of counterterms $H_\mathcal{C}^{(2)}$ in the renormalized Hamiltonians (see equations \eqref{eqn:RenormalizedHamiltonianS}-\eqref{eqn:RenormalizedInteraction}), we can identify the initial state of the system to be a product state between the dressed system $\mathcal{S}$ and the rest of the reservoir, as no additional shift-like term is present.

At this point, we have to stress that we are not claiming that the Lamb-Stark shift does not exist at all. Indeed, the Lamb-Stark shift is an effect that was well verified experimentally many times. We claim instead that the Lamb-Stark shift term that appears in the dynamical equations of the theory of open quantum systems is a mathematical artifact of the Born approximation (coherent component of second-order quantum noise, aka drift). Still, our approach allows incorporating (genuine) Lamb-Stark shift in the form of time-dependent Hamiltonian $H_\mathcal{S}(t)$ of the system $\mathcal{S}$ provided $H_\mathcal{S}(t)$ is determined experimentally or calculated with higher-level theory, for example, quantum electrodynamics (QED).

On the other hand, it would be interesting to consider a total system $\mathcal{S}+\mathcal{R}$ for which the Born approximation is a relevant one~\cite{Gramich_2011,Sekatski_2021}\footnote{Such an experiment should be feasible in solid-state physics  system with phononic reservoir. }. Namely, a total system $\mathcal{S}+\mathcal{R}$ for which the initial state preparation in the product form is experimentally feasible. In this case both $H_\mathcal{S}$ and $H_\mathcal{S}^{(0)}$ should be well defined and measurable. It would be then interesting to compare the measured Lamb-Stark shift with the counterterm $\tilde{H}_\mathcal{C}^{(2)}(t)$ and $\tilde{H}_{mf,\mathcal{C}}^{(2)}(t)$ . 

We remark here that our results may shed new light on the lively contemporary discussion on the presence of the steady-state (equilibrium) coherence  \cite{Guarnieri_2018,Brumer_2018,koyu2021,Goold_2020}. As we note in Section \ref{sec:PartialTrace}, the correction $H_{mf,\mathcal{C}}^{(2)}$ in equation \eqref{eqn:PT:cor2}, does not commute with the bare Hamiltonian $H_\mathcal{S}^{(0)}$. This property implies that the renormalized measurable Hamiltonian $H^{(mf)}_\mathcal{S}$, has a different eigenbasis than $H_\mathcal{S}^{(0)}$; this is important in the case of base-dependent quantities like coherence. Interestingly, the cumulant equation dynamics approaches the Gibbs state with respect to the renormalized Hamiltonian $H_\mathcal{S}$. In this way the cumulant equation reproduces the dynamics of the ``global" (renormalized) Davies-GKSL equation, and therefore in the eigenbasis of the renormalized (measurable) Hamiltonian $H_\mathcal{S}$ all coherence vanish. However, this is not the case for the renormalized Bloch-Redfield equation that exhibits steady-state coherence. \color{black} %\sout{Still, if the eigenbases of $H_\mathcal{S}$ and $H_\mathcal{S}^{(0)}$ do not coincide the steady-state density operator might be non-diagonal in the eigenbasis of the bare Hamiltonian $H_\mathcal{S}^{(0)}$ .}

Finally, we claim (without proof) that the self-consistency condition can be naturally extended to the case of many (not cross-talking) reservoirs, not necessarily in the KMS states. Therefore, the properly renormalized cumulant equation can prove its usefulness in the area of (quantum) thermodynamics, where the questions on the heat flow are the central, contemporary concern~\cite{levy2014local,hofer2017markovian,cattaneo2019local}. 

\section{Conclusions}\label{sec:conclusion}

In conclusion, we have presented a renormalization scheme for the selected dynamical equation of the theory of open quantum systems. The motivation for the renormalization is %\sout{intuition, however confirmed with mathematically rigorous results}
the need for thermodynamical consistency of the reduced dynamics of the open quantum systems. Still, to demonstrate our approach, we consider quite a general physical setup. The justification for the validity of the scheme is twofold. Firstly, in Section \ref{sec:sub:LSterm} we point out the problem of inconsistency in the standard derivation of the Davies-GKSL equation~\cite{Breuer+2006}. The problem of lack of consistency is identified in the presence of the Lamb-Stark shift term and dissipation in the open system $\mathcal{S}$ proceeding with respect to the bare Hamiltonian $H_\mathcal{S}^{(0)}$ Bohr frequencies. However, the system is expected to dissipate with respect to the renormalized Hamiltonian for the consistency to hold. Secondly, as shown in Section \ref{sec:PartialTrace}, Davies-GKSL in its standard form is not only inconsistent with itself but also fails to reproduce the correct reduced stationary state of the open system $\mathcal{S}$. This failure is proved by calculating the partial trace with respect to the reservoir of the KMS state of the total system $\mathcal{S}+\mathcal{R}$. %\sout{Hopefully, in this way, the form of the second correction to the bare Hamiltonian is obtained.}
Furthermore, we show that the bare Hamiltonian of the system renormalized with the Lamb-Stark shift term (for Davies-GKSL or Bloch-Redfield equations) does not reproduce the so-called ''mean-force" Hamiltonian obtained with the partial trace approach. This discrepancy supports the argument on the need of the renormalization discussed in this paper. This is because the Lamb-Stark shift terms seems to be only an mathematical artifact that does not describe any physical process accurately.

In Section \ref{sec:cumulant} we formulate the self-consistency condition that is used in the derivation of the cumulant equation. As it is shown, the self-consistency condition fixes the renormalization procedure and yields dynamical equation for which long-time states are thermodynamicaly consistent with the renormalized (physical) Hamiltonian $H_\mathcal{S}$. We take a pedagogical perspective on the derivation of the cumulant equation; however, we do not hesitate to make minor generalizations and novel contributions. Therefore, we present an extended version of the first derivation in reference~\cite{Alicki1989}, in which we aim to explain all steps and ideas exhaustively.

Finally, in Section \ref{sec:discussion} we discuss how different perspectives on the open quantum systems considered in this manuscript can be reconciled. The reconciliation is done by identifying the renormalized Hamiltonian $H_\mathcal{S}$ with the physical, measurable one while treating $H_\mathcal{S}^{(0)}$ as an auxiliary object. Moreover, we identify the culprit of the confusion present in the literature, and therefore the culprit of the need for renormalization to be the Born approximation. As it is debated, the Born approximation is unlikely to be relevant in a generic case, especially for a quantum optical system. %\sout{ In this way, the presence and the interpretation of the counterterm $H_{\mathcal{C}}^{(2)}$ in explained.}
Moreover, a shortened recipe for renormalization is provided.
\color{black}

\section*{Acknowledgements}
We acknowledge Antonio Mandarino, Marcin {\L}obejko, Gerardo Suarez and Micha{\l} Horodecki for fruitful discussion and support. We also acknowledge Anton Trushechkin for the discussion and communication. MW acknowledges grant PRELUDIUM-20 (grant number: 2021/41/N/ST2/01349) from the National Science Center. This work is partially supported by Foundation for Polish Science (FNP), IRAP project ICTQT, contract no. 2018/MAB/5, co-financed by EU  Smart Growth Operational Programme.

%  MW acknowledges the Faculty of Mathematics, Physics and Informatics of the university of Gda{\'n}k for the financial support.

\bibliographystyle{apsrev4-1}
\bibliography{references}

\newpage
\appendix
\begin{widetext}
~~\\
\begin{center}
		{\Huge Appendix}
\end{center}
%Table of contents entry
\addappheadtotoc  

\section{Properties of interaction operators}\label{sec:app:Properties}

Following the notation in reference~\cite{Kimura_2017}, we define a family of operators $\mathcal{L}_A$:
\begin{align}
    \mathcal{L}_A = \left[A,\cdot\right].
\end{align}

Let us also define $R_i^{(k)}(\Omega)$ operators ($k=0,1$), with the following Fourier transform of interaction picture operators $\tilde{R}_i^{(k)}(s)$:
\begin{align}
    &R_i^{(k)}(\Omega)= \frac{1}{2 \pi}\int_{-\infty}^{+\infty} ds~ e^{i \Omega s} \tilde{R}_i^{(k)}(s),\\
    &\tilde{R}_i^{(k)}(s) = e^{i H_\mathcal{R} s} {R}_i^{(k)} e^{-i H_\mathcal{R} s}.
\end{align}
The above definition yields the following relations:
\begin{align}
    &{R}_i^{(k)}= \int_{-\infty}^{+\infty}d\Omega~R_i^{(k)}(\Omega),~~
    \tilde{R}_i^{(k)}(s)= \int_{-\infty}^{+\infty}d\Omega~e^{-i\Omega s}R_i^{(k)}(\Omega),~~R_i^{(k)}(-\Omega)={R_i^{(k)}}^\dagger(\Omega).
    % \\
    % &H_\mathcal{I}^{(1)}=\sum_i S_i \otimes {R}_i^{(1)} = \sum_i S_i \otimes {R}_i^{(1)} = \int_{-\infty}^{+\infty}d\Omega~ \sum_i S_i \otimes R_i^{(1)}(\Omega).
\end{align}
The definition of $R_i^{(k)}(\Omega)$ can be used to prove that:
\begin{align}
    \mathcal{L}_{H_\mathcal{R}} R_i^{(k)}(\Omega)=\left[H_\mathcal{R},R_i^{(k)}(\Omega)\right]=-\Omega R_i^{(k)}(\Omega).
\end{align}

Analogous relations are true for the standard jump operators $S_i(\omega)$, defined with projections $\Pi(\epsilon_k)$ onto eigenspaces of the Hamiltonians $H_\mathcal{S}^{(k)}$ (in the manuscript we consider $k=0,1,2$) of the system $\mathcal{S}$~\cite{Breuer+2006}.
\begin{align}
&S_i (\omega_k) = \sum_{\epsilon^\prime_k-\epsilon_k=\omega_k} \Pi(\epsilon_k) S_i \Pi(\epsilon^\prime_k), ~~
S_i=\sum_{\omega_k} S_i (\omega_k),\\
&\mathcal{L}_{H_\mathcal{S}^{(k)}} S_i(\omega_k)=\left[H^{(k)}_\mathcal{S},S_i(\omega_k)\right]=-\omega_k S_i(\omega_k),~~\mathcal{L}_{H_\mathcal{S}^{(k)}} S^\dagger_i(\omega_k)=\left[H^{(k)}_\mathcal{S},S_i^\dagger(\omega_k)\right]=\omega_k S_i^\dagger(\omega_k). \label{eqn:Eigenoperators1Type}
\end{align}
The above equation defines the $I$-kind of eigenoperators of superoperator $\mathcal{L}_{H_\mathcal{S}^{(k)}}$. Additionally it is straightforward to show that:
\begin{align}
    &\mathcal{L}_{H_\mathcal{S}^{(k)}}S_i^\dagger(\omega_k)S_j(\omega^\prime_k) =\left[H^{(k)}_\mathcal{S},S_i^\dagger(\omega_k)S_j(\omega^\prime_k)\right] = (\omega_k-\omega^\prime_k) S_i^\dagger(\omega_k)S_j(\omega^\prime_k). \label{eqn:Eigenoperators2Type}
\end{align}
The above equation defines the $II$-kind of eigenoperators of superoperator $\mathcal{L}_{H_\mathcal{S}^{(k)}}$.

The definitions of $R_i^{(k)}(\Omega)$ and $S_i(\omega_k)$ yield (for $k=0,1$):
\begin{align}
    H_\mathcal{I}^{(k)}= \sum_i S_i \otimes {R}_i^{(k)} = \sum_{\omega_k} \sum_i \int_{-\infty}^{+\infty}d\Omega~  S_i(\omega_k) \otimes R_i^{(k)}(\Omega).
\end{align}
Moreover, for the integrand in the equation above, exibits a useful relation:
\begin{align}
    \mathcal{L}_{H_\mathcal{S}^{(k)}+H_\mathcal{R} }   S_i(\omega_k) \otimes R_i^{(k)}(\Omega) = (-\omega_k-\Omega) S_i(\omega_k) \otimes R_i^{(k)}(\Omega),
\end{align}
that can be readily extended, in the following way:
\begin{align}
    &\left( \mathcal{L}_{H_\mathcal{S}^{(k)}+H_\mathcal{R} } \right)^m S_i(\omega_k) \otimes R_i^{(k)}(\Omega) = (-\omega_k-\Omega)^m S_i(\omega_k) \otimes R_i^{(k)}(\Omega).
\end{align}

The properties of the interaction operators can be used to show useful properties (in the case of $\left[H_\mathcal{R},\rho_{\mathcal{R},\beta}\right]=0$) of quantities in equations \eqref{eqn:def:coe1}-\eqref{eqn:def:coe3}:
\begin{align}
    &\gamma_{ij}(\omega) = \Gamma_{ij}(\omega)+\Gamma_{ji}^*(\omega)\\
    &=\int_0^{+\infty}ds~e^{i\omega s}
    \tr \left[{R}^{(1)}_i(s)R^{(1)}_j\rho_{\mathcal{R},\beta}\right]+\int_0^{+\infty}ds~e^{-i\omega s}
    \tr \left[{R}^{(1)}_i(-s)R^{(1)}_j\rho_{\mathcal{R},\beta}\right]\\
    &=\int_0^{+\infty}ds~e^{i\omega s}
    \tr \left[{R}^{(1)}_i(s)R^{(1)}_j\rho_{\mathcal{R},\beta}\right]+\int_{-\infty}^0 ds~e^{i\omega s}
    \tr \left[{R}^{(1)}_i(s)R^{(1)}_j\rho_{\mathcal{R},\beta}\right]\\
    &=\int_{-\infty}^{+\infty}ds~e^{i\omega s}
    \tr \left[{R}^{(1)}_i(s)R^{(1)}_j\rho_{\mathcal{R},\beta}\right]=2\pi
    \tr \left[\left(\frac{1}{2\pi}\int_{-\infty}^{+\infty}ds~e^{i\omega s}{R}^{(1)}_i(s)\right)R^{(1)}_j\rho_{\mathcal{R},\beta}\right]\\
    &=2\pi
    \tr \left[{R}^{(1)}_i(\omega)R^{(1)}_j\rho_{\mathcal{R},\beta}\right],
\end{align}
moreover
\begin{align}
     &\Gamma_{ij}(\omega)=\lim_{\epsilon \to 0^+} \Gamma_{ij}(\omega+i\epsilon)
    =  \lim_{\epsilon \to 0^+} \int_0^{+\infty} ds~ e^{i (\omega+i\epsilon) s} \tr \left[ \tilde{R}_i^{(1)}(s)R_j^{(1)}\rho_{\mathcal{R},\beta} \right] \\
    &=\lim_{\epsilon \to 0^+} \int_0^{+\infty} ds~ e^{i (\omega+i\epsilon) s} \tr \left[ \left( \int_{-\infty}^{+\infty}d\Omega~e^{-i\Omega s}R_i^{(1)}(\Omega) \right) R_j^{(1)}\rho_{\mathcal{R},\beta} \right]\\
    &=\lim_{\epsilon \to 0^+}\int_{-\infty}^{+\infty}d\Omega~ \int_0^{+\infty} ds~ e^{i (\omega+i\epsilon-\Omega) s} \tr \left[  R_i^{(1)}(\Omega)  R_j^{(1)}\rho_{\mathcal{R},\beta} \right]\\
    &=\lim_{\epsilon \to 0^+}\int_{-\infty}^{+\infty}d\Omega~ \frac{i}{\omega+i\epsilon-\Omega} \tr \left[  R_i^{(1)}(\Omega)  R_j^{(1)}\rho_{\mathcal{R},\beta} \right]
    =\lim_{\epsilon \to 0^+}\frac{1}{2\pi}\int_{-\infty}^{+\infty}d\Omega~ \frac{i}{\omega+i\epsilon-\Omega} \gamma_{ij}(\Omega).
% &=
%   \int_{-\infty}^{+\infty}d\Omega~\frac{i}{\omega-\Omega} \tr \left[R_i^{(1)}(\Omega)R_j^{(1)} \rho_{\mathcal{R},\beta} \right]=\frac{1}{2\pi}\int_{-\infty}^{+\infty}d\Omega~\frac{i}{\omega-\Omega}\gamma_{ij}(\Omega),   \\
    %  &\Gamma_{ji}^*(\omega)+\Gamma_{ij}(\omega)=\lim_{\epsilon \to 0^+} 2\pi i~ \mathrm{Res}(\omega+i \epsilon,\frac{i}{\omega+i\epsilon-\Omega}\gamma_{ij}(\Omega))=-2\pi\gamma_{ij}(\omega).
    % = \tr\left[ R_i^{(1)} \left(\int_{-\infty}^{+\infty}d\Omega~ \frac{-i}{\omega+\Omega} R_j^{(1)}(\Omega)\right) \rho_{\mathcal{R},\beta} \right]
\end{align}
From the above it is straightforward, to see that:
\begin{align}
    &\Gamma_{ij}(\omega)=\lim_{\epsilon \to 0^+}\frac{1}{2\pi}\int_{-\infty}^{+\infty}d\Omega~ \frac{i}{\omega+i\epsilon-\Omega} \gamma_{ij}(\Omega), \label{eqn:GammaRegA}\\
    &\Gamma_{ji}^*(\omega)= \lim_{\epsilon \to 0^+} \frac{1}{2\pi} \int_{-\infty}^{+\infty}d\Omega~ \frac{-i}{\omega-i\epsilon-\Omega} \gamma_{ij}(\Omega).\label{eqn:GammaRegB}
\end{align}

Additionally, the quantities in equations \eqref{eqn:def:coe1}-\eqref{eqn:def:coe3} are related via the Sokhotski–Plemelj theorem. 
\begin{align}
    &\Gamma_{ij}(\omega)=\frac{1}{2} \gamma_{ij}(\omega) + i \mathcal{S}_{ij}(\omega),\\
    &\mathcal{S}_{ij}(\omega) = \frac{1}{2\pi}\dashint_{-\infty}^{+\infty}d\Omega~ \frac{1}{\omega-\Omega} \gamma_{ij}(\Omega), \label{eqn:S_IntegralRepresentation}
\end{align}
where $\dashint$ denotes the principal value integral.

Since the asymptotic arc integral (over $s$, with $\omega>0$, if $\omega<0$ we chose lower half-plane) in the upper half-plane is zero (analysis of exponents):
\begin{align}
     &\Gamma_{ji}^*(\omega)+\Gamma_{ij}(\omega)=\lim_{\epsilon \to 0^+} 2\pi i~ \mathrm{Res}(\omega+i \epsilon,\frac{1}{2\pi}\frac{i}{\omega+i\epsilon-\Omega}\gamma_{ij}(\Omega))=\gamma_{ij}(\omega),
\end{align}
and this matches standard things.

\section{Properties of Gammas quantities}\label{sec:app:PropertiesGamma}

On the other hand, we can easily show that, for $\epsilon>0$.
\begin{align}
     &\Gamma_{ij}(\omega)=\lim_{\epsilon \to 0^+} \Gamma_{ij}(\omega+i\epsilon)
    =  \lim_{\epsilon \to 0^+} \int_0^{+\infty} ds~ e^{i (\omega+i\epsilon) s} \tr \left[ \tilde{R}_i^{(1)}(s)R_j^{(1)}\rho_{\mathcal{R},\beta} \right] \\
    &=\lim_{\epsilon \to 0^+} \int_0^{+\infty} ds~ e^{i (\omega+i\epsilon) s} \tr \left[ \left( \int_{-\infty}^{+\infty}d\Omega~e^{-i\Omega s}R_i^{(1)}(\Omega) \right) R_j^{(1)}\rho_{\mathcal{R},\beta} \right]\\
    &=\lim_{\epsilon \to 0^+}\int_{-\infty}^{+\infty}d\Omega~ \int_0^{+\infty} ds~ e^{i (\omega+i\epsilon-\Omega) s} \tr \left[  R_i^{(1)}(\Omega)  R_j^{(1)}\rho_{\mathcal{R},\beta} \right]\\
    &=\lim_{\epsilon \to 0^+}\int_{-\infty}^{+\infty}d\Omega~ \frac{i}{\omega+i\epsilon-\Omega} \tr \left[  R_i^{(1)}(\Omega)  R_j^{(1)}\rho_{\mathcal{R},\beta} \right]
    =\lim_{\epsilon \to 0^+}\frac{1}{2\pi}\int_{-\infty}^{+\infty}d\Omega~ \frac{i}{\omega+i\epsilon-\Omega} \gamma_{ij}(\Omega)
% &=
%   \int_{-\infty}^{+\infty}d\Omega~\frac{i}{\omega-\Omega} \tr \left[R_i^{(1)}(\Omega)R_j^{(1)} \rho_{\mathcal{R},\beta} \right]=\frac{1}{2\pi}\int_{-\infty}^{+\infty}d\Omega~\frac{i}{\omega-\Omega}\gamma_{ij}(\Omega),   \\
    %  &\Gamma_{ji}^*(\omega)+\Gamma_{ij}(\omega)=\lim_{\epsilon \to 0^+} 2\pi i~ \mathrm{Res}(\omega+i \epsilon,\frac{i}{\omega+i\epsilon-\Omega}\gamma_{ij}(\Omega))=-2\pi\gamma_{ij}(\omega).
    % = \tr\left[ R_i^{(1)} \left(\int_{-\infty}^{+\infty}d\Omega~ \frac{-i}{\omega+\Omega} R_j^{(1)}(\Omega)\right) \rho_{\mathcal{R},\beta} \right]
\end{align}
Now we see that:
\begin{align}
    &\Gamma_{ji}^*(\omega)= \lim_{\epsilon \to 0^+}\int_{-\infty}^{+\infty}d\Omega~ \frac{-i}{\omega-i\epsilon-\Omega} \tr \left[  R_i^{(1)} R_j^{(1)}(-\Omega) \rho_{\mathcal{R},\beta} \right]
\end{align}

Since the asymptotic arc integral (over $s$, with $\omega>0$, if $\omega<0$ we chose lower half-plane) in the upper half-plane is zero (analysis of exponents):
\begin{align}
     &\Gamma_{ji}^*(\omega)+\Gamma_{ij}(\omega)=\lim_{\epsilon \to 0^+} 2\pi i~ \mathrm{Res}(\omega+i \epsilon,\frac{1}{2\pi}\frac{i}{\omega+i\epsilon-\Omega}\gamma_{ij}(\Omega))=\gamma_{ij}(\omega),
\end{align}
and this matches standard things.

\section{Renormalization via partial trace}\label{sec:app:PT}

Let $H_\mathcal{S}^{(0)}$, $H_\mathcal{R}$ be the bare Hamiltonians of the system and the Hamiltonian of the reservoir respectively, and let $\lambda H_\mathcal{I}^{(0)}= \lambda \sum_i S_i \otimes R^{(0)}_i$ be the interaction Hamiltonian, in which $\lambda=1$ . The reservoir Hamiltonian does not posses additional suffixes, because it does not go through the renormalization. This is because we assume that the reservoir is too large to be affected by the interaction with the system $\mathcal{S}$ in observable manner (see the main text). The total Hamiltonian of the system, for which we assume it is time independent (this must always be true near the equilibrium), reads\footnote{For the sake of simplicity we do not consider the chemical potential.}:
\begin{align}
    H_{\mathcal{S}+\mathcal{R}} = H_\mathcal{S}^{(0)}+H_\mathcal{R} + \lambda H_\mathcal{I}^{(0)}.
\end{align}
According to the KMS theory, the equilibrium state of the total $\mathcal{S}+\mathcal{R}$ system is 
\begin{align}
    \rho_{\mathcal{S}+\mathcal{R},\beta} = Z_{ \mathcal{S}+\mathcal{R},\beta}^{-1}e^{-\beta H_{\mathcal{S}+\mathcal{R}}}, \\
    Z_{\mathcal{S}+\mathcal{R},\beta}=\tr \left[e^{-\beta H_{\mathcal{S}+\mathcal{R}}}\right].
\end{align}
Therefore the reduced state of system $\mathcal{S}$, at thermal equilibrium, is defined with:
\begin{align}
    \rho_{\mathcal{S}}=\tr_{\mathcal{R}} \left[ \rho_{\mathcal{S}+\mathcal{R},\beta}\right]. \label{eqn:ReducedStateDefinition}
\end{align}
Indeed, the reduced state of the system $\mathcal{S}$ should be determined solely with the bare Hamiltonian $H_\mathcal{S}^{(0)}$, the state of the reservoir $\rho_{\mathcal{R},\beta}$, and the interaction Hamiltonian $H_\mathcal{I}^{(0)}$. We anticipate a bit, and after some manipulation we can rewrite $\rho_{\mathcal{S}}$ in the following form:
\begin{align}
    \rho_{\mathcal{S}} = \frac{\frac{1}{Z_{\mathcal{R},\beta}}\tr_{\mathcal{R}}\left[ e^{-\beta H_{\mathcal{S}+\mathcal{R}}}\right] }{ \frac{1}{Z_{\mathcal{R},\beta}}Z_{\mathcal{S}+\mathcal{R},\beta} },~~Z_{ \mathcal{R},\beta}= \tr \left[e^{-\beta H_{\mathcal{R}}}\right]. \label{eqn:NiceForm}
\end{align}
The form given is convenient for the forthcoming considerations.

Motivated by the zeroth law of the thermodynamics we intend to find $\rho_{\mathcal{S},\beta}$ in the following ''mean-force" form :
\begin{align}
    \rho_{\mathcal{S},\beta}^{(mf)}=\frac{e^{-\beta H^{(mf)}_{\mathcal{S}}}}{Z^{(mf)}_{ \mathcal{S},\beta}},~~Z^{(mf)}_{ \mathcal{S},\beta}= \tr \left[e^{-\beta H^{(mf)}_{\mathcal{S}}}\right],\label{eqn:SthermalExact}
\end{align}
where $H_{\mathcal{S}}$ is the physical (mean-force), renormalized due to interaction Hamiltonian of the system $\mathcal{S}$, ``detectable" by a thermometer. Suppose now, we are able to find $H^{(mf)}_\mathcal{S}$, such that\footnote{This is always possible for $\beta<+\infty$, as $H^{(mf)}_\mathcal{S}$ is an unknown operator (cf. reference \cite{Goold_2020}).}:
\begin{align}
    \frac{1}{Z_{\mathcal{R},\beta}}\tr_{\mathcal{R}}\left[ e^{-\beta H_{\mathcal{S}+\mathcal{R}}}\right] = e^{-\beta H^{(mf)}_{\mathcal{S}}},
\end{align}
then, following equation \eqref{eqn:NiceForm} we have:
\begin{align}
    \rho^{(mf)}_{\mathcal{S},\beta} = \frac{ e^{-\beta H^{(mf)}_{\mathcal{S}}}}{ \frac{1}{Z_{\mathcal{R},\beta}}Z_{\mathcal{S}+\mathcal{R},\beta} }.
\end{align}
Now, because $\tr \left[\rho_S\right]=1$, the above equation implies that:
\begin{align}
    Z^{(mf)}_{\mathcal{S},\beta}=\tr \left[e^{-\beta H^{(mf)}_{\mathcal{S}}}\right]=\frac{1}{Z_{\mathcal{R},\beta}}Z_{\mathcal{S}+\mathcal{R},\beta}.
\end{align}

The above considerations yield the set of two equations, where the second one is automatically satisfied provided we find the solution to the first one.  
% Now, because of the relation:
% \begin{align}
%     Z_{\mathcal{S}+\mathcal{R},\beta}=\tr_\mathcal{S} \left[ \tr_\mathcal{R} \left[e^{-\beta H_{\mathcal{S}+\mathcal{R}}}\right] \right].
% \end{align}
% the choice we made in equation \eqref{eqn:SthermalExact}, together with an intuition on the reservoir's state being not affected by interaction, gives us a hint:
\begin{align}
    e^{-\beta H^{(mf)}_{\mathcal{S}}} = \frac{1}{Z_{\mathcal{R},\beta}} \tr_\mathcal{R}\left[e^{-\beta H_{\mathcal{S}+\mathcal{R}}} \right],~~
    Z^{(mf)}_{\mathcal{S},\beta}=\frac{Z_{\mathcal{S}+\mathcal{R},\beta}}{Z_{\mathcal{R},\beta}}. \label{eqn:hint}
\end{align}
The above formulation, allows us to search for $H^{(mf)}_{\mathcal{S}}$ in the simplest possible form, since the denominators does not contain any terms containing $\lambda$ associated with interaction. Still, the denominator in the reduced state $\rho^{(mf)}_{\mathcal{S,\beta}}$ in equation \eqref{eqn:SthermalExact} does depend on $\lambda$.

Now, to test the validity of the above formulation we study a particular situation. In the case of no interaction, i.e., $\lambda H_\mathcal{I}^{(0)}=0$, the form of $H^{(mf)}_\mathcal{S}$ is particularly simple, since $H_\mathcal{S}^{(0)}$, $H_\mathcal{R}$ commute, and $Z_{\mathcal{S}+\mathcal{R}}$ has a product form. 
\begin{align}
    \rho^{(mf)}_{\mathcal{S},\beta} = Z_{ \mathcal{S},\beta}^{-1}e^{=\beta H^{(0)}_{\mathcal{S}}}, \\
    Z^{(mf)}_{\mathcal{S},\beta}=\tr \left[e^{-\beta H^{(0)}_{\mathcal{S}}}\right].
\end{align}
Therefore, in the case of the system $\mathcal{S}$ interacting weakly with the heat bath we can try to find the reduced state of the system $\mathcal{S}$, using the renormalized Hamiltonian $H^{(mf)}_\mathcal{S}$, in the form: 
\begin{align}
    H^{(mf)}_\mathcal{S}&=H_\mathcal{S}^{(0)}+H_{mf,\mathcal{C}},\\
    H_{mf,\mathcal{C}}&=\sum_{k=1}^{\infty}\lambda^k H_{mf,\mathcal{C}}^{(k)}.
\end{align}
The form above rephrases the intuition that if the interaction is weak the renormalized Hamiltonian $H_\mathcal{S}$ should consist of the bare Hamiltonian $H_\mathcal{S}^{(0)}$ and some corrections $\lambda^k H_{mf,\mathcal{C}}^{(k)}$.

In the following considerations we intend to find the first two corrections $H_{mf,\mathcal{C}}^{(1)}$, and $H_{mf,\mathcal{C}}^{(2)}$ that allow to construct the second-order approximation $H_\mathcal{S}^{(mf,2)}$ to the renormalized Hamiltonian $H^{(mf)}_\mathcal{S}$, and associated reduced state $\rho_{\mathcal{S},\beta}^{(mf,2)} \approx \rho^{(mf)}_\mathcal{S}$.

% The total Hamiltonian can be then expressed as follows:
% \begin{align}
%     H_{\mathcal{S}+\mathcal{R}}=H_{\mathcal{S}}+H_{\mathcal{R}}+ \lambda H_\mathcal{I}^{(0)}-H_\mathcal{C}.
% \end{align}
% The corrections $H_\mathcal{C}^{(k)}$ are to be determined in the following considerations, up to the second-order, i.e., $N=2$.

Let us now find the corrections $H_{mf,\mathcal{C}}^{(k)}$, and therefore the renormalized Hamiltonian $H^{(mf)}_\mathcal{S}$. We start with explicit deception of Zassenhaus formula in reference~\cite{Kimura_2017}:
\begin{align}
    e^{A+B} = \left\{ 1+\sum_{p=1}^{\infty}\sum_{n_1, ...,n_p=1}^{\infty}\frac{n_p \cdots n_1}{n_p(n_p+n_{p-1})\cdots(n_p+\cdots+n_1)}\mathcal{B}_{n_p}\cdots \mathcal{B}_{n_1}\right\}e^A, \label{eqn:ExpansionZassenhous}
\end{align}
where
\begin{align}
    \mathcal{B}_m = \frac{1}{m!} \left(\mathcal{L}_A\right)^{m-1} B,\\
    \mathcal{L}_A \mathcal{O} = \left[A,\mathcal{O}\right].
\end{align}
Let us firstly expand $e^{-\beta H_{\mathcal{S}+\mathcal{R}}}=e^{-\beta\left(H_\mathcal{S}^{(0)}+H_\mathcal{R} + \lambda H_\mathcal{I}^{(0)}\right)}$, by choosing $A=-\beta\left(H_\mathcal{S}^{(0)}+H_\mathcal{R}\right)$, and $B=-\beta \lambda H_\mathcal{I}^{(0)}$. We~have:
\begin{align}
    \mathcal{B}_m = \frac{1}{m!} \left(\mathcal{L}_{-\beta\left(H_\mathcal{S}^{(0)}+H_\mathcal{R}\right)}\right)^{m-1} (-\beta) \lambda H_\mathcal{I}^{(0)} =  \lambda \frac{(-\beta)^m}{m!} \left(\mathcal{L}_{H_\mathcal{S}^{(0)}+H_\mathcal{R}}\right)^{m-1}   H_\mathcal{I}^{(0)} \equiv \lambda \frac{(-\beta)^m}{m!} \mathcal{D}_m,
\end{align}
where we denoted $\mathcal{D}_m=\left(\mathcal{L}_{H_\mathcal{S}^{(0)}+H_\mathcal{R}}\right)^{m-1}   H_\mathcal{I}^{(0)}$.
Hence,
\begin{align}
    e^{-\beta H_{\mathcal{S}+\mathcal{R}}} = \left\{ 1+\sum_{p=1}^{\infty} \lambda^p \sum_{n_1, ...,n_p=1}^{\infty} \frac{n_p \cdots n_1}{n_p(n_p+n_{p-1})\cdots(n_p+\cdots+n_1)}\frac{(-\beta)^{n_p+\cdots+n_1}}{n_p!\cdots n_1!}\mathcal{D}_{n_p}\cdots \mathcal{D}_{n_1}\right\}e^{-\beta H_\mathcal{R}}e^{-\beta H_\mathcal{S}^{(0)}}.
\end{align}
We now denote the elements of the above expansion in the following way:
\begin{align}
    &e^{-\beta H_{\mathcal{S}+\mathcal{R}}}= \sum_{p=0}^{\infty} \lambda^p \kappa^{(p)} e^{-\beta H_\mathcal{R}}e^{-\beta H_\mathcal{S}^{(0)}}, \\
    & \kappa^{(0)}=1, \\
    & \kappa^{(p)}= \sum_{n_1, ...,n_p=1}^{\infty} \frac{n_p \cdots n_1}{n_p(n_p+n_{p-1})\cdots(n_p+\cdots+n_1)}\frac{(-\beta)^{n_p+\cdots+n_1}}{n_p!\cdots n_1!}\mathcal{D}_{n_p}\cdots \mathcal{D}_{n_1}. \label{eqn:expansionBeforePartialTrace}
\end{align}

Let us now calculate the first element in the expansion given by equation \eqref{eqn:expansionBeforePartialTrace}:
\begin{align}
    &\kappa^{(1)}= \sum_{n_1=1}^{\infty} \frac{(-\beta)^{n_1}}{n_1!} \mathcal{D}_{n_1}=  \sum_{n_1=1}^{\infty} \frac{(-\beta)^{n_1}}{n_1!} \left(\mathcal{L}_{H_\mathcal{S}^{(0)}+H_\mathcal{R}}\right)^{n_1-1}   H_\mathcal{I}^{(0)}\\
    &=  \sum_{n_1=1}^{\infty} \frac{(-\beta)^{n_1}}{n_1!} \left(\mathcal{L}_{H_\mathcal{S}^{(0)}+H_\mathcal{R}}\right)^{n_1-1}    \sum_{\omega_0} \int_{-\infty}^{+\infty}d\Omega~ \sum_i S_i(\omega_0) \otimes R_i^{(0)}(\Omega)\\
    &=  \sum_{n_1=1}^{\infty} \frac{(-\beta)^{n_1}}{n_1!}     \sum_{\omega_0} \int_{-\infty}^{+\infty}d\Omega~ \sum_i \left(\mathcal{L}_{H_\mathcal{S}^{(0)}+H_\mathcal{R}}\right)^{n_1-1}S_i(\omega_0) \otimes R_i^{(0)}(\Omega)\\
    &=  \sum_{n_1=1}^{\infty} \frac{(-\beta)^{n_1}}{n_1!}     \sum_{\omega_0} \int_{-\infty}^{+\infty}d\Omega~ \sum_i \left(-{\omega_0}-\Omega\right)^{n_1-1}S_i({\omega_0}) \otimes R_i^{(0)}(\Omega)\\
    &= \left(-{\omega_0}-\Omega\right)^{-1}      \sum_{\omega_0} \int_{-\infty}^{+\infty}d\Omega~ \sum_i\sum_{n_1=1}^{\infty} \frac{(-\beta)^{n_1}}{n_1!} \left(-{\omega_0}-\Omega\right)^{n_1}S_i({\omega_0}) \otimes R_i^{(0)}(\Omega)\\
    &= \left(-{\omega_0}-\Omega\right)^{-1}      \sum_{\omega_0} \int_{-\infty}^{+\infty}d\Omega~ \sum_i \left( e^{\beta({\omega_0}+\Omega)}-1 \right) S_i({\omega_0}) \otimes R_i^{(0)}(\Omega)\\
    &= -  \int_0^\beta dx~   \sum_{\omega_0} \int_{-\infty}^{+\infty}d\Omega~ \sum_i  e^{x({\omega_0}+\Omega)} S_i({\omega_0}) \otimes R_i^{(0)}(\Omega).
\end{align}
Here, $\omega_0$ are the Bohr frequencies of the bare Hamiltonian $H_\mathcal{S}^{(0)}$.

We obtained:
\begin{align}
    \kappa^{(1)} = - \sum_{\omega_0} \int_{-\infty}^{+\infty}d\Omega~ \sum_i \int_{0}^\beta  e^{x ({\omega_0}+\Omega)} S_i({\omega_0}) \otimes  R_i^{(0)}(\Omega). \label{eqn:FirstElementBeforeTrace}
\end{align}

Let us now calculate the second-order element given by equation \eqref{eqn:expansionBeforePartialTrace}.
\begin{align}
    &\kappa^{(2)}= \sum_{n_1,n_2=1}^{\infty} \frac{n_2  n_1}{n_2(n_2+n_{1})}\frac{(-\beta)^{n_2+n_1}}{n_2! n_1!}\mathcal{D}_{n_2} \mathcal{D}_{n_1} \\
    &=  \sum_{n_1,n_2=1}^{\infty} \frac{n_2  n_1}{n_2(n_2+n_{1})}\frac{(-\beta)^{n_2+n_1}}{n_2! n_1!} \nonumber\\
    & \times \sum_{{\omega_0},{\omega_0}^\prime} \int_{-\infty}^{+\infty}d\Omega\int_{-\infty}^{+\infty}d\Omega^\prime~ \sum_{ij} \left(-{\omega_0}-\Omega\right)^{n_2-1}\left(-{\omega_0}^\prime-\Omega^\prime\right)^{n_1-1}   S_i({\omega_0})S_j({\omega_0}^\prime) \otimes R_i^{(0)}(\Omega) R_j^{(0)}(\Omega^\prime)\\
    %%%%%%%%%%%%%%%%%%%
    &=    \sum_{{\omega_0},{\omega_0}^\prime} \int_{-\infty}^{+\infty}d\Omega\int_{-\infty}^{+\infty}d\Omega^\prime~ \sum_{ij}   S_i({\omega_0})S_j({\omega_0}^\prime) \otimes R_i^{(0)}(\Omega) R_j^{(0)}(\Omega^\prime) \nonumber \\
    &\times \sum_{n_1,n_2=1}^{\infty} \frac{n_2  n_1}{n_2(n_2+n_{1})}\frac{(-\beta)^{n_2+n_1}}{n_2! n_1!}\left(-{\omega_0}-\Omega\right)^{n_2-1}\left(-{\omega_0}^\prime-\Omega^\prime\right)^{n_1-1}.
\end{align}
In this case, in order to proceed, we need to perform some more involved algebraic manipulations. 
\begin{align}
    &(a+b)^n=\sum_{k=0}^n \binom{n}{k} a^{n-k} b^k,\\
    &\binom{n}{k} = \frac{n!}{k!(n-k)!}.
\end{align}
Moreover, we make use of the above elementary formula, that is employed to find closed form expressions for the series that appears in the following calculations.

\begin{align}
    &\sum_{n_1,n_2=1}^{\infty}\frac{n_2 n_1}{n_2(n_2+n_1)}\frac{(-\beta)^{n_2+n_1}}{n_2! n_1!}\left(-{\omega_0}-\Omega \right)^{n_2-1}    \left(-{\omega_0}^\prime-\Omega^\prime \right)^{n_1-1}\\
    &=\sum_{n_1,n_2=0}^{\infty}\frac{ 1}{n_2+n_1+2}\frac{(-\beta)^{n_2+n_1+2}}{(n_2+1)! n_1!}\left(-{\omega_0}-\Omega \right)^{n_2}    \left(-{\omega_0}^\prime-\Omega^\prime \right)^{n_1}\\
    &=\sum_{N=0}^{\infty} \sum_{n_1+n_2=N}\frac{ 1}{n_2+n_1+2}\frac{(-\beta)^{n_2+n_1+2}}{(n_2+1)! n_1!}\left(-{\omega_0}-\Omega \right)^{n_2}    \left(-{\omega_0}^\prime-\Omega^\prime \right)^{n_1}\\
    &=\sum_{N=0}^{\infty} \frac{(-\beta)^{N+2}}{N+2}\sum_{n_1=0}^{N}\frac{1}{(N+1-n_1)! n_1!}\left(-{\omega_0}-\Omega \right)^{N-n_1}    \left(-{\omega_0}^\prime-\Omega^\prime \right)^{n_1}\\
    &=\sum_{N=0}^{\infty} \frac{(-\beta)^{N+2}}{N+2}\left(\sum_{n_1=0}^{N+1}\frac{1}{(N+1-n_1)! n_1!}\left(-{\omega_0}-\Omega \right)^{N-n_1}    \left(-{\omega_0}^\prime-\Omega^\prime \right)^{n_1}
    -\frac{1}{ (N+1)!}\left(-{\omega_0}-\Omega \right)^{-1}    \left(-{\omega_0}^\prime-\Omega^\prime \right)^{N+1}\right)\\
    &=\left(-{\omega_0}-\Omega \right)^{-1} \sum_{N=0}^{\infty} \frac{(-\beta)^{N+2}}{(N+2)!}\left(\sum_{n_1=0}^{N+1}\frac{(N+1)!}{(N+1-n_1)! n_1!}\left(-{\omega_0}-\Omega \right)^{N+1-n_1}    \left(-{\omega_0}^\prime-\Omega^\prime \right)^{n_1}
    -   \left(-{\omega_0}^\prime-\Omega^\prime \right)^{N+1}\right)\\
    &=\left(-{\omega_0}-\Omega \right)^{-1} \sum_{N=0}^{\infty} \frac{(-\beta)^{N+2}}{(N+2)!}\left(\left(-{\omega_0}-\Omega-{\omega_0}^\prime-\Omega^\prime \right)^{N+1}
    -   \left(-{\omega_0}^\prime-\Omega^\prime \right)^{N+1}\right)\\
    &=\left(-{\omega_0}-\Omega \right)^{-1}\left(-{\omega_0}-\Omega-{\omega_0}^\prime-\Omega^\prime \right)^{-1} \sum_{N=0}^{\infty} \frac{(-\beta)^{N+2}}{(N+2)!}\left(-{\omega_0}-\Omega-{\omega_0}^\prime-\Omega^\prime \right)^{N+2}
    \nonumber\\
    &-\left(-{\omega_0}-\Omega \right)^{-1} \left(-{\omega_0}^\prime-\Omega^\prime \right)^{-1} \sum_{N=0}^{\infty} \frac{(-\beta)^{N+2}}{(N+2)!}
      \left(-{\omega_0}^\prime-\Omega^\prime \right)^{N+2}\\
    %%%%%%%%%%%%%
    &=\left(-{\omega_0}-\Omega \right)^{-1}\left(-{\omega_0}-\Omega-{\omega_0}^\prime-\Omega^\prime \right)^{-1} \left(e^{-\beta\left(-{\omega_0}-\Omega-{\omega_0}^\prime-\Omega^\prime \right)}-1+\beta\left(-{\omega_0}-\Omega-{\omega_0}^\prime-\Omega^\prime \right) \right) 
    \nonumber\\
    &-\left(-{\omega_0}-\Omega \right)^{-1} \left(-{\omega_0}^\prime-\Omega^\prime \right)^{-1}\left(e^{-\beta\left(-{\omega_0}^\prime-\Omega^\prime \right)}-1+\beta\left(-{\omega_0}^\prime-\Omega^\prime \right)\right) \\
    %%%%%%%%%%%%%
    &=\left(-{\omega_0}-\Omega \right)^{-1}\left(-{\omega_0}-\Omega-{\omega_0}^\prime-\Omega^\prime \right)^{-1} \left(e^{-\beta\left(-{\omega_0}-\Omega-{\omega_0}^\prime-\Omega^\prime \right)}-1 \right) 
    \nonumber\\
    &-\left(-{\omega_0}-\Omega \right)^{-1} \left(-{\omega_0}^\prime-\Omega^\prime \right)^{-1}\left(e^{-\beta\left(-{\omega_0}^\prime-\Omega^\prime \right)}- 1\right) \\
    %%%%%%%%%
    &=\left({\omega_0}+\Omega \right)^{-1}\left({\omega_0}+\Omega+{\omega_0}^\prime+\Omega^\prime \right)^{-1} \left(e^{\beta\left({\omega_0}+\Omega+{\omega_0}^\prime+\Omega^\prime \right)}-1 \right) 
    -\left({\omega_0}+\Omega \right)^{-1} \left({\omega_0}^\prime+\Omega^\prime \right)^{-1}\left(e^{\beta\left({\omega_0}^\prime+\Omega^\prime \right)}- 1\right) \\
    %%%%%%%%%
    &=\left({\omega_0}+\Omega \right)^{-1} \left(\int_0^\beta dx~ e^{x\left({\omega_0}+\Omega+{\omega_0}^\prime+\Omega^\prime \right)}
    - \int_0^\beta dx~ e^{x \left({\omega_0}^\prime+\Omega^\prime \right)} \right).
\end{align}
The above considerations, allow to find the following form of expansion coefficient $\kappa^{(2)}$:
\begin{align}
    &\kappa^{(2)}= \sum_{{\omega_0},{\omega_0}^\prime} \int_{-\infty}^{+\infty}d\Omega\int_{-\infty}^{+\infty}d\Omega^\prime~ \sum_{ij}   S_i({\omega_0})S_j({\omega_0}^\prime) \otimes R_i^{(0)}(\Omega) R_j^{(0)}(\Omega^\prime) \nonumber \\
    &\times \left({\omega_0}+\Omega \right)^{-1} \left(\int_0^\beta dx~ e^{x\left({\omega_0}+\Omega+{\omega_0}^\prime+\Omega^\prime \right)}
    - \int_0^\beta dx~ e^{x \left({\omega_0}^\prime+\Omega^\prime \right)} \right).
\end{align}

Let us now calculate the partial trace with respect to the reservoir $\mathcal{R}$, that induces the following expansion:
\begin{align}
    &Z^{-1}_{\mathcal{R},\beta} \tr_\mathcal{R} \left[e^{-\beta H_{\mathcal{S}+\mathcal{R}}}\right]=  \sum_{p=0}^{\infty} \lambda^p  \tr_\mathcal{R} \left[ \kappa^{(p)} \frac{e^{-\beta H_\mathcal{R}}}{Z_{\mathcal{R},\beta}} \right] e^{-\beta H_\mathcal{S}^{(0)}} = \sum_{p=0}^{\infty} \lambda^p \chi^{(p)} e^{-\beta H_\mathcal{S}^{(0)}},\\
    & \chi^{(p)} =  \tr_\mathcal{R} \left[ \kappa^{(p)} \rho_{\mathcal{R},\beta} \right]. \label{eqn:ExpansionAfterPartialTrace}
\end{align}

Let us now calculate the elements of the expansion  in equation \eqref{eqn:ExpansionAfterPartialTrace}. The zeroth-order coefficient $\chi^{(0)}$ is trivial: 
\begin{align}
    \chi^{(0)} = 1.
\end{align}

The first-order coefficient $\chi^{(1)}$ of the above expansion reads:
\begin{align}
     &\chi^{(1)} = \tr_\mathcal{R} \left[ \kappa^{(1)} \rho_{\mathcal{R},\beta} \right]\\
     &=- \tr_\mathcal{R} \left[  \sum_{\omega_0} \int_{-\infty}^{+\infty}d\Omega~ \sum_i \int_{0}^\beta  e^{x ({\omega_0}+\Omega)} S_i({\omega_0}) \otimes  R_i^{(0)}(\Omega) \rho_{\mathcal{R},\beta} \right]\\
    &=- \sum_{\omega_0} \int_{-\infty}^{+\infty}d\Omega~ \sum_i \int_{0}^\beta  e^{x ({\omega_0}+\Omega)} S_i({\omega_0})\tr_\mathcal{R} \left[ R_i^{(0)}(\Omega) \rho_{\mathcal{R},\beta} \right]\\
    %%%%%%%%%
    &=- \sum_{\omega_0}  \sum_i \int_{0}^\beta  e^{x {\omega_0}} S_i({\omega_0})\int_{-\infty}^{+\infty}d\Omega~\tr_\mathcal{R} \left[e^{x \Omega} R_i^{(0)}(\Omega) \rho_{\mathcal{R},\beta} \right]\\
    %%%%%%%%%
    &=- \sum_{\omega_0}  \sum_i \int_{0}^\beta  e^{x {\omega_0}} S_i({\omega_0})\int_{-\infty}^{+\infty}d\Omega~\tr_\mathcal{R} \left[e^{-x H_\mathcal{R}} R_i^{(0)}(\Omega)e^{+x H_\mathcal{R}}  \rho_{\mathcal{R},\beta} \right]\\
    %%%%%%%%%
    &=- \sum_{\omega_0}  \sum_i \int_{0}^\beta  e^{x {\omega_0}} S_i({\omega_0})\tr_\mathcal{R} \left[\int_{-\infty}^{+\infty}d\Omega~ R_i^{(0)}(\Omega) \rho_{\mathcal{R},\beta} \right]\\
    %%%%%%%%%
    &=- \sum_{\omega_0}  \sum_i   \frac{e^{\beta {\omega_0}}-1}{{\omega_0}} S_i({\omega_0})\tr_\mathcal{R} \left[ R_i^{(0)} \rho_{\mathcal{R},\beta} \right].
\end{align}

The calculations of the second-order coefficient $\chi^{(2)}$ are much more involving:
\begin{align}
    &\chi^{(2)} = \tr_\mathcal{R} \left[\kappa^{(2)} \rho_{\mathcal{R},\beta} \right]= \tr_\mathcal{R} \left[ \kappa_2^{(2)} \rho_{\mathcal{R},\beta} \right]\\
    %%%%%%%%%%%%%%%%%
    &=\sum_{{\omega_0},{\omega_0}^\prime} \int_{-\infty}^{+\infty}d\Omega\int_{-\infty}^{+\infty}d\Omega^\prime~ \sum_{ij}   S_i({\omega_0})S_j({\omega_0}^\prime)\tr_\mathcal{R} \left[  R_i^{(0)}(\Omega) R_j^{(0)}(\Omega^\prime) \rho_{\mathcal{R},\beta} \right] \nonumber\\
    &\times  \left({\omega_0}+\Omega \right)^{-1} \left(\int_0^\beta dx~ e^{x\left({\omega_0}+\Omega+{\omega_0}^\prime+\Omega^\prime \right)}
    - \int_0^\beta dx~ e^{x \left({\omega_0}^\prime+\Omega^\prime \right)} \right)  \\
    %%%%%%%%%%%%%%%%%%%%%%%%%%%
    &\stackrel{(I)}{=}\sum_{{\omega_0},{\omega_0}^\prime}\sum_{ij}  \int_{-\infty}^{+\infty}d\Omega\int_{-\infty}^{+\infty}d\Omega^\prime~   S_i({\omega_0})S_j({\omega_0}^\prime)\tr_\mathcal{R} \left[  R_i^{(0)}(\Omega) R_j^{(0)}(\Omega^\prime) \rho_{\mathcal{R},\beta} \right] \nonumber\\
    &\times \left({\omega_0}+\Omega \right)^{-1} \left(\int_0^\beta dx~ e^{x\left({\omega_0}+{\omega_0}^\prime \right)}
    - \int_0^\beta dx~ e^{x \left({\omega_0}^\prime-\Omega \right)} \right) \\
    %%%%%%%%%%%%%%%%%%%%%%%%%%%
    &=\sum_{{\omega_0},{\omega_0}^\prime} \int_{-\infty}^{+\infty}d\Omega\int_{-\infty}^{+\infty}d\Omega^\prime~ \sum_{ij}   S_i({\omega_0})S_j({\omega_0}^\prime)\tr_\mathcal{R} \left[  R_i^{(0)}(\Omega) R_j^{(0)}(\Omega^\prime) \rho_{\mathcal{R},\beta} \right]   \int_0^\beta dx~ e^{x({\omega_0}+{\omega_0}^\prime)} \int_0^{x}dy~e^{-y({\omega_0}+\Omega)}  \\
    %%%%%%%%%%%%%%%%%%%%%%%%%%%
    &=\sum_{{\omega_0},{\omega_0}^\prime}  \sum_{ij}   S_i({\omega_0})S_j({\omega_0}^\prime)  \int_{-\infty}^{+\infty}d\Omega~\int_0^\beta dx~ e^{x({\omega_0}+{\omega_0}^\prime)}  \int_0^{x}dy~e^{-y({\omega_0}+\Omega)}\tr_\mathcal{R} \left[  R_i^{(0)}(\Omega) R_j^{(0)} \rho_{\mathcal{R},\beta} \right], \label{eqn:Key_Integral}   
    %%%%%%%%%%%%%%%%%%%%%%%%%%%
    % &= \sum_{{\omega_0},{\omega_0}^\prime}  \sum_{ij}   S_i({\omega_0})S_j({\omega_0}^\prime)  \int_0^\beta dx~ e^{x({\omega_0}+{\omega_0}^\prime)}  \int_0^{x}dy~e^{-y{\omega_0}}\int_{-\infty}^{+\infty}d\Omega~ e^{-y\Omega}\tr_\mathcal{R} \left[  R_i^{(0)}(\Omega) R_j^{(0)} \rho_{\mathcal{R},\beta} \right]   \\
    % %%%%%%%%%%%%%%%%%%%%%%%%%%%
    % &\stackrel{(II)}{=}i\sum_{{\omega_0},{\omega_0}^\prime}  \sum_{ij}   S_i({\omega_0})S_j({\omega_0}^\prime)  \int_0^\beta dx~ e^{x({\omega_0}+{\omega_0}^\prime)} \int_0^{-ix}dt~e^{-i{\omega_0} t}\int_{-\infty}^{+\infty}d\Omega~ e^{-i\Omega t}\tr_\mathcal{R} \left[  R_i^{(0)}(\Omega) R_j^{(0)} \rho_{\mathcal{R},\beta} \right]  \\
    % %%%%%%%%%%%%%%%%%%%%%%%%%%%
    % &=i\sum_{{\omega_0},{\omega_0}^\prime}  \sum_{ij}   S_i({\omega_0})S_j({\omega_0}^\prime)  \int_0^\beta dx~ e^{x({\omega_0}+{\omega_0}^\prime)}  \int_0^{-ix}dt~e^{-i{\omega_0} t} \tr_\mathcal{R} \left[  \tilde{R}_i^{(0)}(t) R_j^{(0)} \rho_{\mathcal{R},\beta} \right], \label{eqn:Integral}
\end{align}
where in the step $(I)$ we used the properties of interaction operators described in Section \ref{sec:app:Properties} of the Appendix.

In the next step we proceed with calculating the integral in equation \eqref{eqn:Key_Integral}. The integration involves splitting the well defined integrals (integrand with no poles) into several principal value integrals.
\begin{align}
    &\int_{-\infty}^{+\infty}d\Omega~\int_0^\beta dx~ e^{x({\omega_0}+{\omega_0}^\prime)}  \int_0^{x}dy~e^{-y({\omega_0}+\Omega)}\tr_\mathcal{R} \left[  R_i^{(0)}(\Omega) R_j^{(0)} \rho_{\mathcal{R},\beta} \right]\\
    &=\frac{1}{2\pi}\int_{-\infty}^{+\infty}d\Omega~\int_0^\beta dx~ e^{x({\omega_0}+{\omega_0}^\prime)}  \int_0^{x}dy~e^{-y({\omega_0}+\Omega)} \gamma_{ij} (\Omega)\\
    %%%%%%%%%%%%%%%
    &=\frac{1}{2\pi}\int_{-\infty}^{+\infty}d\Omega~ 
    \left[
    \frac{1}{\omega_0+\omega_0^\prime}\frac{1}{\omega_0^\prime-\Omega}
    +\frac{e^{\beta(\omega_0^\prime-\Omega)}}{\omega_0+\omega_0^\prime} \left(\frac{1}{-\omega_0^\prime+\Omega}-\frac{1}{\omega_0+\Omega}\right)
    -\frac{e^{\beta(\omega_0+\omega_0^\prime)}}{\omega_0+\omega_0^\prime}\frac{1}{-\omega_0-\Omega}
    \right] \gamma_{ij} (\Omega)\\
    %%%%%%%%%%%%%%%%%%%%%%%%%%%%%%
    &=\frac{1}{\omega_0+\omega_0^\prime}\frac{1}{2\pi}\dashint_{-\infty}^{+\infty}d\Omega~ \frac{1}{\omega_0^\prime-\Omega}\gamma_{ij} (\Omega)
    +\frac{e^{\beta \omega_0^\prime}}{\omega_0+\omega_0^\prime}\frac{1}{2\pi}\dashint_{-\infty}^{+\infty}d\Omega~ \left(\frac{1}{-\omega_0^\prime+\Omega}-\frac{1}{\omega_0+\Omega}\right)\gamma_{ij} (\Omega)e^{-\beta \Omega}\\
    &-\frac{e^{\beta(\omega_0+\omega_0^\prime)}}{\omega_0+\omega_0^\prime}  \frac{1}{2\pi}\dashint_{-\infty}^{+\infty}d\Omega~ \frac{1}{-\omega_0-\Omega} \gamma_{ij} (\Omega)\\
    %%%%%%%%%%%%%%%%%%%%%%%%%%%%%%
    &\stackrel{(I)}{=}\frac{1}{\omega_0+\omega_0^\prime}\frac{1}{2\pi}\dashint_{-\infty}^{+\infty}d\Omega~ \frac{1}{\omega_0^\prime-\Omega}\gamma_{ij} (\Omega)
    +\frac{e^{\beta \omega_0^\prime}}{\omega_0+\omega_0^\prime}\frac{1}{2\pi}\dashint_{-\infty}^{+\infty}d\Omega~ \left(\frac{1}{-\omega_0^\prime-\Omega}-\frac{1}{\omega_0-\Omega}\right)\gamma_{ji} (\Omega)\\
    &-\frac{e^{\beta(\omega_0+\omega_0^\prime)}}{\omega_0+\omega_0^\prime}  \frac{1}{2\pi}\dashint_{-\infty}^{+\infty}d\Omega~ \frac{1}{-\omega_0-\Omega} \gamma_{ij} (\Omega)\\
    %%%%%%%%%%%%%%%%%%%%%%%%%%%%%%
    &=\frac{1}{\omega_0+\omega_0^\prime}\frac{1}{2\pi}\dashint_{-\infty}^{+\infty}d\Omega~ \frac{1}{\omega_0^\prime-\Omega}\gamma_{ij} (\Omega)
    +\frac{e^{\beta \omega_0^\prime}}{\omega_0+\omega_0^\prime}\left(\frac{1}{2\pi}\dashint_{-\infty}^{+\infty}d\Omega~ \frac{1}{-\omega_0^\prime-\Omega}\gamma_{ji} (\Omega)-\frac{1}{2\pi}\dashint_{-\infty}^{+\infty}d\Omega~\frac{1}{\omega_0-\Omega}\gamma_{ji} (\Omega)\right)\\
    &-\frac{e^{\beta(\omega_0+\omega_0^\prime)}}{\omega_0+\omega_0^\prime}  \frac{1}{2\pi}\dashint_{-\infty}^{+\infty}d\Omega~ \frac{1}{-\omega_0-\Omega} \gamma_{ij} (\Omega)\\
    %%%%%%%%%%%%%%%%%%%%%%%%%%%%%%
    &\stackrel{(II)}{=}\frac{1}{\omega_0+\omega_0^\prime} \mathcal{S}^{(0)}_{ij}(\omega_0^\prime)
    +\frac{e^{\beta \omega_0^\prime}}{\omega_0+\omega_0^\prime}\left(\mathcal{S}^{(0)}_{ji}(-\omega_0^\prime)-\mathcal{S}^{(0)}_{ji}(\omega_0)\right)-\frac{e^{\beta(\omega_0+\omega_0^\prime)}}{\omega_0+\omega_0^\prime}  \mathcal{S}^{(0)}_{ij}(-\omega_0)\\
    %%%%%%%%%%%%%%%%%%%%%%%%%%%%%%
    &=-\frac{1}{\omega_0^\prime+\omega_0} \left(e^{\beta(\omega_0+\omega_0^\prime)}  \mathcal{S}^{(0)}_{ij}(-\omega_0) 
    -e^{\beta \omega_0^\prime}\left(\mathcal{S}^{(0)}_{ji}(-\omega_0^\prime)-\mathcal{S}^{(0)}_{ji}(\omega_0)\right)-\mathcal{S}^{(0)}_{ij}(\omega_0^\prime) \right),
\end{align}
where in $(I)$ we used the relation:
\begin{align}
    \gamma_{ji} (-\Omega)=\gamma_{ij} (\Omega)e^{-\beta \Omega},
\end{align}
and subsequently we changed the variable of integration. In the step $(II)$ we used the formula for $\mathcal{S}_{ij}(\omega)$ in equation \eqref{eqn:S_IntegralRepresentation}.

We obtain:
\begin{align}
    \chi^{(2)} &=-\sum_{{\omega_0},{\omega_0}^\prime}  \sum_{ij}   S_i({\omega_0})S_j({\omega_0}^\prime)  \frac{1}{\omega_0^\prime+\omega_0} \left(e^{\beta(\omega_0+\omega_0^\prime)}  \mathcal{S}^{(0)}_{ij}(-\omega_0) 
    -e^{\beta \omega_0^\prime}\left(\mathcal{S}^{(0)}_{ji}(-\omega_0^\prime)-\mathcal{S}^{(0)}_{ji}(\omega_0)\right)-\mathcal{S}^{(0)}_{ij}(\omega_0^\prime) \right)\\
    %%%%%%%%%%%%%
    &=-\sum_{{\omega_0},{\omega_0}^\prime}  \sum_{ij}   S^\dagger_i({\omega_0})S_j({\omega_0}^\prime)  \frac{1}{\omega_0^\prime-\omega_0} \left(e^{\beta(\omega_0^\prime-\omega_0)}  \mathcal{S}^{(0)}_{ij}(\omega_0) 
    -e^{\beta \omega_0^\prime}\left(\mathcal{S}^{(0)}_{ji}(-\omega_0^\prime)-\mathcal{S}^{(0)}_{ji}(-\omega_0)\right)-\mathcal{S}^{(0)}_{ij}(\omega_0^\prime) \right),
\end{align}
where suffix ``${}^{(0)}$" corresponds to quantities evaluated with $\tilde{R}_i^{(0)}$ that are not centered.

% We obtain:
% \begin{align}
%     &\int_0^{-ix}dt~e^{-i{\omega_0} t} \tr_\mathcal{R} \left[  \tilde{R}_i^{(0)}(t) R_j^{(0)} \rho_{\mathcal{R},\beta} \right] =  i\mathcal{S}^{(0)}_{ij}(-{\omega_0}), 
%     % \\ &\int_0^{-i(\beta-x)}dt~ e^{i{\omega_0}^\prime t} \tr_\mathcal{R} \left[ \tilde{R}_i^{(0)}(t) R_j^{(0)} \rho_{\mathcal{R},\beta} \right] = i\mathcal{S}^{(0)}_{ij}({\omega_0}^\prime),
% \end{align}
% where suffix ``${}^{(0)}$" corresponds to quantities evaluated with $\tilde{R}_i^{(0)}$ that are not centered. Hence
% \begin{align}
%     &\chi^{(2)}=-  \sum_{{\omega_0},{\omega_0}^\prime}  \sum_{ij}   S_i({\omega_0})S_j({\omega_0}^\prime)  \int_0^\beta dx~ e^{x({\omega_0}+{\omega_0}^\prime)}  \mathcal{S}^{(0)}_{ij}(-{\omega_0}) \\
%     &=-  \sum_{{\omega_0},{\omega_0}^\prime}  \sum_{ij}   S_i^\dagger({\omega_0})S_j({\omega_0}^\prime)  \int_0^\beta dx~ e^{x({\omega_0}^\prime-{\omega_0})}  \mathcal{S}^{(0)}_{ij}({\omega_0}) ,\\
%     &=- \sum_{{\omega_0},{\omega_0}^\prime}  \sum_{ij}   S_i^\dagger({\omega_0})S_j({\omega_0}^\prime)  \frac{e^{\beta({\omega_0}^\prime-{\omega_0})}-1}{{\omega_0}^\prime-{\omega_0}}   \mathcal{S}^{(0)}_{ij}({\omega_0}) .
% \end{align}
We obtain:
\begin{align}
    &Z^{-1}_{\mathcal{R},\beta} \tr_\mathcal{R} \left[e^{-\beta H_{\mathcal{S}+\mathcal{R}}}\right]=  \sum_{p=0}^{\infty} \lambda^p \chi^{(p)} e^{-\beta H_\mathcal{S}^{(0)}},\\
    &\chi^{(0)} = 1,\\
    &\chi^{(1)} = -\sum_{\omega_0}  \sum_i   \frac{e^{\beta {\omega_0}}-1}{{\omega_0}} S_i({\omega_0})\tr_\mathcal{R} \left[ R_i^{(0)} \rho_{\mathcal{R},\beta} \right], \label{eqn:ExpansionTotalst1Order}\\
    &\chi^{(2)} = -\sum_{{\omega_0},{\omega_0}^\prime}  \sum_{ij}   S^\dagger_i({\omega_0})S_j({\omega_0}^\prime)  \frac{1}{\omega_0^\prime-\omega_0} \left(e^{\beta(\omega_0^\prime-\omega_0)}  \mathcal{S}^{(0)}_{ij}(\omega_0) 
    -e^{\beta \omega_0^\prime}\left(\mathcal{S}^{(0)}_{ji}(-\omega_0^\prime)-\mathcal{S}^{(0)}_{ji}(-\omega_0)\right)-\mathcal{S}^{(0)}_{ij}(\omega_0^\prime) \right), \label{eqn:ExpansionTota2ndOrder} \\
    &\dots,\\
    & \chi^{(p)} =  \tr_\mathcal{R} \left[ \kappa^{(p)} \rho_{\mathcal{R},\beta} \right].
\end{align}

In the next step we proceed to find the corrections $H_{mf,\mathcal{C}}^{(k)}$. Let us remind a part of equation \eqref{eqn:hint}.
\begin{align}
    &e^{-\beta H^{(mf)}_\mathcal{S}} = e^{-\beta \left(H_\mathcal{S}^{(0)}(t)+H_{mf,\mathcal{C}}\right)} = \frac{1}{Z_{\mathcal{R},\beta}} \tr_\mathcal{R}\left[e^{-\beta H_{\mathcal{S}+\mathcal{R}}} \right].
\end{align}

In this place, we use expansion in equation \eqref{eqn:ExpansionZassenhous} again, and we denote with $\varpi^{(k)}$ its coefficients.
\begin{align}\label{eqn:ExpansionVaraible}
    &e^{-\beta H^{(mf)}_\mathcal{S}} = \left\{ 1+\sum_{p=1}^{\infty}\sum_{n_1, ...,n_p=1}^{\infty}\frac{n_p \cdots n_1}{n_p(n_p+n_{p-1})\cdots(n_p+\cdots+n_1)}\frac{(-\beta)^{n_p+\cdots+n_1}}{n_p!\cdots n_1!}\mathcal{C}_{n_p}(\lambda)\cdots \mathcal{C}_{n_1}(\lambda)\right\}e^{-\beta H^{(0)}_\mathcal{S}} \\
    &=\sum_{p=0}^{\infty} \lambda^p \varpi^{(p)} e^{-\beta H^{(0)}_\mathcal{S}},
\end{align}
where $\varpi^{(0)}=1$, and: 
\begin{align}
    &H_{mf,\mathcal{C}}=\sum_{k=1}^{\infty}\lambda^k H_{mf,\mathcal{C}}^{(k)},\\
    & \mathcal{C}_m (\lambda)=\left(\mathcal{L}_{ H^{(0)}_\mathcal{S}}\right)^{m-1}  H_{mf,\mathcal{C}}.
\end{align}

We find the coefficients $\varpi^{(p)}$.
\begin{align}
    &\varpi^{(1)}= \sum_{n_1=1}^{\infty}\frac{(-\beta)^{n_1}}{ n_1!} \left(\mathcal{L}_{ H^{(0)}_\mathcal{S}}\right)^{n_1-1}  H_{mf,\mathcal{C}}^{(1)}, \\
    &\varpi^{(2)}=\sum_{n_1=1}^{\infty}\frac{(-\beta)^{n_1}}{ n_1!} \left(\mathcal{L}_{ H^{(0)}_\mathcal{S}}\right)^{n_1-1}  H_{mf,\mathcal{C}}^{(2)}
    +\sum_{n_1,n_2=1}^{\infty}\frac{n_2  n_1}{n_2(n_2+n_{1})}\frac{(-\beta)^{n_2+n_1}}{n_2! n_1!}\left(\mathcal{L}_{ H^{(0)}_\mathcal{S}}\right)^{n_2-1}  H_{mf,\mathcal{C}}^{(1)}\left(\mathcal{L}_{ H^{(0)}_\mathcal{S}}\right)^{n_1-1}  H_{mf,\mathcal{C}}^{(1)},\\
    &\vdots \nonumber\\
    &\varpi^{(p)}=\cdots.
\end{align}

By comparing equations (\ref{eqn:ExpansionTotalst1Order}) and (\ref{eqn:ExpansionVaraible}), we find that at the first-order:
\begin{align}
    \sum_{n_1=1}^{\infty}\frac{(-\beta)^{n_1}}{ n_1!} \left(\mathcal{L}_{ H^{(0)}_\mathcal{S}}\right)^{n_1-1}  H_{mf,\mathcal{C}}^{(1)}=-\sum_{\omega_0}  \sum_i   \frac{e^{\beta {\omega_0}}-1}{{\omega_0}} S_i({\omega_0})\tr_\mathcal{R} \left[ R_i^{(0)} \rho_{\mathcal{R},\beta} \right]. \label{eqn:Comparison1stOrder}
\end{align}
To solve the equation above for the first unknown correction $H_{mf,\mathcal{C}}^{(1)}$, we expand the correction into eigenoperators of $\mathcal{L}_{ H^{(0)}_\mathcal{S}}$ of a $I$-kind in equation \eqref{eqn:Eigenoperators1Type}.
\begin{align}
    H_{mf,\mathcal{C}}^{(1)} =\sum_{{\omega_0},i} \Upsilon^{(1)}_i({\omega_0}) S_i({\omega_0}).
\end{align}

The l.h.s. of equation \eqref{eqn:Comparison1stOrder} is equivalent to:
\begin{align}
    &\sum_{n_1=1}^{\infty}\frac{(-\beta)^{n_1}}{ n_1!} \left(\mathcal{L}_{ H^{(0)}_\mathcal{S}}\right)^{n_1-1}  H_{mf,\mathcal{C}}^{(1)} 
    =\sum_{n_1=1}^{\infty}\frac{(-\beta)^{n_1}}{ n_1!}  \sum_{{\omega_0},i} \Upsilon^{(1)}_i({\omega_0}) \left(\mathcal{L}_{ H^{(0)}_\mathcal{S}}\right)^{n_1-1} S_i({\omega_0})\\
    &=\sum_{n_1=1}^{\infty}\frac{(-\beta)^{n_1}}{ n_1!}  \sum_{{\omega_0},i} \Upsilon^{(1)}_i({\omega_0}) \left(-{\omega_0}\right)^{n_1-1} S_i({\omega_0})
    =-\frac{1}{{\omega_0}}\sum_{{\omega_0},i}\sum_{n_1=1}^{\infty}\frac{(\beta{\omega_0})^{n_1}}{ n_1!}   \Upsilon^{(1)}_i({\omega_0})  S_i({\omega_0})\\
    &=-\sum_{{\omega_0},i}\frac{e^{\beta {\omega_0}}-1}{{\omega_0}}  \Upsilon^{(1)}_i({\omega_0})  S_i({\omega_0}),
\end{align}
therefore we can rewrite equation \eqref{eqn:Comparison1stOrder}
\begin{align}
    -\sum_{{\omega_0},i}\frac{e^{\beta {\omega_0}}-1}{{\omega_0}}  \Upsilon^{(1)}_i({\omega_0})  S_i({\omega_0}) = -\sum_{\omega_0}  \sum_i   \frac{e^{\beta {\omega_0}}-1}{{\omega_0}} S_i({\omega_0})\tr_\mathcal{R} \left[ R_i^{(0)} \rho_{\mathcal{R},\beta} \right].
\end{align}
By comparing l.h.s. and r.h.s. of the equation above we find the coefficients $\Upsilon^{(1)}_i({\omega_0})$
\begin{align}
    \Upsilon^{(1)}_i({\omega_0})=\tr_\mathcal{R} \left[ R_i^{(0)} \rho_{\mathcal{R},\beta} \right].
\end{align}
Finally:
\begin{align}
    H_{mf,\mathcal{C}}^{(1)} =\sum_{{\omega_0}}\sum_i S_i({\omega_0}) \tr_\mathcal{R} \left[ R_i^{(0)} \rho_{\mathcal{R},\beta} \right] =\sum_i S_i \tr_\mathcal{R} \left< R_i^{(0)}  \right>_{\rho_{\mathcal{R},\beta}},
\end{align}
and
\begin{align}
    H_\mathcal{S}^{(mf,1)}=H_\mathcal{S}^{(0)}+\lambda H_{mf,\mathcal{C}}^{(1)} \equiv H_\mathcal{S}^{(1)}.
\end{align}
The correction $H_{mf,\mathcal{C}}^{(1)}$ in the equation above is equivalent to centering of interaction correction known from Davies-GKSL equation (see Section \ref{sec:MME} of the main text).

Let us recall now the definition of centered interaction operators $R_i^{(1)}$ (see the main text):
\begin{align}
    R_i^{(1)}=R_i^{(0)}-  \tr_\mathcal{R} \left[ R_i^{(0)} \rho_{\mathcal{R},\beta} \right].
\end{align}
We assume now, without the loss of generality and for the sake of ease of calculations, that from the very beginning we worked with the centralized operators. In this case $H_{mf,\mathcal{C}}^{(1)}=0$, and $\varpi^{(2)}$ has a simplified form. 
\begin{align}\label{eqn:expansionVariable2}
    \varpi^{(2)}=\sum_{n_1=1}^{\infty}\frac{(-\beta)^{n_1}}{ n_1!} \left(\mathcal{L}_{ H^{(mf,1)}_\mathcal{S}}\right)^{n_1-1}  H_{mf,\mathcal{C}}^{(2)}. 
\end{align}
We compare now equations (\ref{eqn:ExpansionTota2ndOrder}) and (\ref{eqn:expansionVariable2}). We note that now, the quantites $\mathcal{S}_{ij}(\omega)$ contain the interaction operators that are already centered.
\begin{align}
    &\sum_{n_1=1}^{\infty}\frac{(-\beta)^{n_1}}{ n_1!} \left(\mathcal{L}_{ H^{(mf,1)}_\mathcal{S}}\right)^{n_1-1}  H_{mf,\mathcal{C}}^{(2)} \nonumber \\
    &=-\sum_{{\omega_1},{\omega_1}^\prime}  \sum_{ij}   S^\dagger_i({\omega_1})S_j({\omega_1}^\prime)  \frac{1}{\omega_1^\prime-\omega_1} \left(e^{\beta(\omega_1^\prime-\omega_1)}  \mathcal{S}_{ij}(\omega_1) 
    -e^{\beta \omega_1^\prime}\left(\mathcal{S}_{ji}(-\omega_1^\prime)-\mathcal{S}_{ji}(-\omega_1)\right)-\mathcal{S}_{ij}(\omega_1^\prime) \right).
    \label{eqn:Comparison2ndtOrder}
\end{align}
Let us search for $H_{mf,\mathcal{C}}^{(2)}$, by expanding it into eigenoperators of $\mathcal{L}_{ H^{(mf,1)}_\mathcal{S}}$ of a $II$-kind in equation \eqref{eqn:Eigenoperators2Type}.
\begin{align}
    &H_{mf,\mathcal{C}}^{(2)} = \sum_{{\omega_1},{\omega_1}^\prime} \sum_{ij} \Upsilon^{(2)}_{ij}({\omega_1},{\omega_1}^\prime) S_i^\dagger({\omega_1})S_j({\omega_1}^\prime).
    % \\
    % &\Upsilon^{(2)}_{ij}({\omega_1},{\omega_1}^\prime) = \frac{1}{e^{\beta (\omega_1^\prime-\omega_1)}-1} \left( e^{\beta (\omega_1^\prime-\omega_1)} \mathcal{S}_{ij}(\omega_1)-e^{\beta \omega_1^\prime}\mathcal{S}_{ji}(-\omega_1) +e^{\beta \omega_1^\prime}\mathcal{S}_{ji}(-\omega_1^\prime)- \mathcal{S}_{ij}(\omega_1^\prime)\right)
\end{align}
The l.h.s. of the equation above is equivalent to:
\begin{align}
    &\sum_{n_1=1}^{\infty}\frac{(-\beta)^{n_1}}{ n_1!} \left(\mathcal{L}_{ H^{(mf,1)}_\mathcal{S}}\right)^{n_1-1} H_{mf,\mathcal{C}}^{(2)}
    =\sum_{{\omega_1},{\omega_1}^\prime} \sum_{ij}\sum_{n_1=1}^{\infty}\frac{(-\beta)^{n_1}}{ n_1!}\Upsilon^{(2)}_{ij}({\omega_1},{\omega_1}^\prime) \left(\mathcal{L}_{ H^{(mf,1)}_\mathcal{S}}\right)^{n_1-1}   S_i^\dagger({\omega_1})S_j({\omega_1}^\prime)\\
    &=\sum_{{\omega_1},{\omega_1}^\prime} \sum_{ij}\sum_{n_1=1}^{\infty}\frac{(-\beta)^{n_1}}{ n_1!}\Upsilon^{(2)}_{ij}({\omega_1},{\omega_1}^\prime) \left({\omega_1}-{\omega_1}^\prime\right)^{n_1-1}   S_i^\dagger({\omega_1})S_j({\omega_1}^\prime)\\
    &=-\frac{1}{{\omega_1}^\prime-{\omega_1}}\sum_{{\omega_1},{\omega_1}^\prime} \sum_{ij}\sum_{n_1=1}^{\infty}\frac{\left(\beta ({\omega_1}^\prime-{\omega_1})\right)^{n_1}}{ n_1!}\Upsilon^{(2)}_{ij}({\omega_1},{\omega_1}^\prime) S_i^\dagger({\omega_1})S_j({\omega_1}^\prime)\\
    %%%%%%%%%%%%%%%%%%%%%%%%%%%%%%%%%
    &=-\sum_{{\omega_1},{\omega_1}^\prime}\sum_{ij}\Upsilon^{(2)}_{ij}({\omega_1},{\omega_1}^\prime)\frac{e^{\beta ({\omega_1}^\prime-{\omega_1})}-1}{{\omega_1}^\prime-{\omega_1}} S_i^\dagger({\omega_1})S_j({\omega_1}^\prime).
\end{align}
By substituting the result above to equation \eqref{eqn:Comparison2ndtOrder} we obtain:
\begin{align}
    &-\sum_{{\omega_1},{\omega_1}^\prime}\sum_{ij}\Upsilon^{(2)}_{ij}({\omega_1},{\omega_1}^\prime)\frac{e^{\beta ({\omega_1}^\prime-{\omega_1})}-1}{{\omega_1}^\prime-{\omega_1}} S_i^\dagger({\omega_1})S_j({\omega_1}^\prime)\nonumber\\
    &= -\sum_{{\omega_1},{\omega_1}^\prime}  \sum_{ij}   S^\dagger_i({\omega_1})S_j({\omega_1}^\prime)  \frac{1}{\omega_1^\prime-\omega_1} \left(e^{\beta(\omega_1^\prime-\omega_1)}  \mathcal{S}_{ij}(\omega_1) 
    -e^{\beta \omega_1^\prime}\left(\mathcal{S}_{ji}(-\omega_1^\prime)-\mathcal{S}_{ji}(-\omega_1)\right)-\mathcal{S}_{ij}(\omega_1^\prime) \right),
\end{align}
we obtain the coefficients $\Upsilon^{(2)}_{ij}({\omega_1},{\omega_1}^\prime)$
\begin{align}
    \Upsilon^{(2)}_{ij}({\omega_1},{\omega_1}^\prime) =
    \frac{1}{e^{\beta ({\omega_1}^\prime-{\omega_1})}-1} \left(e^{\beta(\omega_1^\prime-\omega_1)}  \mathcal{S}_{ij}(\omega_1) 
    -e^{\beta \omega_1^\prime}\left(\mathcal{S}_{ji}(-\omega_1^\prime)-\mathcal{S}_{ji}(-\omega_1)\right)-\mathcal{S}_{ij}(\omega_1^\prime) \right),
\end{align}
where the elements for which $\omega_1^\prime=\omega_1$ are determined via a limiting procedure $\omega_1^\prime \to \omega_1$.
\begin{align}
    &\Upsilon^{(2)}_{ij}({\omega_1},{\omega_1})= \lim_{\omega_1^\prime \to \omega_1} \Upsilon^{(2)}_{ij}({\omega_1},{\omega_1}^\prime) \\
    &= \lim_{\omega_1^\prime \to \omega_1} \frac{1}{e^{\beta ({\omega_1}^\prime-{\omega_1})}-1} \left(e^{\beta(\omega_1^\prime-\omega_1)}  \mathcal{S}_{ij}(\omega_1) 
    -e^{\beta \omega_1^\prime}\left(\mathcal{S}_{ji}(-\omega_1^\prime)-\mathcal{S}_{ji}(-\omega_1)\right)-\mathcal{S}_{ij}(\omega_1^\prime) \right) \\
    &=\lim_{\omega_1^\prime \to \omega_1} \frac{1}{e^{\beta ({\omega_1}^\prime-{\omega_1})}-1} \left( e^{\beta \omega_1^\prime} \left(e^{-\beta\omega_1}  \mathcal{S}_{ij}(\omega_1) -e^{-\beta \omega_1^\prime}\mathcal{S}_{ij}(\omega_1^\prime)\right)
    -e^{\beta \omega_1^\prime}\left(\mathcal{S}_{ji}(-\omega_1^\prime)-\mathcal{S}_{ji}(-\omega_1)\right) \right) \\
    &=\lim_{\omega_1^\prime \to \omega_1}  \left( \frac{e^{\beta \omega_1^\prime} (\omega_1-\omega_1^\prime)}{e^{\beta ({\omega_1}^\prime-{\omega_1})}-1} \frac{e^{-\beta\omega_1}  \mathcal{S}_{ij}(\omega_1) -e^{-\beta \omega_1^\prime} \mathcal{S}_{ij}(\omega_1^\prime)}{\omega_1-\omega_1^\prime}
    -\frac{e^{\beta \omega_1^\prime}(\omega_1-\omega_1^\prime)}{e^{\beta ({\omega_1}^\prime-{\omega_1})}-1} \frac{\mathcal{S}_{ji}(-\omega_1^\prime)-\mathcal{S}_{ji}(-\omega_1)}{\omega_1-\omega_1^\prime} \right)\\
    &=-\frac{e^{\beta \omega_1}}{\beta} \frac{\partial}{\partial \omega} \mathcal{S}_{ij}(\omega_1) e^{-\beta \omega_1} - \frac{e^{\beta \omega_1}}{\beta} \frac{\partial}{\partial \omega} \mathcal{S}_{ji}(-\omega_1) =\mathcal{S}_{ij}(\omega_1) - \frac{1}{\beta} \left(\frac{\partial}{\partial \omega} \mathcal{S}_{ij}(\omega_1)+e^{\beta \omega_1}\frac{\partial}{\partial \omega} \mathcal{S}_{ji}(-\omega_1)  \right).
\end{align}
Therefore:
\begin{align}
    &\Upsilon^{(2)}_{ij}({\omega_1},{\omega_1})= \mathcal{S}_{ij}(\omega_1) - \frac{1}{\beta} \left(\frac{\partial}{\partial \omega} \mathcal{S}_{ij}(\omega_1)+e^{\beta \omega_1}\frac{\partial}{\partial \omega} \mathcal{S}_{ji}(-\omega_1)  \right).
\end{align}
Using formula in equation \eqref{eqn:S_IntegralRepresentation} and detailed-balance conditions $\gamma_{ji}(-\Omega)=e^{-\beta \Omega} \gamma_{ij}(\Omega)$ we find that:
\begin{align}
\frac{\partial}{\partial \omega} \mathcal{S}_{ij}(\omega_1)+e^{\beta \omega_1}\frac{\partial}{\partial \omega} \mathcal{S}_{ji}(-\omega_1)
=\frac{1}{2\pi} \dashint_{-\infty}^{+\infty}d\Omega~\frac{e^{\beta(\omega_1-\Omega)}-1}{(\omega_1-\Omega)^2} \gamma_{ij}(\Omega).
\end{align}
% \color{red}
% If then the temperature is high enough (comparing to the cut-off frequency) such that an approximation (inside the integral) $e^{\beta(\omega_1-\Omega)}-1 \approx \beta(\omega_1-\Omega)$ is justified, we have:
% \begin{align}
%     \frac{\partial}{\partial \omega} \mathcal{S}_{ij}(\omega_1)+e^{\beta \omega_1}\frac{\partial}{\partial \omega} \mathcal{S}_{ji}(-\omega_1) \approx \frac{\beta}{2 \pi} \dashint_{-\infty}^{+\infty}d\Omega~ \frac{1}{\omega_1-\Omega} \gamma_{ij}(\Omega) = \beta \mathcal{S}_{ij}(\omega_1).
% \end{align}
% Therefore, we infer that for temperatures (small $\beta$) large enough
% \begin{align}
%     \Upsilon^{(2)}_{ij}({\omega_1},{\omega_1}) \approx 0.
% \end{align}
% The above result suggest that at temperatures high enough, the shift of the energy levels due to an interaction with the environment is negligible. This agrees with the predictions of QFT suggesting that at energy-scales that are high enough particles behave as bare species.

\color{black}
In this way, we obtain the form of the second correction $H_{mf,\mathcal{C}}^{(2)}$:
\begin{align}
    H_{mf,\mathcal{C}}^{(2)} = \sum_{{\omega_1},{\omega_1}^\prime} \sum_{ij} \frac{1}{e^{\beta ({\omega_1}^\prime-{\omega_1})}-1} \left(e^{\beta(\omega_1^\prime-\omega_1)}  \mathcal{S}_{ij}(\omega_1) 
    -e^{\beta \omega_1^\prime}\left(\mathcal{S}_{ji}(-\omega_1^\prime)-\mathcal{S}_{ji}(-\omega_1)\right)-\mathcal{S}_{ij}(\omega_1^\prime) \right) S_i^\dagger({\omega_1})S_j({\omega_1}^\prime), \label{eqn:2ndCorrection0}
\end{align}
where $\mathcal{S}_{ij}(\omega)$ follows the standard definition in equation \eqref{eqn:def:coe1}.

To check that the obtained form of the second-order correction $H_{mf,\mathcal{C}}^{(2)}$ is Hermitian, it is enough to see that:
\begin{align}
    {\Upsilon^{(2)}_{ji}}^*({\omega_1}^\prime,{\omega_1})=\Upsilon^{(2)}_{ij}({\omega_1},{\omega_1}^\prime).
\end{align}

We finish this Section by writing down the second-order approximation to the reduced state of the system $\mathcal{S}$ obtained with partial trace:
\begin{align}
    \rho_{\mathcal{S}}=\tr_{\mathcal{R}} \left[ \rho_{\mathcal{S}+\mathcal{R},\beta}\right]\approx \rho_{\mathcal{S},\beta}^{(mf,2)} =\frac{e^{-\beta H^{(mf,2)}_{\mathcal{S}}}}{Z^{(mf,2)}_{ \mathcal{S},\beta}},
\end{align}
where:
\begin{align}
    &H^{(mf,2)}_{\mathcal{S}} = H_\mathcal{S}^{(0)}+H_{mf,\mathcal{C}}^{(1)}+H_{mf,\mathcal{C}}^{(2)}, \\
    &Z^{(mf,2)}_{ \mathcal{S},\beta}=\tr \left[e^{-\beta H^{(mf,2)}_{\mathcal{S}}}\right].
\end{align}

\newpage
\section{The derivation of the cumulant equation}\label{sec:app:derivationCumulantEquation}

In this Section we rederive the cumulant equation. This type of dynamical equations was firstly in reference~\cite{Alicki1989}, therefore the presence of this Section has rather pedagogical aspect in which we aim to carefully explain all ideas and step the derivation bases on. Still, we do not hesitate to make some generalisations. It is important to mention that the cumulant equation (up to the correct renormalization) has been recently, independently discovered by A. Rivas in reference, where the name {\it refined weak coupling limit} is used. However, the derivation in reference~\cite{Rivas_2017} is based on different technique, in which the need for the renormalization is not manifested.

Suppose we have a quantum system $\mathcal{S}$, weakly interacting with a reservoir $\mathcal{R}$. The reservoir is assumed to be a large system, modelled with a numerous (ideally infinite) collection of bosonic or fermionic harmonic oscillators, and therefore well defined in thermodynamic limit. Presence of a number of weak, independent perturbations can be regarded a Gaussian stochastic process. Such a situation especially fits the scenario of weak coupling, and particularly quantum optics. The total system $\mathcal{S}+\mathcal{R}$ is assumed to be a closed one. Furthermore, we assume the reservoir to be in stationary state $\rho_\mathcal{R}$, i.e., $\left[H_\mathcal{R},\rho_\mathcal{R}\right]=0$. The Hamiltonian of the total system $\mathcal{S}+\mathcal{R}$, that we consider is of the following form:
\begin{align}\label{eqn:H_tot_NR}
    H_{\mathcal{S}+\mathcal{R}}=H_\mathcal{S}^{(0)}(t)+H_\mathcal{R}+\lambda H_\mathcal{I}^{(0)}(t),
\end{align}
where $H_\mathcal{S}^{(0)}(t)$ is the ``bare" Hamiltonian of the system, $H_\mathcal{R}$ is the free Hamiltonian of the reservoir, and $\lambda H_\mathcal{I}^{(0)} (t)$ is the interaction Hamiltonian, in the form $\lambda H_\mathcal{I}^{(0)}(t)= \lambda \sum_i S_i \otimes R^{(0)}_i(t)$, with $S_i$, $R_i^{(0)}(t)$ being hermitian operators and $\lambda=1$\footnote{The meaning of $\lambda$ is enumerating elements in perturbative expansion with respect to powers of the coupling constant.}. The treatment based on time-dependent Hamiltonians $H_\mathcal{S}^{(0)}(t)$ and $H_\mathcal{I}^{(0)}(t)$ is the most adequate in realistic situations in which effects of state preparation and finite time of initiating the interaction have to be considered. The proposed relaxation helps us to bypass the aforementioned issues in the spirit of the (quantum) field-theoretical treatment.

Despite of the time dependence at the r.h.s. of equation (\ref{eqn:H_tot_NR}), we assume the total Hamiltonian of the system $H_{\mathcal{S}+\mathcal{R}}$ is approximately time-independent. Here, by approximately time independent we mean that for the time evolution of the total system $\mathcal{S}+\mathcal{R}$ the description employing Von Neumann equation without time-ordering operator (when equation is integrated) is sufficient. However, we still have to keep in mind that due to initial state incompatible with the interaction (e.g. product state due to Born approximation) dressing processes assosiated with the reorganization of the sysetm (alike in the polaron transformation that changes the partition between the system and its environment) can occur. The presence of such processes changes in general the properties of the open system $\mathcal{S}$, and leads to the renormalization of its Hamiltonian. The assertion on Hamiltonian of the total system being time-independent is a good approximation at the considered energy scale, since the total system $\mathcal{S}+\mathcal{R}$ is closed and the interaction between $\mathcal{S}$ and $\mathcal{R}$ is assumed to be weak.

In this derivation we strive to describe dynamics of the open quantum system $\mathcal{S}$, using its physical (measurable) Hamiltonian $H_\mathcal{S}$, which is in general different than the ``bare" one. The bare Hamiltonian may appear infeasible to be determined, because some systems can not be detached from their environment. Here, we assume that the measurable Hamiltonian $H_\mathcal{S}$ (alike system's $\mathcal{S}$ interaction operators $S_i$) does not depend on time, however, the generalization to time dependent case is straightforward\footnote{This is a relevant case when, for example, the system $\mathcal{S}$ interacts with quasi-classical external fields.}.  This discrepancy is due to the fact that interaction with the reservoir $\mathcal{R}$ can modify energy levels of system $\mathcal{S}$, e.g., Lamb-Stark shift effect can occur. It is temping to ask why does the reservoir $\mathcal{R}$ Hamiltonian $H_\mathcal{R}$ does not undergo similar procedure. This is due to assumption that the reservoir is large enough with respect to system $\mathcal{S}$, that its state is not or negligibly affected by the system $\mathcal{S}$ (see Section \ref{sec:intuition} of the  main text). 

In our derivation we employ a usual assumption on the initial state of the joined system.
\begin{align}\label{eqn:BornApprox}
    {\rho}_{\mathcal{S}+\mathcal{R}} (0) = {\rho}_{\mathcal{S}} (0) \otimes {\rho}_{\mathcal{R}}.
\end{align}
The above condition on the initial state is not relevant for all weakly interacting physical systems, since some systems can not be detached from their environment. In this way the initial state in the product form is not compatible with interaction. On the other hand, this condition is necessary for the dynamics to be state-to-state mapping. Interestingly, treatments of open quantum systems going beyond this assumption are of contemporary interest~\cite{Alipour_2020,Merkli_2021,Trushechkin_2021}. Furthermore, the possible consequence of assumption in equation (\ref{eqn:BornApprox}) is an abrupt dressing of system $\mathcal{S}$ with excitations of the reservoir at the initial stage of the evolution ($t>0$)\footnote{Such an abrupt process must be out of weak coupling regime.}.

The physical, measurable Hamiltonian $H_\mathcal{S}$ is an effective one and contains the corrections due to interaction between system $\mathcal{S}$ and the reservoir $\mathcal{R}$. In other words, it governs the evolution of the dressed state. Therefore, we compensate aftermaths of assumption (\ref{eqn:BornApprox}) with the time dependent ``bare" Hamiltonian and the proper renormalization procedure. The renormalized form of the Hamiltonian $H_{\mathcal{S}+\mathcal{R}}$ reads:
\begin{align}\label{eqn:H_tot_R}
    H_{\mathcal{S}+\mathcal{R}}=H_\mathcal{S}+H_\mathcal{R}+\lambda H_\mathcal{I}^{(0)}(t)-H_\mathcal{C}(t),
\end{align}
where
\begin{align}
    H_\mathcal{S}&=H_\mathcal{S}^{(0)}(t)+H_\mathcal{C}(t),\\
    H_\mathcal{C}(t)&=\sum_{k=1}^{\infty}\lambda^k H_\mathcal{C}^{(k)}(t),
\end{align}
are the corrections to the ``bare" Hamiltonian, due to the interaction between the system and the reservoir. The above operation, provided  we chose $H_\mathcal{C}(t)$ wisely, corresponds to the intuition that the dressing takes place at some distant past ($t \to - \infty$), and at $t=0$ we ``measure" the dressed system $\mathcal{S}$ with its effective, physical Hamiltonian $H_\mathcal{S}$.  What fixes our renormalization is the requirement for the interaction picture (with respect to the physical Hamiltonian $H_\mathcal{S}$) evolution of system $\mathcal{S}$, to be ``purely dissipative", i.e., to contain to Hamiltonian terms. We refer to the aforementioned requirement as to the self-consistency condition.\\
~~\\
{\it In the interaction picture with respect to the physical (renormalized) Hamiltonian $H_\mathcal{S}$, the reduced dynamics of the system $\mathcal{S}$ should be purely dissipative, i.e., it should not include any Hamiltonian-like terms.}\\
~~\\
Furthermore, if the renormalization succeeds we can understand the condition in equation (\ref{eqn:BornApprox}), as factorization between dressed state of system $\mathcal{S}$, and (the rest of) reservoir $\mathcal{R}$. In the following consideration we identify the physical, effective Hamiltonian $H_\mathcal{S}$ with $H_\mathcal{S}^{(2))}$ that contains the corrections up to the second-oder in $\lambda$. Still, we remain consistent with the Born approximation (see the main text).

In the next step, we perform the transition to the interaction picture with respect to free part $H_\mathcal{S}+H_\mathcal{R}$ of the renormalized Hamiltonian $H_{\mathcal{S}+\mathcal{R}}$. The interaction picture interaction Hamiltonian (including the contribution of the corrections) has the following form:
\begin{align}
    %\lambda \tilde{H}_\mathcal{I} (s)=
    \lambda \tilde{H}_\mathcal{I}^{(0)} (t)-\tilde{H}_\mathcal{C}(t) = \lambda \sum_i \tilde{S}_i(t) \otimes \tilde{R}_i^{(0)} (t) -\tilde{H}_\mathcal{C}(t),
\end{align}
with
\begin{align}
    \tilde{S}_i(t) = e^{i H_\mathcal{S} t} S_i e^{-i H_\mathcal{S} t},~~\tilde{R}_i(t) = e^{i H_\mathcal{R} t} R_i^{(0)}(t) e^{-i H_\mathcal{R} t},\\
    \tilde{H}_\mathcal{C}(t) = \sum_{k=1}^{\infty} \lambda^k \tilde{H}_\mathcal{C}^{(k)}(t),~~\tilde{H}_\mathcal{C}^{(k)}(t)= e^{i (H_\mathcal{S}+H_\mathcal{R}) t} H_\mathcal{C}^{(k)}(t) e^{-i (H_\mathcal{S}+H_\mathcal{R}) t}.
\end{align}

The reduced dynamics of $\mathcal{S}$ in the interaction picture is govern by a dynamical map $\tilde{W}_t$. On the other hand, the below expression is derived by the formal integration of the von Neumann equation and performing the partial trace over the reservoir $\mathcal{R}$:
\begin{align}
    \tilde{\rho}_\mathcal{S} (t) = \tilde{W}_t \tilde{\rho}_\mathcal{S} (0) = \tr_\mathcal{R} \left \{\mathbb{T} \exp \left[-i  \int_{0}^t ds~ \left[\lambda \tilde{H}_\mathcal{I}^{(0)} (s)-\tilde{H}_\mathcal{C}(s),\cdot \right]\right] \tilde{\rho}_{\mathcal{S}} (0) \otimes \tilde{\rho}_{\mathcal{R}} \right\},\label{eqn:Dmap1}
    %&= -i\tr_\mathcal{R} \left\{ \int_{0}^t ds~ \left[ \sum_i \tilde{S}_i(s) \otimes \tilde{R}_i^{(0)} (s) ,\tilde{\rho}_{\mathcal{S}} (0) \otimes \tilde{\rho}_{\mathcal{R}} \right] \right\},
\end{align}
where $\mathbb{T}$ is the time-ordering operator, and $\tr_\mathcal{R}$ is the partial trace over the bath subsystem. 

On the other hand side, due to quantum analogs of central limit theorem (CLT)~\cite{Goderis_1989,Goderis_1990,Accardi_1990,Accardi_2002}, we know that the interaction between system $\mathcal{S}$ and the reservoir $\mathcal{R}$, being a sum of numerous\footnote{See Section \ref{sec:setup} of the main text.} (ideally infinite) independent and weak perturbations, can be considered a Gaussian stochastic process~\cite{Kubo1963}. Similarly as in reference~\cite{Alicki1989} we use a form of generalized cumulant expansion~\cite{Kubo1962,Fox1976,Bianucci2020}:
\begin{align}\label{eqn:Dmap2}
    \tilde{W}_t = \exp_M \left(\sum_{n=1}^{\infty} \lambda^n \tilde{K}^{(n)}(t)\right),
\end{align}
where $\exp_M$ is a generalized exponential, and $\tilde{K}(t)=\sum_{n=1}^{\infty} \lambda^n \tilde{K}^{(n)}(t)$ is the so-called cumulant generating function~\cite{Bianucci2020}.  As we will show in the following lines the exact meaning of $\exp_M$ is not important for this work, and plays here only a formal role. However, we refer an interested reader to a work or Marco Bianucci, and Mauro Bologna~\cite{Bianucci2020}, that clarifies a discussion about generalized cumulants started by R. Kubo~\cite{Kubo1962} and R. Fox~\cite{Fox1976}.

We consider the cumulant generating function truncated to only two first terms, i.e.,  $\tilde{K}(t)=\sum_{n=1}^{2} \lambda^n \tilde{K}^{(n)}(t)$~\cite{Kubo1963}\footnote{This approximation is consistent with the Bohn approximation.}. This approximation, corresponds to the evolution of open quantum system $\mathcal{S}$ being guided by the classical Gaussian noise for which all higher-order cumulants $n>2$ vanish. However, this approximation allows to reproduce the well known mathematical structure of dynamical map in GKSL form. Moreover, the generalized exponential $\exp_M$, can be unambiguously expanded in power series up to the second-order in $\lambda$, as no ordering of terms needs to be considered at this stage ($M$-ordering map of~\cite{Bianucci2020}).

In order to proceed with the derivation we confront the Hamiltonian dynamics given by equation \eqref{eqn:Dmap1} with the statistical description of noise due to the reservoir in equation \eqref{eqn:Dmap2}. Therefore, in the next step we expand the exponentials in r.h.s.'s of equations (\ref{eqn:Dmap1}) and (\ref{eqn:Dmap2}). At this point we are ready to apply the perturbation theory by comparing terms with the same powers of $\lambda$. We explicitly write down terms up to the second-order in $\lambda$:
\begin{align}\label{eqn:compareK}
    % &\left(\mathds{1}+\sum_{n=1}^{\infty} \lambda^n \tilde{K}^{(n)}(t)+\frac 12 \left(\sum_{n=1}^{\infty} \lambda^n \tilde{K}^{(n)}(t)\right)^2+\dots\right) \tilde{\rho}_\mathcal{S} (0) \nonumber\\
    &\left(\mathds{1}+\lambda \tilde{K}^{(1)}(t)+\lambda^2 \left(\frac 12 \left( \tilde{K}^{(1)}(t)\right)^2+\tilde{K}^{(2)}(t)\right)+\dots \right) \tilde{\rho}_\mathcal{S} (0)
    \nonumber\\
    &= \tr_\mathcal{R} \left\{ \tilde{\rho}_{\mathcal{S}} (0) \otimes \tilde{\rho}_{\mathcal{R}}-i  \lambda \int_{0}^t ds~ \left[ \tilde{H}_\mathcal{I}^{(0)} (s)-\tilde{H}^{(1)}_\mathcal{C}(s),\tilde{\rho}_{\mathcal{S}} (0) \otimes \tilde{\rho}_{\mathcal{R}} \right]
    +i \lambda^2 \int_{0}^t ds~ \left[\tilde{H}^{(2)}_\mathcal{C}(s),\tilde{\rho}_{\mathcal{S}} (0) \otimes \tilde{\rho}_{\mathcal{R}} \right]
    \right.\nonumber\\
    &\left.-\frac{ \lambda^2}{2} \mathbb{T}\int_{0}^t ds\int_{0}^t dw~ \left[ \tilde{H}_\mathcal{I}^{(0)} (s)-\tilde{H}^{(1)}_\mathcal{C}(s),  \left[ \tilde{H}_\mathcal{I}^{(0)} (w)-\tilde{H}^{(1)}_\mathcal{C}(w),\tilde{\rho}_{\mathcal{S}} (0) \otimes \tilde{\rho}_{\mathcal{R}} \right] \right]+\dots \right\}.
\end{align}
In the r.h.s. of the equation above we used $\tilde{H}_\mathcal{C}(s) = \sum_{k=1}^{\infty} \lambda^k \tilde{H}_\mathcal{C}^{(k)}(s)$ consistently\footnote{The above simple expansion is due to Definition A, Definition D, and Remark H in reference~\cite{Bianucci2020}.}. Let us now compare the terms in the perturbative expansion in order to separate the coherent evolution from incoherent (noisy) processes. The zeroth-order is trivial, and in the first-order we have:
\begin{align}
    \tilde{K}^{(1)}(t) \tilde{\rho}_\mathcal{S} (0) =  \tr_\mathcal{R} \left\{  -i \int_{0}^t ds~ \left[ \tilde{H}_\mathcal{I}^{(0)} (s)-\tilde{H}^{(1)}_\mathcal{C}(s),\tilde{\rho}_{\mathcal{S}} (0) \otimes \tilde{\rho}_{\mathcal{R}} \right] \right\}.
\end{align}
However, accordingly to the self-consistency condition the interaction picture dynamical equations should be ``purely dissipative" (see Section \ref{sec:cumulant} of the main text), and contain no Hamiltonian-like terms, as the one in the equation above. This imposes on superoperator $\tilde{K}^{(1)}(t)$ the following requirement:
\begin{align}
    \tilde{K}^{(1)}(t) \tilde{\rho}_\mathcal{S} (0) = 0.
\end{align}
Consequently, the equation above imposes constraints on the correction $\tilde{H}_\mathcal{C}^{(1)}(t)$, and helps us to find explicit form of it. Namely,
\begin{align}
    \tilde{H}_\mathcal{C}^{(1)}(t) = \sum_i \tilde{S}_i(t)  \left<\tilde{R}_i^{(0)}(t)\right>_{\tilde{\rho}_\mathcal{R}},
\end{align}
that after the transformation to the interaction yields the form:
\begin{align}
    {H}_\mathcal{C}^{(1)}(t) = \sum_i {S}_i \left<\tilde{R}_i^{(0)}(t)\right>_{{\rho}_\mathcal{R}}.\label{eqn:Cumulant1stCorrectionSchroefinger}
\end{align}
%\sout{ The form of the correction is not unique, still we consider the above choice as a canonical one.}

In a generic case, the noise due to environment is not exactly the classical Gaussian stochastic process. Therefore, truncating the cumulant generating function up to two terms, what we performed, is only the second-order approximation to the dynamics \cite{Goderis_1989,Goderis_1990,Accardi_1990,Accardi_2002}. In this manner, we introduce a new dynamical map $\tilde{W}^{(2)}_t$ that that accounts for the second-order approximation for the exact dynamics $\tilde{W}_t$
\begin{align}
    \tilde{W}^{(2)}_t : = e^{\lambda^2 \tilde{K}^{(2)}(t)} \stackrel{\mathcal{O}\left(\lambda^3\right)}{\approx}  \tilde{W}_t,~~t\ge 0,
\end{align}
that is defined with the standard exponential (of q-number). Moreover, the approximate dynamical map $\tilde{W}^{(2)}_t$ is equal to the exact one $\tilde{W}_t$ in the case of noise being classically Gaussian. The above approximation is especially well justified for weak coupling setting, where the coupling constant is small. Despite $\tilde{W}^{(2)}_t$, is an approximation, from now on we treat it as a central object of our investigation, and we allow ourselves for skipping approximation symbols in dynamical equations where it appears.

% \begin{remark}
% Interestingly, one can construct a renormalization procedure in which $\tilde{K}^{2}(t)$ is the only non-zero element of the cumulant expansion, and $\tilde{W}_t^{(2)}$ is exact, i.e., $\tilde{W}_t^{(2)} \equiv \tilde{W}_t$. However, in this case the interpretation of the higher-order corrections to the interaction is unclear.
% \end{remark}

Before we determine $\tilde{K}^{(2)}(t)$, and the second-order correction $\tilde{H}^{(2)}_\mathcal{C}(t)$, let us introduce a new object for the sake of conciseness, i.e., the centered interaction Hamiltonian:
\begin{align}
   \tilde{H}_\mathcal{I}^{(1)} (t)= \tilde{H}_\mathcal{I}^{(0)} (t)-\tilde{H}^{(1)}_\mathcal{C}(t)=\sum_i \tilde{S}_i(s) \otimes \tilde{R}_i^{(1)} (s),
\end{align}
where $\tilde{R}_i^{(1)} (s)=\tilde{R}_i^{(0)} (s)-\left<\tilde{R}_i^{(0)}(s)\right>_{\tilde{\rho}_\mathcal{R}}$. The meaning of the first correction becomes clear now, i.e., it is ``centering of interaction", that is a usually an assumption on reservoir operators $R^{(0)}_i(t)$ (see Remark \ref{rem:centering}). Equation (\ref{eqn:compareK}), takes now a simplified form:
% \begin{align}
%     \tilde{R}_i^{(1)} (s)=\tilde{R}_i^{(0)} (s)-\left<\tilde{R}_i^{(0)}(t)\right>_{\tilde{\rho}_\mathcal{R}}
% \end{align}
\begin{align}\label{eqn:compareK2}
    &\left(\mathds{1}+\lambda^2 \tilde{K}^{(2)}(t)+\dots \right) \tilde{\rho}_\mathcal{S} (0) 
    = \tr_\mathcal{R} \left\{ \tilde{\rho}_{\mathcal{S}} (0) \otimes \tilde{\rho}_{\mathcal{R}}
    +i \lambda^2 \int_{0}^t ds~ \left[\tilde{H}^{(2)}_\mathcal{C}(s),\tilde{\rho}_{\mathcal{S}} (0) \otimes \tilde{\rho}_{\mathcal{R}} \right]
    \right.\nonumber\\
    &\left.-\frac{ \lambda^2}{2} \mathbb{T}\int_{0}^t ds\int_{0}^t dw~ \left[ \tilde{H}_\mathcal{I}^{(1)} (s),  \left[ \tilde{H}_\mathcal{I}^{(1)} (w),\tilde{\rho}_{\mathcal{S}} (0) \otimes \tilde{\rho}_{\mathcal{R}} \right] \right]+\dots \right\}.
\end{align}
The second-order reads:
\begin{align}
    \tilde{K}^{(2)}(t) \tilde{\rho}_\mathcal{S} (0) =\tr_\mathcal{R} \left\{i \int_{0}^t ds~ \left[\tilde{H}^{(2)}_\mathcal{C}(s),\tilde{\rho}_{\mathcal{S}} (0) \otimes \tilde{\rho}_{\mathcal{R}} \right]\right\} - \tr_\mathcal{R} \left\{ \int_{0}^t ds\int_{0}^s dw~ \left[\tilde{H}_\mathcal{I}^{(1)} (s),  \left[\tilde{H}_\mathcal{I}^{(1)} (w),\tilde{\rho}_{\mathcal{S}} (0) \otimes \tilde{\rho}_{\mathcal{R}} \right] \right] \right\},
\end{align}
where we time-ordered the expression at r.h.s..  The second-order correction $\tilde{H}^{(2)}_\mathcal{C}(t)$ will be used to compensate any other Hamiltonian terms, that can be hidden in the second term at r.h.s. of the equation above, as required by the self-consistency condition. Let us firstly, calculate the second term in the r.h.s. of the equation above.
\begin{align}
    &- \tr_\mathcal{R} \left\{ \int_{0}^t ds\int_{0}^s dw~ \left[\tilde{H}_\mathcal{I}^{(1)} (s),  \left[\tilde{H}_I^{(1)} (w),\tilde{\rho}_{\mathcal{S}} (0) \otimes \tilde{\rho}_{\mathcal{R}} \right] \right] \right\}  \nonumber \\
    %%%%%%%%%%%%%%%%%%%%%%%
    &=-\sum_{ij}\int_{0}^t ds\int_{0}^s dw~ \tr_\mathcal{R} \left\{ 
    \tilde{S}_j (s)\tilde{S}_i (w)\tilde{\rho}_{\mathcal{S}} (0) \otimes \tilde{R}^{(1)}_j (s)\tilde{R}^{(1)}_i (w)\tilde{\rho}_{\mathcal{R}} 
    -\tilde{S}_j (s)\tilde{\rho}_{\mathcal{S}} (0)\tilde{S}_i (w) \otimes \tilde{R}^{(1)}_j (s)\tilde{\rho}_{\mathcal{R}}\tilde{R}^{(1)}_i (w) \right.\nonumber \\
    &\left.-\tilde{S}_i (w)\tilde{\rho}_{\mathcal{S}} (0)\tilde{S}_j (s) \otimes \tilde{R}^{(1)}_i (w)\tilde{\rho}_{\mathcal{R}}\tilde{R}^{(1)}_j (s)
    +\tilde{\rho}_{\mathcal{S}} (0)\tilde{S}_i (w)\tilde{S}_j (s) \otimes \tilde{\rho}_{\mathcal{R}}\tilde{R}^{(1)}_i (w)\tilde{R}^{(1)}_j (s)
    \right\} \\
    %%%%%%%%%%%%%%%%%%%%%%%%%
    &=-\sum_{ij}\int_{0}^t ds\int_{0}^s dw~  \left( 
    \tilde{S}_j (s)\tilde{S}_i (w)\tilde{\rho}_{\mathcal{S}} (0) \left< \tilde{R}^{(1)}_j (s)\tilde{R}^{(1)}_i (w)\right>_{\tilde{\rho}_{\mathcal{R}} }
    -\tilde{S}_j (s)\tilde{\rho}_{\mathcal{S}} (0)\tilde{S}_i (w) \left< \tilde{R}^{(1)}_i (w)\tilde{R}^{(1)}_j (s)\right>_{\tilde{\rho}_{\mathcal{R}} } \right.\nonumber \\
    &\left.-\tilde{S}_i (w)\tilde{\rho}_{\mathcal{S}} (0)\tilde{S}_j (s) \left< \tilde{R}^{(1)}_j (s)\tilde{R}^{(1)}_i (w)\right>_{\tilde{\rho}_{\mathcal{R}} }
    +\tilde{\rho}_{\mathcal{S}} (0)\tilde{S}_i (w)\tilde{S}_j (s) \left<\tilde{R}^{(1)}_i (w)\tilde{R}^{(1)}_j (s)\right>_{\tilde{\rho}_{\mathcal{R}} }
    \right) \\
    %%%%%%%%%%%%%%%%%%%%%%%%
    &\stackrel{(I)}{=}
    -\sum_{ij}\int_{0}^t ds\int_{0}^s dw~ \tilde{S}_j (s)\tilde{S}_i (w)\tilde{\rho}_{\mathcal{S}} (0) \left< \tilde{R}^{(1)}_j (s)\tilde{R}^{(1)}_i (w)\right>_{\tilde{\rho}_{\mathcal{R}} }
    +\sum_{ij}\int_{0}^t dw\int_{0}^w ds~ \tilde{S}_i (w)\tilde{\rho}_{\mathcal{S}} (0)\tilde{S}_j (s) \left< \tilde{R}^{(1)}_j (s)\tilde{R}^{(1)}_i (w)\right>_{\tilde{\rho}_{\mathcal{R}} } \nonumber \\
    &+\sum_{ij}\int_{0}^t ds\int_{0}^s dw~ \tilde{S}_i (w)\tilde{\rho}_{\mathcal{S}} (0)\tilde{S}_j (s) \left< \tilde{R}^{(1)}_j (s)\tilde{R}^{(1)}_i (w)\right>_{\tilde{\rho}_{\mathcal{R}} }
    -\sum_{ij}\int_{0}^t dw\int_{0}^w ds~ \tilde{\rho}_{\mathcal{S}} (0)\tilde{S}_j (s)\tilde{S}_i (w) \left<\tilde{R}^{(1)}_j (s)\tilde{R}^{(1)}_i (w)\right>_{\tilde{\rho}_{\mathcal{R}} }
    \label{eqn:long-time_start}\\
    %%%%%%%%%%%%%%%%%%%%%%%%
    &\stackrel{(II)}{=}
    \sum_{ij}\int_{0}^t ds\int_{0}^t dw~ \tilde{S}_i (w)\tilde{\rho}_{\mathcal{S}} (0)\tilde{S}_j (s) \left< \tilde{R}^{(1)}_j (s)\tilde{R}^{(1)}_i (w)\right>_{\tilde{\rho}_{\mathcal{R}} } 
    -\sum_{ij}\int_{0}^t ds\int_{0}^s dw~ \tilde{S}_j (s)\tilde{S}_i (w)\tilde{\rho}_{\mathcal{S}} (0) \left< \tilde{R}^{(1)}_j (s)\tilde{R}^{(1)}_i (w)\right>_{\tilde{\rho}_{\mathcal{R}} } \nonumber \\
    &-\sum_{ij}\int_{0}^t dw\int_{0}^w ds~ \tilde{\rho}_{\mathcal{S}} (0)\tilde{S}_j (s)\tilde{S}_i (w) \left<\tilde{R}^{(1)}_j (s)\tilde{R}^{(1)}_i (w)\right>_{\tilde{\rho}_{\mathcal{R}} } ,
    \label{eqn:CFnotsplitted}
\end{align}
where in $(I)$ we interchanged variables of integration and summation indices for 2-nd and 4-th term, and in $(II)$ we firstly connected 2-nd and 3-rd terms from previous lines, then we rearranged the terms. In the next lines, the first term of equation (\ref{eqn:CFnotsplitted}) stays unchanged, we split 2-nd and 3-rd terms of the same equation into halves (2-nd and 4-th line below), and finally we add and subtract a useful auxiliary term (3-rd and 5-th line below).

\begin{align}
    &\eqref{eqn:CFnotsplitted}=\sum_{ij}\int_{0}^t ds\int_{0}^t dw~ \tilde{S}_i (w)\tilde{\rho}_{\mathcal{S}} (0)\tilde{S}_j (s) \left< \tilde{R}^{(1)}_j (s)\tilde{R}^{(1)}_i (w)\right>_{\tilde{\rho}_{\mathcal{R}} } \nonumber \\
    &-\frac 12 \sum_{ij}\int_{0}^t ds\int_{0}^s dw~ \tilde{S}_j (s)\tilde{S}_i (w)\tilde{\rho}_{\mathcal{S}} (0) \left< \tilde{R}^{(1)}_j (s)\tilde{R}^{(1)}_i (w)\right>_{\tilde{\rho}_{\mathcal{R}} }
    -\frac 12 \sum_{ij}\int_{0}^t ds\int_{0}^s dw~ \tilde{S}_j (s)\tilde{S}_i (w)\tilde{\rho}_{\mathcal{S}} (0) \left< \tilde{R}^{(1)}_j (s)\tilde{R}^{(1)}_i (w)\right>_{\tilde{\rho}_{\mathcal{R}} } \nonumber \\
    &-\frac 12 \sum_{ij}\int_{0}^t dw \int_{0}^w ds~ \tilde{S}_j (s)\tilde{S}_i (w)\tilde{\rho}_{\mathcal{S}} (0) \left< \tilde{R}^{(1)}_j (s)\tilde{R}^{(1)}_i (w)\right>_{\tilde{\rho}_{\mathcal{R}} }
    +\frac 12 \sum_{ij}\int_{0}^t dw\int_{0}^w ds~ \tilde{S}_j (s)\tilde{S}_i (w)\tilde{\rho}_{\mathcal{S}} (0) \left< \tilde{R}^{(1)}_j (s)\tilde{R}^{(1)}_i (w)\right>_{\tilde{\rho}_{\mathcal{R}} } \nonumber \\
    &-\frac 12 \sum_{ij}\int_{0}^t dw\int_{0}^w ds~ \tilde{\rho}_{\mathcal{S}} (0)\tilde{S}_j (s)\tilde{S}_i (w) \left<\tilde{R}^{(1)}_j (s)\tilde{R}^{(1)}_i (w)\right>_{\tilde{\rho}_{\mathcal{R}} }
    -\frac 12 \sum_{ij}\int_{0}^t dw\int_{0}^w ds~ \tilde{\rho}_{\mathcal{S}} (0)\tilde{S}_j (s)\tilde{S}_i (w) \left<\tilde{R}^{(1)}_j (s)\tilde{R}^{(1)}_i (w)\right>_{\tilde{\rho}_{\mathcal{R}} } \nonumber\\
    &-\frac 12 \sum_{ij}\int_{0}^t ds\int_{0}^s dw~ \tilde{\rho}_{\mathcal{S}} (0)\tilde{S}_j (s)\tilde{S}_i (w) \left<\tilde{R}^{(1)}_j (s)\tilde{R}^{(1)}_i (w)\right>_{\tilde{\rho}_{\mathcal{R}} }
    +\frac 12 \sum_{ij}\int_{0}^t ds\int_{0}^s dw~ \tilde{\rho}_{\mathcal{S}} (0)\tilde{S}_j (s)\tilde{S}_i (w) \left<\tilde{R}^{(1)}_j (s)\tilde{R}^{(1)}_i (w)\right>_{\tilde{\rho}_{\mathcal{R}} } \label{eqn:CFnotsplitted2} \\
    %%%%%%%%%%%%%%%%%%%%%%%%%%%%%%%
    &\stackrel{(III)}{=} \sum_{ij}\int_{0}^t ds\int_{0}^t dw~ \tilde{S}_i (w)\tilde{\rho}_{\mathcal{S}} (0)\tilde{S}_j (s) \left< \tilde{R}^{(1)}_j (s)\tilde{R}^{(1)}_i (w)\right>_{\tilde{\rho}_{\mathcal{R}} } \nonumber \\
    &-\frac 12 \sum_{ij}\int_{0}^t ds\int_{0}^t dw~ \tilde{S}_j (s)\tilde{S}_i (w)\tilde{\rho}_{\mathcal{S}} (0) \left< \tilde{R}^{(1)}_j (s)\tilde{R}^{(1)}_i (w)\right>_{\tilde{\rho}_{\mathcal{R}} } 
    -\frac 12 \sum_{ij}\int_{0}^t ds\int_{0}^s dw~ \tilde{\rho}_{\mathcal{S}} (0)\tilde{S}_j (s)\tilde{S}_i (w) \left<\tilde{R}^{(1)}_j (s)\tilde{R}^{(1)}_i (w)\right>_{\tilde{\rho}_{\mathcal{R}} } \nonumber\\
    & -\frac 12 \sum_{ij}\int_{0}^t ds\int_{0}^t dw~ \mathrm{sgn}(s-w) \tilde{S}_j (s)\tilde{S}_i (w)\tilde{\rho}_{\mathcal{S}} (0) \left< \tilde{R}^{(1)}_j (s)\tilde{R}^{(1)}_i (w)\right>_{\tilde{\rho}_{\mathcal{R}} } \nonumber\\
    &+\frac 12 \sum_{ij}\int_{0}^t ds\int_{0}^t dw~ \mathrm{sgn}(s-w) \tilde{\rho}_{\mathcal{S}} (0)\tilde{S}_j (s)\tilde{S}_i (w) \left<\tilde{R}^{(1)}_j (s)\tilde{R}^{(1)}_i (w)\right>_{\tilde{\rho}_{\mathcal{R}} } \\
    %%%%%%%%%%%%%%%%%%%%%%%%%%%%%%%
    &\stackrel{(III)}{=} \sum_{ij}\int_{0}^t ds\int_{0}^t dw~ \left( \tilde{S}_i (w)\tilde{\rho}_{\mathcal{S}} (0)\tilde{S}_j (s)-\frac 12 \left\{ \tilde{S}_j (s)\tilde{S}_i (w),\tilde{\rho}_{\mathcal{S}} (0)\right\}\right) \left< \tilde{R}^{(1)}_j (s)\tilde{R}^{(1)}_i (w)\right>_{\tilde{\rho}_{\mathcal{R}} } \nonumber \\
    & -\frac 12 \sum_{ij}\int_{0}^t ds\int_{0}^t dw~ \mathrm{sgn}(s-w) \left[\tilde{S}_j (s)\tilde{S}_i (w),\tilde{\rho}_{\mathcal{S}} (0)\right] \left< \tilde{R}^{(1)}_j (s)\tilde{R}^{(1)}_i (w)\right>_{\tilde{\rho}_{\mathcal{R}} }, \label{eqn:Cumulant_derivation_intermediate_step}
\end{align}
where in $(III)$ we connected integrals from equation (\ref{eqn:CFnotsplitted2}) in the following way. Terms in equation (\ref{eqn:CFnotsplitted2}) form two columns. We connected first two and two last terms, from each column  pairwise. In the consecutive step we simplified the expression with anticommutator and commutator.

To proceed with the next step we  use the so-called jump operators (see Section \ref{sec:app:Properties}):
\begin{align}
    	&{S}_i(\omega_2)=\sum_{\epsilon_2^\prime - \epsilon_2=\omega_2} \Pi(\epsilon_2) {S}_i \Pi(\epsilon_2^\prime),
\end{align}
where the energies $\{\epsilon_2\}$ are eigenvalues of the renormalized Hamiltonian $H_\mathcal{S}$. For these operators we have:
\begin{align}
    &\tilde{S}_k (t) = \sum_{\omega_2} e^{-i \omega_2 t} S_k (\omega_2),\label{eqn:jumpOperatorsA}\\
    &\tilde{S}_k^\dagger (t) =\tilde{S}_k (t)=   \sum_{\omega_2} e^{ i \omega_2 t} S_k^\dagger (\omega_2).\label{eqn:jumpOperatorsB}
\end{align}
We use now the formula in equation (\ref{eqn:jumpOperatorsA}) for operators with indices $(i,w)$ and the formula in equation (\ref{eqn:jumpOperatorsB}) for operators with indices $(j,s)$, introducing $\omega_2$, $\omega_2^\prime$ respectively.

\begin{align}
    &\eqref{eqn:Cumulant_derivation_intermediate_step} = 
    \sum_{ij}\sum_{\omega_2,\omega_2^\prime}\int_{0}^t ds\int_{0}^t dw~e^{i(\omega_2^\prime s - \omega_2 w)} \left( S_i (\omega_2)\tilde{\rho}_{\mathcal{S}} (0)S_j^\dagger (\omega_2^\prime)-\frac 12 \left\{ S_j^\dagger (\omega_2^\prime) S_i (\omega_2),\tilde{\rho}_{\mathcal{S}} (0)\right\}\right) \left< \tilde{R}^{(1)}_j (s)\tilde{R}^{(1)}_i (w)\right>_{\tilde{\rho}_{\mathcal{R}} } \nonumber \\
    & -\frac 12 \sum_{ij}\sum_{\omega_2,\omega_2^\prime} \int_{0}^t ds\int_{0}^t dw~ e^{i(\omega_2^\prime s - \omega_2 w)} \mathrm{sgn}(s-w) \left[S_j^\dagger (\omega_2^\prime)S_i (\omega_2),\tilde{\rho}_{\mathcal{S}} (0)\right] \left< \tilde{R}^{(1)}_j (s)\tilde{R}^{(1)}_i (w)\right>_{\tilde{\rho}_{\mathcal{R}} } \\
    %%%%%%%%%%%%%%%%%%%%%%%%%%%%%%%%%%%%%%%%%%%%%%%%%%%%%%%%%%%
    &= - i\left[\frac{1}{2i} \sum_{ij}\sum_{\omega_2,\omega_2^\prime} \int_{0}^t ds\int_{0}^t dw~ e^{i(\omega_2^\prime s - \omega_2 w)} \mathrm{sgn}(s-w)\left< \tilde{R}^{(1)}_j (s)\tilde{R}^{(1)}_i (w)\right>_{\tilde{\rho}_{\mathcal{R}} }S_j^\dagger (\omega_2^\prime)S_i (\omega_2),\tilde{\rho}_{\mathcal{S}} (0)\right]  \nonumber \\
    &+\sum_{ij}\sum_{\omega_2,\omega_2^\prime}\int_{0}^t ds\int_{0}^t dw~e^{i(\omega_2^\prime s - \omega_2 w)} \left< \tilde{R}^{(1)}_j (s)\tilde{R}^{(1)}_i (w)\right>_{\tilde{\rho}_{\mathcal{R}} } \left( S_i (\omega_2)\tilde{\rho}_{\mathcal{S}} (0)S_j^\dagger (\omega_2^\prime)-\frac 12 \left\{ S_j^\dagger (\omega_2^\prime) S_i (\omega_2),\tilde{\rho}_{\mathcal{S}} (0)\right\}\right),
\end{align}
where in the last step we did some rearranging. 

By making some assignments we obtain expressions that resembles the known forms~\cite{Alicki1989,Rivas_2017}.
\begin{align}\label{eqn:AofK}
    % &\tilde{\rho}_\mathcal{S} (t)= \tilde{W}_t^{(2)}\tilde{\rho}_\mathcal{S} (0) =e^{\lambda^2 \tilde{K}^{(2)}(t)}\tilde{\rho}_\mathcal{S} (0),\\ 
	&\tilde{K}^{(2)}(t) \tilde{\rho}_\mathcal{S} (0) = \tr_\mathcal{R} \left\{i \int_{0}^t ds~ \left[\tilde{H}^{(2)}_\mathcal{C}(s),\tilde{\rho}_{\mathcal{S}} (0) \otimes \tilde{\rho}_{\mathcal{R}} \right]\right\} -i \left[\Lambda(t),\tilde{\rho}_{\mathcal{S}} (0) \right] \nonumber \\
	&+\sum_{i,j} \sum_{\omega_2, \omega_2^\prime} \gamma_{ij}(\omega_2,\omega_2^\prime,t)  \left(S_i (\omega_2) \tilde{\rho}_\mathcal{S}(0) S_j^\dagger (\omega_2^\prime) - \frac{1}{2} \left\{S_j^\dagger (\omega_2^\prime) S_i (\omega_2), \tilde{\rho}_\mathcal{S} (0) \right\} \right),
\end{align}
with:
\begin{align}
	&\gamma_{ij} (\omega,\omega^\prime,t) = \int_0^t ds \int_0^t dw~ e^{i (\omega^\prime s - \omega w)} \left< \tilde{R}^{(1)}_j (s) \tilde{R}^{(1)}_i (w) \right>_{\tilde{\rho}_\mathcal{R}},\\
	&\Lambda(t)= \sum_{ij}\sum_{\omega_2,\omega_2^\prime} \Xi_{ij} (\omega_2,\omega_2^\prime,t) S_j^\dagger (\omega_2^\prime) S_i(\omega_2),\\
	&\Xi_{ij} (\omega,\omega^\prime,t) = \frac{1}{2i} \int_0^t ds \int_0^t dw~ \mathrm{sgn} (s-w) e^{i(\omega^\prime s - \omega w)} \left<\tilde{R}^{(1)}_j (s) \tilde{R}^{(1)}_i (w)\right>_{\tilde{\rho}_\mathcal{R}}.
\end{align}
If we use the results in Section \ref{sec:app:Properties}, then $\gamma_{ij} (\omega,\omega^\prime,t)$ can be alternatively expressed as (see reference~\cite{Winczewski_2021}):
\begin{align}\label{eqn:gammaPreIntegrated}
    &\gamma_{ij} (\omega,\omega^\prime,t) = e^{i \frac{\omega^\prime-\omega}{2}t} \int_{-\infty}^{\infty} d\Omega~
	    \left[t~ \mathrm{sinc} \left(\frac{\omega^\prime-\Omega}{2}t\right)\right]  \left[t~ \mathrm{sinc} \left(\frac{\omega-\Omega}{2}t\right)\right]  R_{ji} (\Omega),
\end{align}
where $\mathrm{sinc}(x) \equiv \frac{\sin (x)}{x}$, and $R_{ji} (\Omega)=\left< {R}^{(1)}_j (\Omega) {R}^{(1)}_i \right>_{\tilde{\rho}_B}$. As we observe the dissipative part of the $\tilde{K}^{(2)}(t) $ superoperator incorporates the the (time-independent) Bohr's frequencies of the renormalized (physical) Hamiltonian $H_\mathcal{S}$. This situation (informally) confirms that the renormalization shifts any dressing processes to the distant past, and therefore the renormalized equation describes the evolution of the dressed system.  %\sout{The above formula neglects the time dependence of (Schr{\"o}dinger picture) interaction operators $R^{(0)}_i(t)$. Still, this approximation is valid for all times $t$ except super-short time scale, and can be understood in terms of locally measurable quantities (see the main text).  }

Let us remark now, that if we set $\tilde{H}^{(2)}_\mathcal{C}(s)=0$ (and obviously $\lambda=1$) then, we obtain exactly the form known from reference~\cite{Rivas_2017}. However, the physical requirement for ``purely dissipative dynamics" in interaction picture with respect to measurable Hamiltonian $H_\mathcal{S}$, imposes constraints on the correction $\tilde{H}^{(2)}_\mathcal{C}(s)$, and therefore fixes our renormalization procedure. 
\begin{align}\label{eqn:NoLS}
    \tr_\mathcal{R} \left\{i \int_{0}^t ds~ \left[\tilde{H}^{(2)}_\mathcal{C}(s),\tilde{\rho}_{\mathcal{S}} (0) \otimes \tilde{\rho}_{\mathcal{R}} \right]\right\} -i \left[\Lambda(t),\tilde{\rho}_{S} (0) \right]= 0.
\end{align}
Again, we are able to find the explicit form of the correction %\sout{the form of the correction is not unique, but we are able to find a satisfying expression:}
\begin{align}
    &\tilde{H}_\mathcal{C}^{(2)} (t)= \sum_{ij}\sum_{\omega_2,\omega_2^\prime}  S_j^\dagger (\omega_2^\prime) S_i(\omega_2)  \frac{d}{dt} \Xi_{ij}(\omega_2,\omega_2^\prime,t).
\end{align}
Finally imposing $\lambda=1$, we write down:
\begin{align}\label{eqn:AofKreduced}
    &\tilde{\rho}_\mathcal{S} (t)= \tilde{W}_t^{(2)}\tilde{\rho}_\mathcal{S} (0) =e^{ \tilde{K}^{(2)}(t)}\tilde{\rho}_\mathcal{S} (0),\\ 
    &\tilde{K}^{(2)}(t) \tilde{\rho}_\mathcal{S} (0) = \sum_{i,j} \sum_{\omega_2, \omega_2^\prime} \gamma_{ij}(\omega_2,\omega_2^\prime,t)  \left(S_i (\omega_2) \tilde{\rho}_\mathcal{S} (0) S_j^\dagger (\omega_2^\prime) - \frac{1}{2} \left\{S_j^\dagger (\omega_2^\prime) S_i (\omega_2), \tilde{\rho}_\mathcal{S} (0) \right\} \right),
    % &\gamma_{ij} (\omega_2,\omega_2^\prime,t) = \int_0^t ds \int_0^t dw~ e^{i (\omega_2^\prime s - \omega_2 w)} \left< \tilde{R}^{(1)}_j (s) \tilde{R}^{(1)}_i (w) \right>_{\tilde{\rho}_\mathcal{R}},
\end{align}
%\sout{With the aid of Bochner's Theorem}
The functions $\gamma_{ij}(\omega,\omega^\prime,t)$ can be shown to be elements of a positive semi-definite matrix~\cite{AlickiLendi1987,Alicki1989,Rivas_2012} (see Section \ref{sec:app:CPTP}). This is sufficient for $\tilde{W}^{(2)}_t$ to be completely positive and trace preserving (CPTP), since $\tilde{K}^{(2)}(t)$ is in the GKSL form \cite{gorini1976completely,lindblad1976generators}. 

At this point we want to remark that the procedure described above is not an involution in the following sense. One can not start the procedure with writing the total Hamiltonian $H_{\mathcal{S}+\mathcal{R}}$, using the physical Hamiltonian $H_{\mathcal{S}}$, and obtain correct result with corrections $\tilde{H}_\mathcal{C}^{(2)}(t)$ equal to zero. This is because the renormalized interaction is not of the form of $\lambda H_\mathcal{I}^{(0)}(t)= \lambda \sum_i S_i \otimes R^{(0)}_i(t)$, and contains terms bilinear in $S_i$, due to $\tilde{H}_\mathcal{C}^{(2)}(t)$.

We conclude this part with saying that ``$\tilde{W}^{(2)}_t$ is a one-parameter family of completely positive trace preserving maps and provides mathematically consistent weak-coupling non-Markovian approximation for the reduced dynamics $\tilde{W}_t$"~\cite{Alicki1989}, that governs the evolution (with respect to physical Hamiltonian $H_\mathcal{S}$) of density operator $\tilde{\rho}_\mathcal{S} (0)$ describing the dressed state of open quantum system $\mathcal{S}$.

The treatment described in this Section allows us to find the ``zeroth-order'' approximation (not consistent with Born approximation) for the stationary state state of the cumulant equation\footnote{The general (any order) proof for the same state will be showed in \cite{MW_preparation}.}. This state is a Gibbs (thermal) state with respect to the physical Hamiltonian $H_\mathcal{S}$. To show this it is enough to replace $\gamma_{ij} (\omega,\omega^\prime,t)$, with its long time approximation:
\begin{align}
    \lim_{t\to\infty} \frac{\gamma_{ij} (\omega,\omega^\prime,t)}{t} =\gamma_{ij}(\omega) \delta_{\omega,\omega^\prime}~~\implies~~\gamma_{ij} (\omega,\omega^\prime,t) \approx t \gamma_{ij}(\omega),~~\text{for}~~t \gg 0, \label{eqn:gammaAsymptotic}
\end{align}
and therefore $\tilde{K}^{(2)}(t) \tilde{\rho}_\mathcal{S} (0) \stackrel{\mathcal{O}(\lambda^0)}{\approx} t \tilde{L} \tilde{\rho}_\mathcal{S} (0),~t\gg 0$\footnote{This limit can be readily calculated using properties of models of Dirac delta distribution~\cite{Winczewski_2021}}. We recognize now $\tilde{L}$, to be a generator of quantum dynamical semigroup~\cite{AlickiLendi1987} (see equation \eqref{eqn:MME}), and thus $\tilde{\rho}_\mathcal{S} (t) \approx e^{t \tilde{L}}\tilde{\rho}_\mathcal{S} (0)$, for $\left(t\gg 0\right)$, is an integrated form of the (correctly renormalized) Davies-GKSL in Section \ref{sec:MME}. 

The limiting procedure in equation \eqref{eqn:gammaAsymptotic} reproduces the long-time limit of the cumulant equation (up to the renormalization) proposed by A. Rivas in \cite{Rivas_2017}. However, the above limit neglects any constant or oscillatory terms present in $\gamma_{ij} (\omega,\omega^\prime,t)$. Therefore, the above result is consistent only to the zeroth-order in the coupling constant. In Sections \ref{sec:app:subsec:Relatioon_BR} and \ref{sec:app:subsec:long-time} we find the long-time state of the cumulant equation consistent with the second-order, and the long-time limit of $\tilde{K}^{(2)}(t)$ the superoperator.
 
%\sout{Now, if the reservoir $\mathcal{R}$ is a thermal heat bath at inverse temperature $\beta$, obvious identification of $\gamma_{ij}(\omega)$ with the one from equation (\ref{eqn:MME}), tells us that a KMS condition $\gamma_{ij}(-\omega)=e^{-\beta \omega}\gamma_{ji}(\omega)$ holds. Therefore approach developed in this Section reproduces the correct stationary state, and is consistent with the thermodynamics (cf.~\cite{Rivas_2017} where this connection does not hold). }
\color{black}

\begin{remark}
Because we are interested solely in the two first correction to the bare Hamiltonian $H_\mathcal{S}^{(0)}$, we can identify the renormalized $H_\mathcal{S}^{(2)}=H_\mathcal{S}^{(0)}+H_\mathcal{C}^{(1)}+H_\mathcal{C}^{(2)}$, with the physical, measurable Hamiltonian $H_\mathcal{S}$.
\end{remark}

\subsection{The cumulant equation in the Schr{\"o}dinger picture}\label{sec:app:subsec:SchroedingerPicture}

The cumulant equation was derived in the interaction picture with respect to renormalized, physical Hamiltonian~$H_\mathcal{S}$. The transformation to the Schr{\"o}dinger picture can be performed with aid of Baker–Campbell–Hausdorff (BCH) formula. 
\begin{align}
    e^X e^Y = \exp{X+Y+\frac{1}{2} [X,Y]+\frac{1}{12}[X,[X,Y]]-\frac{1}{12}[Y,[X,Y]]+\cdots}.\label{eqn:BCHformula}
\end{align}
We start with noticing that:
\begin{align}
    {\rho}_\mathcal{S}(t)= e^{-i H_\mathcal{S} t}
    \tilde{\rho}_\mathcal{S}(t)e^{i H_\mathcal{S} t} 
     = e^{-it[H_\mathcal{S},\cdot]} e^{\tilde{K}^{(2)}(t)}{\rho}_\mathcal{S}(0) =: e^{{K}^{(2)}(t)}{\rho}_\mathcal{S}(0),
\end{align}
where we used the fact that at $t=0$ interaction and Schr{\"o}dinger pciture operators coincide. The last equality defines ${K}^{(2)}(t)$, i.e., the Schr{\"o}dinger picture generator of the dynamics. Unfortunately, finding the ${K}^{(2)}(t)$ superoperator is cumbersome if even possible due to the higher order commutators. Still, we want to propose an approximate form.

In order to obtain the the approximation of ${K}^{(2)}(t)$, we shall calculate the first commutator in the r.h.s. of the BCH formula in equation \eqref{eqn:BCHformula}, in which we substitute $X=-it[H_\mathcal{S},\cdot]$, and $Y=\tilde{K}^{(2)}(t)$.
\begin{align}
    &\left[-it[H_\mathcal{S},\cdot],\tilde{K}^{(2)}(t)\right]\rho=-it\left[[H_\mathcal{S},\cdot],\tilde{K}^{(2)}(t)\right]\rho\\
    %%%%%
    &=-it \left([H_\mathcal{S},\tilde{K}^{(2)}(t)\rho] - \tilde{K}^{(2)}(t)\left( [H_\mathcal{S},\rho]\right) \right)\\
    %%%%%%%%%%%%%%%%%%%%%%%
    &=-it \left(H_\mathcal{S}  \tilde{K}^{(2)}(t)\rho - \left(\tilde{K}^{(2)}(t)\rho \right) H_\mathcal{S}   - \tilde{K}^{(2)}(t) \left(H_\mathcal{S}\rho-\rho H_\mathcal{S} \right)\right)\\
    %%%%%%%%%%%%%%%%%%%%%%%
    &=-it \sum_{i,j} \sum_{\omega, \omega^\prime} \gamma_{ij}(\omega,\omega^\prime,t)\left(H_\mathcal{S}    \left(S_i (\omega) \rho  S_j^\dagger (\omega^\prime) - \frac{1}{2} \left\{S_j^\dagger (\omega^\prime) S_i (\omega), \rho  \right\} \right) -  \left(S_i (\omega) \rho  S_j^\dagger (\omega^\prime) - \frac{1}{2} \left\{S_j^\dagger (\omega^\prime) S_i (\omega), \rho  \right\} \right) H_\mathcal{S}  \nonumber \right.\\
    &\left.-   \left(S_i (\omega) H_\mathcal{S}\rho  S_j^\dagger (\omega^\prime) - \frac{1}{2} \left\{S_j^\dagger (\omega^\prime) S_i (\omega), H_\mathcal{S}\rho \right\} \right) +  \left(S_i (\omega) \rho H_\mathcal{S}  S_j^\dagger (\omega^\prime) - \frac{1}{2} \left\{S_j^\dagger (\omega^\prime) S_i (\omega), \rho H_\mathcal{S} \right\} \right)\right)\\
    %%%%%%%%%%%%%%%%%%%%
    &=-it \sum_{i,j} \sum_{\omega, \omega^\prime} \gamma_{ij}(\omega,\omega^\prime,t)\left( [H_\mathcal{S} ,S_i (\omega)]\rho S_j^\dagger (\omega^\prime) +S_i (\omega)\rho [H_\mathcal{S},S_j^\dagger (\omega^\prime)]-\frac{1}{2}[H_\mathcal{S},S_j^\dagger (\omega^\prime) S_i (\omega)]\rho - \frac{1}{2}\rho[H_\mathcal{S},S_j^\dagger (\omega^\prime) S_i (\omega)] \right)\\
    %%%%%%%%%%%%%%%%%%%%
    &\stackrel{(I)}{=}-it \sum_{i,j} \sum_{\omega, \omega^\prime} \gamma_{ij}(\omega,\omega^\prime,t) (\omega^\prime-\omega)\left( S_i (\omega)\rho S_j^\dagger (\omega^\prime) -\frac{1}{2} \left\{S_j^\dagger (\omega^\prime) S_i (\omega),\rho\right\}\right),
\end{align}
where in ($I$) we used the properties descibred in Section \ref{sec:app:Properties} of the Appendix.

If we now skip the higher-order commutators in the formula \eqref{eqn:BCHformula} and use the above calculations we obtain:
\begin{align}
    {K}^{(2)}(t) \rho &\approx  -it[H_\mathcal{S},\rho] +  \sum_{i,j} \sum_{\omega, \omega^\prime} \gamma_{ij}(\omega,\omega^\prime,t) \left(1-\frac{it}{2}(\omega^\prime-\omega)\right) \left( S_i (\omega)\rho S_j^\dagger (\omega^\prime) -\frac{1}{2} \left\{S_j^\dagger (\omega^\prime) S_i (\omega),\rho\right\}\right) \\
    &\approx -it[H_\mathcal{S},\rho] +  \sum_{i,j} \sum_{\omega, \omega^\prime} e^{-\frac{it}{2}(\omega^\prime-\omega)} \gamma_{ij}(\omega,\omega^\prime,t) \left( S_i (\omega)\rho S_j^\dagger (\omega^\prime) -\frac{1}{2} \left\{S_j^\dagger (\omega^\prime) S_i (\omega),\rho\right\}\right),
\end{align}
Because the form of $\gamma_{ij}(\omega,\omega^\prime,t)$ in equation \eqref{eqn:gammaPreIntegrated}, we can introduce a new object $\bar{\gamma}_{ij}(\omega,\omega^\prime,t)$:
\begin{align}
    &\bar{\gamma}_{ij}(\omega,\omega^\prime,t) = e^{-\frac{it}{2}(\omega^\prime-\omega)} \gamma_{ij}(\omega,\omega^\prime,t)  \\
    &= e^{-\frac{it}{2}(\omega^\prime-\omega)}  e^{i \frac{\omega^\prime-\omega}{2}t} \int_{-\infty}^{\infty} d\Omega~\left[t~ \mathrm{sinc} \left(\frac{\omega^\prime-\Omega}{2}t\right)\right]  \left[t~ \mathrm{sinc} \left(\frac{\omega-\Omega}{2}t\right)\right]  R_{ji} (\Omega)\\
	&= \int_{-\infty}^{\infty} d\Omega~\left[t~ \mathrm{sinc} \left(\frac{\omega^\prime-\Omega}{2}t\right)\right]  \left[t~ \mathrm{sinc} \left(\frac{\omega-\Omega}{2}t\right)\right]  R_{ji} (\Omega),
\end{align}
which is a positive semi-definite matrix, what follows from the the fact that it is a Hadamard product of two positive semi-definite matrices (see Section \ref{sec:app:CPTP})
\begin{align}
% _{(\omega,i),(\omega^\prime,j)}
    &\left(e^{-\frac{it}{2}(\omega^\prime-\omega)} \gamma_{ij}(\omega,\omega^\prime,t)\right) = \left( e^{-\frac{it}{2}(\omega^\prime-\omega)} \right) \circ \Big( \gamma_{ij}(\omega,\omega^\prime,t)\Big),\\ 
    &e^{-\frac{it}{2}(\omega^\prime-\omega)} =  \left( e^{-\frac{it}{2}(\omega^\prime-\omega)} \right)_{(\omega,i),(\omega^\prime,j)},~~
    \gamma_{ij}(\omega,\omega^\prime,t) = \Big(\gamma_{ij}(\omega,\omega^\prime,t)\Big)_{(\omega,i),(\omega^\prime,j)}.
\end{align}

Finally, we can write down the approximate form of the cumulant equation in the Schr{\"o}dinger picture
\begin{align}\label{eqn:AofKreducedSchroedinger}
    &{\rho}_\mathcal{S} (t)= e^{ {K}^{(2)}(t)}{\rho}_\mathcal{S} (0),\\ 
    &{K}^{(2)}(t) {\rho}_\mathcal{S} (0) =-it[H_\mathcal{S},\rho] + \sum_{i,j} \sum_{\omega, \omega^\prime} \bar{\gamma}_{ij}(\omega,\omega^\prime,t)  \left(S_i (\omega) {\rho}_\mathcal{S} (0) S_j^\dagger (\omega^\prime) - \frac{1}{2} \left\{S_j^\dagger (\omega^\prime) S_i (\omega), {\rho}_\mathcal{S} (0) \right\} \right),\\
    &\bar{\gamma}_{ij}(\omega,\omega^\prime,t) =\int_{-\infty}^{\infty} d\Omega~\left[t~ \mathrm{sinc} \left(\frac{\omega^\prime-\Omega}{2}t\right)\right]  \left[t~ \mathrm{sinc} \left(\frac{\omega-\Omega}{2}t\right)\right]  R_{ji} (\Omega).
\end{align}
%\sout{Similarly, as the formula in equation \eqref{eqn:gammaPreIntegrated} the formula above is valid for all times except super-short time scale.}

\newpage
\subsection{The detailed analysis of the second correction}\label{sec:app:subsec:detailed}

In this Section we intend to find long-time limit for the second correction $\tilde{H}_\mathcal{C}^{(2)} (t)$. %\sout{For the sake of simplicity (and ease of notation) in this Section that Sch{\"o}dinger picture interaction operators $R_i^{(0)}$ are time-independent. }
\begin{align}
    &\tilde{H}_\mathcal{C}^{(2)} (t)= \sum_{ij}\sum_{\omega_2,\omega_2^\prime}  S_j^\dagger (\omega_2^\prime) S_i(\omega_2)  \frac{d}{dt} \Xi_{ij}(\omega_2,\omega_2^\prime,t), \\
    &\Xi_{ij} (\omega,\omega^\prime,t) = \frac{1}{2i} \int_0^t ds \int_0^t dw~ \mathrm{sgn} (s-w) e^{i(\omega^\prime s - \omega w)} \left<\tilde{R}^{(1)}_j (s) \tilde{R}^{(1)}_i (w)\right>_{\tilde{\rho}_\mathcal{R}}.
\end{align}
Let us start with rephrasing the formula above.
\begin{align}
    &\Xi_{ij} (\omega,\omega^\prime,t) = \frac{1}{2i} \int_0^t ds \int_0^t dw~ \mathrm{sgn} (s-w) e^{i(\omega^\prime s - \omega w)} \left<\tilde{R}^{(1)}_j (s) \tilde{R}^{(1)}_i (w)\right>_{\tilde{\rho}_\mathcal{R}}\\
    &=\frac{1}{2i} \left( \int_0^t ds \int_0^s dw~  e^{i(\omega^\prime s - \omega w)} \left<\tilde{R}^{(1)}_j (s) \tilde{R}^{(1)}_i (w)\right>_{\tilde{\rho}_\mathcal{R}}-\int_0^t ds \int_s^t dw~  e^{i(\omega^\prime s - \omega w)} \left<\tilde{R}^{(1)}_j (s) \tilde{R}^{(1)}_i (w)\right>_{\tilde{\rho}_\mathcal{R}}\right) \\
    &=\frac{1}{2i} \left( \int_0^t ds \int_0^s dw~  e^{i(\omega^\prime s - \omega w)} \left<\tilde{R}^{(1)}_j (s) \tilde{R}^{(1)}_i (w)\right>_{\tilde{\rho}_\mathcal{R}}-\int_0^t dw \int_0^w ds~  e^{i(\omega^\prime s - \omega w)} \left<\tilde{R}^{(1)}_j (s) \tilde{R}^{(1)}_i (w)\right>_{\tilde{\rho}_\mathcal{R}}\right) \\
    &=\frac{1}{2i} \left( \int_0^t ds \int_0^s dw~  e^{i(\omega^\prime s - \omega w)} \left<\tilde{R}^{(1)}_j (s) \tilde{R}^{(1)}_i (w)\right>_{\tilde{\rho}_\mathcal{R}}-\int_0^t ds \int_0^s dw~  e^{i(\omega^\prime w - \omega s)} \left<\tilde{R}^{(1)}_j (w) \tilde{R}^{(1)}_i (s)\right>_{\tilde{\rho}_\mathcal{R}}\right) \\
    &=\frac{1}{2i} \int_0^t ds \left( \int_0^s dw~  e^{i(\omega^\prime s - \omega w)} \left<\tilde{R}^{(1)}_j (s) \tilde{R}^{(1)}_i (w)\right>_{\tilde{\rho}_\mathcal{R}}- \int_0^s dw~  e^{i(\omega^\prime w - \omega s)} \left<\tilde{R}^{(1)}_j (w) \tilde{R}^{(1)}_i (s)\right>_{\tilde{\rho}_\mathcal{R}}\right).
\end{align}
At this place we can calculate derivative very easily in the second step.
\begin{align}
    \frac{d}{dt} \Xi_{ij}(\omega,\omega^\prime,t) = 
    \frac{1}{2i}  \left( \int_0^t dw~  e^{i(\omega^\prime t - \omega w)} \left<\tilde{R}^{(1)}_j (t) \tilde{R}^{(1)}_i (w)\right>_{\tilde{\rho}_\mathcal{R}}- \int_0^t dw~  e^{i(\omega^\prime w - \omega t)} \left<\tilde{R}^{(1)}_j (w) \tilde{R}^{(1)}_i (t)\right>_{\tilde{\rho}_\mathcal{R}}\right).
\end{align}
To obtain convenient form. Let us now perform some manipulations:
\begin{align}
    &\frac{d}{dt} \Xi_{ij}(\omega,\omega^\prime,t) = 
    \frac{1}{2i}  \left( \int_0^t dw~  e^{i(\omega^\prime t - \omega w)} \left<\tilde{R}^{(1)}_j (t) \tilde{R}^{(1)}_i (w)\right>_{\tilde{\rho}_\mathcal{R}}- \int_0^t dw~  e^{i(\omega^\prime w - \omega t)} \left<\tilde{R}^{(1)}_j (w) \tilde{R}^{(1)}_i (t)\right>_{\tilde{\rho}_\mathcal{R}}\right)\\
    &=\frac{1}{2i}  \left( \int_0^t dw~  e^{i(\omega^\prime t - \omega w)} \left<\tilde{R}^{(1)}_j (t-w) {R}^{(1)}_i \right>_{\tilde{\rho}_\mathcal{R}}- \int_0^t dw~  e^{i(\omega^\prime w - \omega t)} \left<\tilde{R}^{(1)}_j (w-t) {R}^{(1)}_i \right>_{\tilde{\rho}_\mathcal{R}}\right)\\
    &=\frac{e^{i(\omega^\prime  - \omega )t}}{2i}  \left( \int_0^t du~  e^{i\omega u} \left<\tilde{R}^{(1)}_j (u) {R}^{(1)}_i \right>_{\tilde{\rho}_\mathcal{R}}- \int_0^t du~  e^{-i\omega^\prime u} \left<\tilde{R}^{(1)}_j (-u) {R}^{(1)}_i \right>_{\tilde{\rho}_\mathcal{R}}\right).
\end{align}

Let us introduce a new object
\begin{align}
    &\Gamma_{ji}^{(t)}(\omega)= \int_0^t du~  e^{i\omega u} \left<\tilde{R}^{(1)}_j (u) {R}^{(1)}_i \right>_{\rho_{\mathcal{R}}},
\end{align}
which has the property that:
\begin{align}
    \lim_{t \to \infty}  \Gamma_{ji}^{(t)}(\omega) = \int_0^\infty du~  e^{i\omega u} \left<\tilde{R}^{(1)}_j (u) {R}^{(1)}_i \right>_{\rho_{\mathcal{R}}} = \Gamma_{ji}(\omega).
\end{align}
We have:
\begin{align}
    &\frac{d}{dt} \Xi_{ij}(\omega,\omega^\prime,t) = \frac{e^{i(\omega^\prime  - \omega )t}}{2i} \left(\Gamma_{ji}^{(t)}(\omega)-{\Gamma_{ij}^{(t)}}^*(\omega^\prime)\right).
\end{align}

Using the above considerations, we obtain the following form of the second correction $\tilde{H}_\mathcal{C}^{(2)} (t)$ in the interaction picture:
\begin{align}
     &\tilde{H}_\mathcal{C}^{(2)} (t)= \sum_{ij}\sum_{\omega_2,\omega_2^\prime}  e^{i(\omega_2^\prime  - \omega_2 )t}  \frac{\Gamma_{ji}^{(t)}(\omega_2)-{\Gamma_{ij}^{(t)}}^*(\omega_2^\prime)}{2i}S_j^\dagger (\omega_2^\prime) S_i(\omega_2) ,  
\end{align}
if we now transform the above formula to the Schr{\"o}dinger picture, and reshuffle the indices we obtain:
\begin{align}
     &H_\mathcal{C}^{(2)} (t)= \sum_{ij}\sum_{\omega_2,\omega_2^\prime}  \frac{\Gamma_{ij}^{(t)}(\omega_2^\prime)-{\Gamma_{ji}^{(t)}}^*(\omega_2)}{2i}   S_i^\dagger (\omega_2) S_j(\omega_2^\prime),
\end{align}
and in the long time limit,
\begin{align}
     &H_\mathcal{C}^{(2)} = \lim_{t \to +\infty}H_\mathcal{C}^{(2)} (t)= \sum_{ij}\sum_{\omega_{2},\omega_{2}^\prime}  \frac{\Gamma_{ij} (\omega_{2}^\prime)-\Gamma_{ji}^*(\omega_{2})}{2i}   S_i^\dagger (\omega_{2}) S_j(\omega_{2}^\prime)  \label{egn:SecondCorrectionCumulantAsymptoic}
\end{align}
In the above formula we obtain an agreement with the form of Lamb-Stark shift for the Bloch-Redfield equation in Reference~\cite{cattaneo2019local}. 

In order make the second corrections $H_\mathcal{C}^{(2)}$ obtained form the cumulant equation comparable with the second correction derived from the partial trace $H_{mf,\mathcal{C}}^{(2)}$ (see Section \ref{sec:PartialTrace}) we firstly have to assume the following: \color{black}
\begin{align}
    \lim_{t \to + \infty } H_\mathcal{S}^{(0)}(t)=H_\mathcal{S}^{(0)},
\end{align}
where the r.h.s. of the equation above coincides with the bare Hamiltonian of the system $\mathcal{S}$ in equation \eqref{eqn:Htot}. We recall here, that $H_\mathcal{S}^{(0)}(t)$ was an auxiliary object in the derivation of the cumulant equation, therefore for the comparison the above condition is necessary.

We proceed with the following calculations:
\begin{align}
    &H_\mathcal{C}^{(2)} = \sum_{ij}\sum_{\omega_{2},\omega_{2}^\prime}  \frac{\Gamma_{ij} (\omega_{2}^\prime)-\Gamma_{ji}^*(\omega_{2})}{2i}   S_i^\dagger (\omega_{2}) S_j(\omega_{2}^\prime) \\
    &=  \frac{1}{2i}\sum_{ij}\sum_{\omega_{2}^\prime} \Gamma_{ij} (\omega_{2}^\prime) S_i  S_j(\omega_{2}^\prime)-\frac{1}{2i}\sum_{ij}\sum_{\omega_{2}} \Gamma_{ji}^* (\omega_{2}) S_i(\omega_{2})  S_j. \label{eqn:Coincidence}
\end{align}
Let us consider now the first term in equation \eqref{eqn:Coincidence} above. 
\begin{align}
    &\frac{1}{2i}\sum_{ij}\sum_{\omega_{2}^\prime} \Gamma_{ij} (\omega_{2}^\prime) S_i  S_j(\omega_{2}^\prime) = \frac{1}{2i}\sum_{ij}\sum_{\omega_{2}^\prime} 
    \int_0^{+\infty}ds~e^{i\omega_2^\prime s} \left<\tilde{R}^{(1)}_i(s)R^{(1)}_j\right>_{\rho_{\mathcal{R},\beta}} S_i  S_j(\omega_{2}^\prime) \\
    %%%%%%%%%%%%%%%%%%%%%%%%%%
    &= \frac{1}{2i}\sum_{ij}
    \int_0^{+\infty}ds~ \left<\tilde{R}^{(1)}_i(s)R^{(1)}_j\right>_{\rho_{\mathcal{R},\beta}} S_i  \sum_{\omega_{2}^\prime} e^{i\omega_2^\prime s}S_j(\omega_{2}^\prime) 
    =\frac{1}{2i}\sum_{ij}
    \int_0^{+\infty}ds~ \left<\tilde{R}^{(1)}_i(s)R^{(1)}_j\right>_{\rho_{\mathcal{R},\beta}} S_i  \tilde{S}_j(-s) \\
    %%%%%%%%%%%%%%%%%%%%%%%%%%
    &=\frac{1}{2i}\sum_{ij}
    \int_0^{+\infty}ds~ \left<\tilde{R}^{(1)}_i(s)R^{(1)}_j\right>_{\rho_{\mathcal{R},\beta}} S_i  e^{-i H_\mathcal{S}^{(2)}s}S_je^{i H_\mathcal{S}^{(2)}s} \\
    &=\frac{1}{2i}\sum_{ij}
    \int_0^{+\infty}ds~ \left<\tilde{R}^{(1)}_i(s)R^{(1)}_j\right>_{\rho_{\mathcal{R},\beta}} S_i  e^{-i H_\mathcal{S}^{(2)}s}\sum_{\omega_1^\prime}S_j(\omega_1^\prime)e^{i H_\mathcal{S}^{(2)}s} \\
    %%%%%%%%%%%%%%%%%%%%%%%%%%
    &=\frac{1}{2i}\sum_{ij}\sum_{\omega_1^\prime}
    \int_0^{+\infty}ds~ \left<\tilde{R}^{(1)}_i(s)R^{(1)}_j\right>_{\rho_{\mathcal{R},\beta}} S_i  e^{-i H_\mathcal{S}^{(2)}s}e^{i H_\mathcal{S}^{(1)}s}e^{-i H_\mathcal{S}^{(1)}s}S_j(\omega_1^\prime)e^{i H_\mathcal{S}^{(1)}s}e^{-i H_\mathcal{S}^{(1)}s}e^{i H_\mathcal{S}^{(2)}s} \\
    %%%%%%%%%%%%%%%%%%%%%%%%%%
    &=\frac{1}{2i}\sum_{ij}\sum_{\omega_1^\prime}
    \int_0^{+\infty}ds~e^{i\omega_1^\prime s} \left<\tilde{R}^{(1)}_i(s)R^{(1)}_j\right>_{\rho_{\mathcal{R},\beta}} S_i  e^{-i H_\mathcal{S}^{(2)}s}e^{i H_\mathcal{S}^{(1)}s}S_j(\omega_1^\prime)e^{-i H_\mathcal{S}^{(1)}s}e^{i H_\mathcal{S}^{(2)}s}. \label{eqn:Coreections_comparison}
\end{align}

The product of exponents including $H_\mathcal{S}^{(2)}$ and $H_\mathcal{S}^{(1)}$ can be treated most easily with noticing that
\begin{align}
    &e^{-i H_\mathcal{S}^{(2)}s}e^{i H_\mathcal{S}^{(1)}s} \stackrel{\mathcal{O}(\lambda^2)}{\approx} e^{-i H_\mathcal{S}^{(2)}s}e^{i (H_\mathcal{S}^{(1)}+\lambda^2 H_\mathcal{C}^{(2)})s} = e^{-i H_\mathcal{S}^{(2)}s}e^{i H_\mathcal{S}^{(2)} s} =\mathds{1}.
\end{align}
% treated with Baker–Campbell–Hausdorff (BCH) formula (see equation \eqref{eqn:BCHformula}):
% \begin{align}
%     &e^{-i H_\mathcal{S}^{(2)}s}e^{i H_\mathcal{S}^{(1)}s} = e^{-i \left(H_\mathcal{S}^{(1)}+\lambda^2H_\mathcal{C}^{(2)}\right)s}e^{i H_\mathcal{S}^{(1)}s}\\ 
%     &= \exp\left(-i\lambda^2H_\mathcal{C}^{(2)}s+{\frac 12 \lambda^2 \left[H_\mathcal{C}^{(2)},H_\mathcal{S}^{(1)}\right]s^2+\cdots}\right) \stackrel{\mathcal{O}(\lambda^2)}{\approx} \mathds{1}.
% \end{align}
Because any term proportional to $\lambda^2$ in the second-order correction $\lambda^2 H_\mathcal{C}^{(2)}$ is the fourth-order correction to the bare Hamiltonian $H_\mathcal{S}^{(0)}$, it can be neglected. Therefore we have, 
\begin{align}
    &\eqref{eqn:Coreections_comparison} \stackrel{\mathcal{O}(\lambda^2)}{=} \frac{1}{2i}\sum_{ij}\sum_{\omega_1^\prime}
    \int_0^{+\infty}ds~e^{i\omega_1 s} \left<\tilde{R}^{(1)}_i(s)R^{(1)}_j\right>_{\rho_{\mathcal{R},\beta}} S_i  S_j(\omega_1^\prime) = \frac{1}{2i}\sum_{ij}\sum_{\omega_{1}^\prime} \Gamma_{ij} (\omega_{1}^\prime) S_i  S_j(\omega_{1}^\prime),
\end{align}
where the first equality is up to terms of order $\lambda^2$ ($\lambda^4$ with respect to $H_\mathcal{S}^{(0)}$) for which Born approximation is insensitive.

Analogous analysis performed for the second term in equation \eqref{eqn:Coincidence} yields:
\begin{align}
    \lambda^2 H_\mathcal{C}^{(2)} = \lambda^2 \sum_{ij}\sum_{\omega_{2},\omega_{2}^\prime}  \frac{\Gamma_{ij} (\omega_{2}^\prime)-\Gamma_{ji}^*(\omega_{2})}{2i}   S_i^\dagger (\omega_{2}) S_j(\omega_{2}^\prime) \stackrel{\mathcal{O}(\lambda^4)}{=} \lambda^2 \sum_{ij}\sum_{\omega_{1},\omega_{1}^\prime}  \frac{\Gamma_{ij} (\omega_{1}^\prime)-\Gamma_{ji}^*(\omega_{1})}{2i}   S_i^\dagger (\omega_{1}) S_j(\omega_{1}^\prime). 
\end{align}
The above results allows to compare the second correction from the cumulant equation $H_\mathcal{C}^{(2)}$ and the second correction $H_{mf,\mathcal{C}}^{(2)}$ from the mean-force Hamiltonian approach.
In this way we find that the both corrections lead to different renormalization of the energy levels of $H_\mathcal{S}^{(0)}$. This is done via considering the difference between diagonal elements of $H_{mf,\mathcal{C}}^{(2)}$ and  $H_\mathcal{C}^{(2)}$ in the eigenbasis of $H_\mathcal{S}^{(1)}$
\begin{align}
    \mathrm{diag}_1\left(H_{mf,\mathcal{C}}^{(2)}-H_\mathcal{C}^{(2)}\right) \stackrel{\mathcal{O}(\lambda^4)}{\approx}  -\frac{1}{\beta} \sum_{\omega_{1} } \sum_{ij}  \left(\frac{\partial}{\partial \omega} \mathcal{S}_{ij}(\omega_1)+e^{\beta \omega_1}\frac{\partial}{\partial \omega} \mathcal{S}_{ji}(-\omega_1)  \right)S_i^\dagger (\omega_{1}) S_j(\omega_{1}),
\end{align}
where $\mathrm{diag}_1(\cdot)$ is the projection on the diagonal part of the operator in the $H_\mathcal{S}^{(1)}$ basis.
\begin{align}
    \mathrm{diag}_1(A) = \sum_{\epsilon_1} \Pi(\epsilon_1) A \Pi(\epsilon_1).
\end{align}
\color{black}

%\sout{The above results shows how the self-consistency condition provides the same second correction $\lambda^2 H_\mathcal{C}^{(2)}$ (within second-order approximation for the dynamics), and consequently the same stationary state as the treatment based on the partial trace considered in Section \ref{sec:PartialTrace}.}

\subsection{Complete Positive and Trace Preserving (CPTP) dynamics of the cumulant equation}\label{sec:app:CPTP}

The CPTP property of the cumulant equations dynamics follows from the GKSL form of the $\tilde{K}^{(2)}(t)$ superoperator~\cite{gorini1976completely,lindblad1976generators}, and the positive semi-definiteness of the $\big( \gamma_{ij}(\omega,\omega^\prime,t) \big)$ matrix. For the sake of completeness of the paper, in this section we prove the later property.
\begin{align}
    \forall_g~~ &\sum_{ij} \sum_{\omega,\omega^\prime} g^*_i(\omega)g_j(\omega^\prime)\gamma_{ij}(\omega,\omega^\prime,t) = \sum_{ij} \sum_{\omega,\omega^\prime} g^*_i(\omega)g_j(\omega^\prime)\int_0^t ds \int_0^t dw~ e^{i (\omega^\prime s - \omega w)} \left< \tilde{R}^{(1)}_j (s) \tilde{R}^{(1)}_i (w) \right>_{\tilde{\rho}_\mathcal{R}} \\
    &=\sum_{ij} \sum_{\omega,\omega^\prime}  \left<\left(\int_0^t ds~g_j(\omega^\prime) e^{i \omega^\prime s } \tilde{R}^{(1)}_j (s)\right)\left( \int_0^t dw~g_i(\omega) e^{i \omega w} \tilde{R}^{(1)}_i (w) \right)^\dagger\right>_{\tilde{\rho}_\mathcal{R}}\\
    &=\sum_{ij} \sum_{\omega,\omega^\prime}  \tr\left[\left(\int_0^t ds~g_j(\omega^\prime) e^{i \omega^\prime s } \tilde{R}^{(1)}_j (s)\right)\left( \int_0^t dw~g_i(\omega) e^{i \omega w} \tilde{R}^{(1)}_i (w) \right)^\dagger \tilde{\rho}_\mathcal{R}\right]\\
    &=\int d\epsilon~   \bra{\phi_\epsilon}\left(\sum_{j} \sum_{\omega^\prime}\int_0^t ds~g_j(\omega^\prime) e^{i \omega^\prime s } \tilde{\rho}^{\frac 12}_\mathcal{R}\tilde{R}^{(1)}_j (s)  \right)\left( \sum_{i} \sum_{\omega} \int_0^t dw~g_i(\omega) e^{i \omega w} \tilde{\rho}^{\frac 12}_\mathcal{R}\tilde{R}^{(1)}_i (w)  \right)^\dagger \ket{\phi_\epsilon} \\
    &=\int d\epsilon \int  d\epsilon^\prime~   \bra{\phi_\epsilon}\left(\sum_{j} \sum_{\omega^\prime}\int_0^t ds~g_j(\omega^\prime) e^{i \omega^\prime s } \tilde{\rho}^{\frac 12}_\mathcal{R}\tilde{R}^{(1)}_j (s)  \right) \ket{\phi_{\epsilon^\prime}}\bra{\phi_{\epsilon^\prime}}\left( \sum_{i} \sum_{\omega} \int_0^t dw~g_i(\omega) e^{i \omega w} \tilde{\rho}^{\frac 12}_\mathcal{R}\tilde{R}^{(1)}_i (w)  \right)^\dagger \ket{\phi_\epsilon} \\
    &=\int d\epsilon \int  d\epsilon^\prime~   \left(\bra{\phi_\epsilon}\sum_{j} \sum_{\omega^\prime}\int_0^t ds~g_j(\omega^\prime) e^{i \omega^\prime s } \tilde{\rho}^{\frac 12}_\mathcal{R}\tilde{R}^{(1)}_j (s)   \ket{\phi_{\epsilon^\prime}}\right)\left(\bra{\phi_{\epsilon^\prime}} \sum_{i} \sum_{\omega} \int_0^t dw~g_i(\omega) e^{i \omega w} \tilde{\rho}^{\frac 12}_\mathcal{R}\tilde{R}^{(1)}_i (w)   \ket{\phi_\epsilon}\right)^* \\
    &=\int d\epsilon \int  d\epsilon^\prime~   \abs{\bra{\phi_\epsilon}\left(\sum_{j} \sum_{\omega^\prime}\int_0^t ds~g_j(\omega^\prime) e^{i \omega^\prime s } \tilde{\rho}^{\frac 12}_\mathcal{R}\tilde{R}^{(1)}_j (s)  \right) \ket{\phi_{\epsilon^\prime}} }^2 \ge 0,
\end{align}
where $\{\phi_\epsilon\}$ are elements of the complete orthonormal basis. The above proves that indeed  $\big( \gamma_{ij}(\omega,\omega^\prime,t) \big)$ is a positive semi-definite matrix.

\subsection{Relation to the Bloch-Redfield equation}\label{sec:app:subsec:Relatioon_BR}

In this Section of the Appendix we elaborate on the relation between the cumulant equation and the Bloch-Redfield equation~\cite{Maniscalco_2004,Blum_2012}. Let us start with showing that 
\begin{align}
    \frac{d}{dt} \tilde{K}^{(2)}(t) = \tilde{\mathcal{L}}_{BR}(t),
\end{align}
where the $\tilde{\mathcal{L}}_{BR}(t)$ is the generator of the (interaction picture, renormalized) Bloch-Redfield equation
\begin{align}
    &\frac{d}{dt} \tilde{\rho}_{S}(t)=\tilde{\mathcal{L}}_{BR}(t)\tilde{\rho}_{S}(t), \label{eqn:BR}\\
    &\tilde{\mathcal{L}}_{BR}(t) \rho_\mathcal{S} = \sum_{i,j} \sum_{\omega_2, \omega^\prime_2} \tilde{\gamma}^{BR}_{ij}(\omega_2,\omega^\prime_2,t)  \left(S_i (\omega_2) {\rho}_\mathcal{S}  S_j^\dagger (\omega^\prime_2) - \frac{1}{2} \left\{S_j^\dagger (\omega^\prime_2) S_i (\omega_2), {\rho}_\mathcal{S}  \right\} \right).
\end{align}

For the proof let us notice that:
\begin{align}
    &\gamma_{ij} (\omega,\omega^\prime,t) = \int_0^t ds \int_0^t dw~ e^{i (\omega^\prime s - \omega w)} \left< \tilde{R}^{(1)}_j (s) \tilde{R}^{(1)}_i (w) \right>_{\tilde{\rho}_\mathcal{R}}\\
    &=\int_0^t ds \int_0^s dw~ e^{i (\omega^\prime s - \omega w)} \left< \tilde{R}^{(1)}_j (s) \tilde{R}^{(1)}_i (w) \right>_{\tilde{\rho}_\mathcal{R}}
    +\int_0^t ds \int_s^t dw~ e^{i (\omega^\prime s - \omega w)} \left< \tilde{R}^{(1)}_j (s) \tilde{R}^{(1)}_i (w) \right>_{\tilde{\rho}_\mathcal{R}}\\
    &=\int_0^t ds \int_0^s dw~ e^{i (\omega^\prime s - \omega w)} \left< \tilde{R}^{(1)}_j (s) \tilde{R}^{(1)}_i (w) \right>_{\tilde{\rho}_\mathcal{R}}
    +\int_0^t dw\int_0^w ds ~ e^{i (\omega^\prime s - \omega w)} \left< \tilde{R}^{(1)}_j (s) \tilde{R}^{(1)}_i (w) \right>_{\tilde{\rho}_\mathcal{R}}.
\end{align}
Therefore:
\begin{align}
    &\frac{d}{dt} \gamma_{ij} (\omega,\omega^\prime,t)= \int_0^t dw~ e^{i (\omega^\prime t - \omega w)} \left< \tilde{R}^{(1)}_j (t) \tilde{R}^{(1)}_i (w) \right>_{\tilde{\rho}_\mathcal{R}}
    +\int_0^t ds ~ e^{i (\omega^\prime s - \omega t)} \left< \tilde{R}^{(1)}_j (s) \tilde{R}^{(1)}_i (t) \right>_{\tilde{\rho}_\mathcal{R}}\\
    &= e^{i \omega^\prime t } \int_0^t ds~ e^{-i \omega s} \left< \tilde{R}^{(1)}_j (t-s) {R}^{(1)}_i  \right>_{\tilde{\rho}_\mathcal{R}}
    +e^{ -i \omega t} \int_0^t ds ~ e^{i \omega^\prime s } \left< \tilde{R}^{(1)}_j (s-t) {R}^{(1)}_i  \right>_{\tilde{\rho}_\mathcal{R}} \\
    &= e^{i (\omega^\prime-\omega) t } \int_{0}^t ds~ e^{i \omega s} \left< \tilde{R}^{(1)}_j (s) {R}^{(1)}_i  \right>_{\tilde{\rho}_\mathcal{R}}
    +e^{ i(\omega^\prime - \omega) t} \int_{0}^t ds ~ e^{-i \omega^\prime s } \left< {R}^{(1)}_j  \tilde{R}^{(1)}_i(s)  \right>_{\tilde{\rho}_\mathcal{R}} \\
    &= e^{i (\omega^\prime-\omega) t } \int_{0}^t ds~ e^{i \omega s} \left< \tilde{R}^{(1)}_j (s) {R}^{(1)}_i  \right>_{\tilde{\rho}_\mathcal{R}}
    +e^{ i(\omega^\prime - \omega) t} \left( \int_{0}^t ds ~ e^{i \omega^\prime s } \left<   \tilde{R}^{(1)}_i(s) {R}^{(1)}_j \right>_{\tilde{\rho}_\mathcal{R}}  \right)^* \\
    &= e^{i (\omega^\prime-\omega) t } \left(\Gamma_{ji}^{(t)}(\omega)+{\Gamma_{ij}^{(t)}}^*(\omega^\prime)\right).
\end{align}
In the above formula we recognise the elements of the $\tilde{\gamma}_{ij}^{BR}(\omega,\omega^\prime,t)$ (see \cite{cattaneo2019local} for the agreement with the long-time form). 

Using Lie-algebraic methods \cite{Wulf_2008}\footnote{See the Theorem 5. on page 15. The proof here is analogous up to some minor modifications.}, the cumulant equation can be transformed into a differential equation of the following form
\begin{align}
\frac{d}{dt} \tilde{\rho}_S(t)= \left(\frac{e^{[\tilde{K}^{(2)}(t),\cdot]}-1}{[\tilde{K}^{(2)}(t),\cdot]}\frac{d \tilde{K}^{(2)}(t)}{dt}\right)\tilde{\rho}_S(t).
\end{align}
If we now truncate the above equation up to the second-order (in the coupling constant) on the r.h.s. we obtain the Bloch-Redfield equation \eqref{eqn:BR}. As we see above the Bloch-Redfield equation is an approximation of the cumulant equation. 

% \textcolor{red}{Finding the stationary state of the cumulant equation (if exists) is a formidable task out of scope of this paper. In order to find an approximation of the long-time state of the cumulant equation we observe firstly that the Bloch-Redfield equation (that approximates the cumulant equation) has a stationary state in the Schr{\"o}dinger picture.} 

We infer that the long-time limit, renormalized Bloch Redfield equation in the Schr{\"o}dinger picture has the following form
\begin{align}
    &\frac{d}{dt} {\rho}_{S}(t)={\mathcal{L}^{\infty}_{BR}}{\rho}_{S}(t), \label{eqn:BR_Schroedinger}\\
    &{\mathcal{L}^{\infty}_{BR}} \rho_\mathcal{S} =-i\left[H_\mathcal{S},\rho_\mathcal{S}\right] +\sum_{i,j} \sum_{\omega_2, \omega^\prime_2} {\gamma}^{BR}_{ij}(\omega_2,\omega^\prime_2)  \left(S_i (\omega_2) {\rho}_\mathcal{S}  S_j^\dagger (\omega^\prime_2) - \frac{1}{2} \left\{S_j^\dagger (\omega^\prime_2) S_i (\omega_2), {\rho}_\mathcal{S}  \right\} \right), \\
    &{\gamma}^{BR}_{ij}(\omega,\omega^\prime) =\Gamma_{ji}(\omega)+{\Gamma^*_{ij}}(\omega^\prime),
\end{align}
were ${\mathcal{L}^{\infty}_{BR}}$ denotes the long-time limit of ${\mathcal{L}_{BR}}$, i.e.,
\begin{align}
    {\mathcal{L}^{\infty}_{BR}} = \lim_{t\to\infty} {\mathcal{L}_{BR}}(t).
\end{align}

We now look for the stationary state in the following form
\begin{align}
    \rho^{(2,ss)}_{\mathcal{S},BR} \sim e^{-\beta (H_\mathcal{S}^{(2)}+\delta H)}, 
\end{align}
under the condition 
\begin{align}
    \mathcal{L}^{\infty}_{BR}(t) \rho^{(2,ss)}_{\mathcal{S},BR}=0, \label{eqn:SS_full}
\end{align}
where we assume that $\delta H$ is second-order in the coupling constant.

We now use equation \eqref{eqn:SS_full} to find the stationary state $\rho^{(2,ss)}_{\mathcal{S},BR}$. Firstly we approximate the $\rho^{(2,ss)}_{\mathcal{S},BR}$ state with the Dyson expansion:
\begin{align}
    \rho^{(2,ss)}_{\mathcal{S},BR} &\sim e^{-\beta (H_\mathcal{S}^{(2)}+\delta H)}
    =e^{-\beta H_\mathcal{S}^{(2)}} \left(\mathds{1}-\int_0^\beta dx~ \delta\tilde{H}(-ix) e^{x H_\mathcal{S}^{(2)}} e^{-x (H_\mathcal{S}^{(2)}+\delta H)} \right)\\
    &\stackrel{\mathcal{O}(\lambda^4)}{\approx} e^{-\beta H_\mathcal{S}^{(2)}} -e^{-\beta H_\mathcal{S}^{(2)}} \int_0^\beta dx~ \delta\tilde{H}(-ix) , \label{eqn:ss_leading_orders}
\end{align}
where
\begin{align}
    \delta\tilde{H}(-ix) = e^{x H_\mathcal{S}} \delta H e^{-x H_\mathcal{S}}. 
\end{align}
Let us recall that for long times $H_\mathcal{S}^{(2)}=H_\mathcal{S}^{(0)}+H_\mathcal{C}^{(2)}$ (we assume $H_\mathcal{C}^{(1)}=0$ for the sake of simplicity). However, the splitting of the state in equation \eqref{eqn:ss_leading_orders} into two terms is not accidental. The first term in equation \eqref{eqn:ss_leading_orders} has zeroth-order as leading order, and the second-term has the second-order as the leading one.

We substitute now the form of the stationary state in equation \eqref{eqn:ss_leading_orders} to equation \eqref{eqn:SS_full}, and we truncate all terms for which the leading order is higher than two.
\begin{align}
    i\left[H_\mathcal{S}^{(2)},e^{-\beta H_\mathcal{S}^{(2)}} \int_0^\beta dx~ \delta\tilde{H}(-ix)\right]+\sum_{i,j} \sum_{\omega_2, \omega^\prime_2} {\gamma}^{BR}_{ij}(\omega_2,\omega^\prime_2)  \left(S_i (\omega_2) e^{-\beta H_\mathcal{S}^{(2)}}  S_j^\dagger (\omega^\prime_2) - \frac{1}{2} \left\{S_j^\dagger (\omega^\prime_2) S_i (\omega_2), e^{-\beta H_\mathcal{S}^{(2)}}  \right\} \right)=0.
\end{align}

Let us now expand the correction $\delta H$ in the following form
\begin{align}
    \delta H = \sum_{\omega_2,\omega_2^\prime} \sum_{i,j} f_{ij}(\omega_2,\omega_2^\prime) S_j^\dagger (\omega^\prime_2) S_i (\omega_2).
\end{align}
Using the properties described in section \ref{sec:app:Properties} of the Appendix we immediately obtain that:
\begin{align}
    \int_0^\beta dx~\delta \tilde{H} (-ix) &= \int_0^\beta dx~\sum_{\omega_2,\omega_2^\prime} \sum_{ij} f_{ij}(\omega_2,\omega_2^\prime) e^{x (\omega^\prime_2-\omega_2)} S_j^\dagger (\omega^\prime_2) S_i (\omega_2) \\
    &= \sum_{\omega_2,\omega_2^\prime} \sum_{ij} f_{ij}(\omega_2,\omega_2^\prime) \frac{e^{\beta (\omega^\prime_2-\omega_2)}-1}{\omega^\prime_2-\omega_2} S_j^\dagger (\omega^\prime_2) S_i (\omega_2),
\end{align}
and consequently
\begin{align}
    i\left[H_\mathcal{S}^{(2)}, \int_0^\beta dx~ \delta\tilde{H}(-ix)\right] = i  \sum_{\omega_2,\omega_2^\prime} \sum_{ij} f_{ij}(\omega_2,\omega_2^\prime) \left(e^{\beta (\omega^\prime_2-\omega_2)}-1\right) e^{-\beta H_\mathcal{S}^{(2)}} S_j^\dagger (\omega^\prime_2) S_i (\omega_2).
\end{align}
Similarly for the second term we have:
\begin{align}
    &\sum_{\omega_2, \omega^\prime_2}\sum_{i,j}  {\gamma}^{BR}_{ij}(\omega_2,\omega^\prime_2)  \left(S_i (\omega_2) e^{-\beta H_\mathcal{S}^{(2)}}  S_j^\dagger (\omega^\prime_2) - \frac{1}{2} \left\{S_j^\dagger (\omega^\prime_2) S_i (\omega_2), e^{-\beta H_\mathcal{S}^{(2)}}  \right\} \right) \\
    &= 
    \sum_{\omega_2,\omega_2^\prime} \sum_{ij} \left( e^{\beta \omega_2^\prime}\gamma_{ji}^{BR}(-\omega_2^\prime,-\omega_2)-\frac{1}{2}\left(e^{-\beta(\omega_2-\omega_2^\prime)}+1\right)\gamma_{ij}^{BR}(\omega_2,\omega_2^\prime)\right)  e^{-\beta H_\mathcal{S}\textbf{}} S_j^\dagger (\omega^\prime_2) S_i (\omega_2).
\end{align}
We obtain an equation
\begin{align}
    &i  \sum_{\omega_2,\omega_2^\prime} \sum_{ij} f_{ij}(\omega_2,\omega_2^\prime) \left(e^{\beta (\omega^\prime_2-\omega_2)}-1\right) e^{-\beta H_\mathcal{S}^{(2)}} S_j^\dagger (\omega^\prime_2) S_i (\omega_2) \nonumber \\
    &+ \sum_{\omega_2,\omega_2^\prime} \sum_{ij} \left( e^{\beta \omega_2^\prime}\gamma_{ji}^{BR}(-\omega_2^\prime,-\omega_2)-\frac{1}{2}\left(e^{-\beta(\omega_2-\omega_2^\prime)}+1\right)\gamma_{ij}^{BR}(\omega_2,\omega_2^\prime)\right)  e^{-\beta H_\mathcal{S}^{(2)}} S_j^\dagger (\omega^\prime_2) S_i (\omega_2). \label{eqn:Redfield_intermediate_step}
\end{align}
As we observe the above matrix equation does not provide a solution for $f_{ij}(\omega_2,\omega_2)$ coefficients. This is because for $\omega_2^\prime=\omega_2$ the matrix elements ${\gamma}^{BR}_{ij}(\omega_2,\omega^\prime_2)$ satisfy the detailed-balance condition
\begin{align}
    {\gamma}^{BR}_{ji}(-\omega,-\omega) = e^{-\beta \omega} {\gamma}^{BR}_{ij}(\omega,\omega),
\end{align}
and the equation for the diagonal elements ($\omega^\prime=\omega$) that can be extracted from equation \eqref{eqn:Redfield_intermediate_step} is satisfied identically\footnote{See equation \eqref{eqn:DetailedBalance} for the argument why these elements vanish in the case of the cumulant equation.}. Still, this is only possible due to initial Dyson series approximation of $\rho^{(2,ss)}_{\mathcal{S},BR}$ around $e^{-\beta H_\mathcal{S}^{(2)}}$, therefore $f_{ij}(\omega_2,\omega_2)=0$.

Finally for $\omega_2^\prime \neq \omega_2$ we obtain
\begin{align}
    f_{ij}(\omega_2,\omega_2^\prime)=\frac{i}{e^{\beta (\omega^\prime_2-\omega_2)}-1}\left( e^{\beta \omega_2^\prime}\gamma_{ji}^{BR}(-\omega_2^\prime,-\omega_2)-\frac{1}{2}\left(e^{-\beta(\omega_2-\omega_2^\prime)}+1\right)\gamma_{ij}^{BR}(\omega_2,\omega_2^\prime)\right),
\end{align}
and therefore:
\begin{align}
     \delta H = \sum_{\omega_2 \neq \omega_2^\prime} \sum_{i,j} \frac{i}{e^{\beta (\omega^\prime_2-\omega_2)}-1}\left( e^{\beta \omega_2^\prime}\gamma_{ji}^{BR}(-\omega_2^\prime,-\omega_2)-\frac{1}{2}\left(e^{-\beta(\omega_2-\omega_2^\prime)}+1\right)\gamma_{ij}^{BR}(\omega_2,\omega_2^\prime)\right) S_j^\dagger (\omega^\prime_2) S_i (\omega_2).
\end{align}
It is easy to check that the above correction $\delta H$ is hermitian as required. 

\color{black}

\subsection{The long-time limit for the cumulant equation}\label{sec:app:subsec:long-time}

In this Section we aim to find the long-time limit of the cumulant equation. We target to express $\gamma_{ij}(\omega,\omega^\prime,t)$ and $\Xi_{ij}(\omega,\omega^\prime,t)$ present in the cumulant equation in terms of the quantities known from the Davies-GKSL equation, i.e., these in equations \eqref{eqn:def:coe1}, \eqref{eqn:def:coe2} and \eqref{eqn:def:coe3}. In this way we correct the result in \cite{Rivas_2017}.

Let us notice that the formula in equation \eqref{eqn:long-time_start} can be transformed into the following form\footnote{We skip the renormalization indices ($\omega_2 \to \omega$) for the purposes of this Section.}
\begin{align}
    \eqref{eqn:long-time_start}&=\sum_{\omega,\omega^\prime}\sum_{i,j} \int_{-\infty}^{+\infty}d\Omega \int_0^t ds \int_0^s dw~ e^{i(\Omega-\omega)w} e^{-i(\Omega+\omega^\prime)s} \left({S}_i(\omega) \tilde{\rho}_S (0) {S}_j(\omega^\prime) -  {S}_j(\omega^\prime) {S}_i(\omega) \tilde{\rho}_S (0) \right) R_{ji}(\Omega) \nonumber \\
    & +\sum_{\omega,\omega^\prime}\sum_{i,j} \int_{-\infty}^{+\infty}d\Omega\int_0^t ds \int_0^s dw~e^{-i(\Omega+\omega)w} e^{i(\Omega-\omega^\prime)s} \left( {S}_i(\omega^\prime) \tilde{\rho}_S (0) {S}_j(\omega) - \tilde{\rho}_S (0)  {S}_j(\omega) {S}_i(\omega^\prime)  \right)  R_{ji}(\Omega)\\
    &=\sum_{\omega,\omega^\prime}\sum_{i,j} A_{ij}(\omega,\omega^\prime,t) \left({S}_i(\omega) \tilde{\rho}_S (0) {S}_j(\omega^\prime) -  {S}_j(\omega^\prime) {S}_i(\omega) \tilde{\rho}_S (0) \right) R_{ji}(\Omega) \nonumber \\
    & +\sum_{\omega,\omega^\prime}\sum_{i,j} B_{ij}(\omega^\prime,\omega,t) \left( {S}_i(\omega^\prime) \tilde{\rho}_S (0) {S}_j(\omega) - \tilde{\rho}_S (0)  {S}_j(\omega) {S}_i(\omega^\prime)  \right)  R_{ji}(\Omega)\\
    &=-i  \left[\sum_{\omega,-\omega^\prime}\sum_{i,j} \Xi_{ij}(\omega,\omega^\prime,t){S}_j^\dagger(\omega^\prime) {S}_i(\omega),\tilde{\rho}_S (0)\right] \nonumber\\
    &+\sum_{\omega,\omega^\prime}\sum_{i,j} \gamma_{ij}(\omega,\omega^\prime,t) \left({S}_i(\omega) \tilde{\rho}_S (0) {S}^\dagger_j(\omega^\prime) -  \frac 12 \left\{{S}^\dagger_j(\omega^\prime) {S}_i(\omega) ,\tilde{\rho}_S (0)\right\} \right),
\end{align}
where $R_{ji}(\Omega)=\frac{1}{2\pi} \gamma_{ji}(\Omega)$ and
\begin{align}
    &A_{ij}(\omega,\omega^\prime,t)= \int_{-\infty}^{+\infty}d\Omega \int_0^t ds \int_0^s dw~ e^{i(\Omega-\omega)w} e^{-i(\Omega+\omega^\prime)s} R_{ji}(\Omega),\\
    &B_{ij}(\omega^\prime,\omega,t)\equiv \int_{-\infty}^{+\infty}d\Omega\int_0^t ds \int_0^s dw~e^{-i(\Omega+\omega)w} e^{i(\Omega-\omega^\prime)s} R_{ji}(\Omega), \\
    &\Xi_{ij}(\omega,\omega^\prime,t)=\frac{A_{ij}(\omega,-\omega^\prime,t)-B_{ij}(\omega,-\omega^\prime,t)}{2i}, \\
    &\gamma_{ij}(\omega,\omega^\prime,t)=A_{ij}(\omega,-\omega^\prime,t)+B_{ij}(\omega,-\omega^\prime,t).
\end{align}
The above can readily be made with relations in the Section \ref{sec:app:Properties} of the Appendix.

It can be shown that:
\begin{align}
    &\int_0^t ds \int_0^s dw~ e^{i(\Omega-\omega)w} e^{-i(\Omega+\omega^\prime)s} = \frac{1-e^{-it(\omega+\omega^\prime)}}{(\omega+\omega^\prime)(\omega-\Omega)}+\frac{e^{-it(\Omega+\omega^\prime)}-1}{(\omega-\Omega)(\omega^\prime+\Omega)}, \label{eqn:Intermediat_int1}\\
    &\int_0^t ds \int_0^s dw~e^{-i(\Omega+\omega)w} e^{i(\Omega-\omega^\prime)s}=\frac{1-e^{-it(\omega+\omega^\prime)}}{(\omega+\omega^\prime)(\omega+\Omega)}+\frac{e^{it(\Omega-\omega^\prime)}-1}{(\omega+\Omega)(\omega^\prime-\Omega)}
\end{align}

We use the form in equation \eqref{eqn:Intermediat_int1} to compute the following.
\begin{align}
    &\int_{-\infty}^{+\infty}d\Omega \int_0^t ds \int_0^s dw~ e^{i(\Omega-\omega)w} e^{-i(\Omega+\omega^\prime)s} R_{ji}(\Omega)= \int_{-\infty}^{+\infty}d\Omega \left( \frac{1-e^{-it(\omega+\omega^\prime)}}{(\omega+\omega^\prime)(\omega-\Omega)}+\frac{e^{-it(\Omega+\omega^\prime)}-1}{(\omega-\Omega)(\omega^\prime+\Omega)} \right)R_{ji}(\Omega) \\
    &=\frac{1-e^{-it(\omega+\omega^\prime)}}{(\omega+\omega^\prime)} \dashint_{-\infty}^{+\infty}d\Omega~  \frac{1}{\omega-\Omega}R_{ji}(\Omega)
    +\frac{1}{\omega^\prime+\omega}\dashint_{-\infty}^{+\infty}d\Omega~\left(e^{-it(\Omega+\omega^\prime)}-1\right) (\frac{1}{\omega-\Omega}+\frac{1}{\Omega+\omega^\prime}) R_{ji}(\Omega) \\
    &=\frac{1-e^{-it(\omega+\omega^\prime)}}{(\omega+\omega^\prime)} \dashint_{-\infty}^{+\infty}d\Omega~  \frac{1}{\omega-\Omega}R_{ji}(\Omega)
    +\frac{1}{\omega^\prime+\omega}\dashint_{-\infty}^{+\infty}d\Omega~e^{-it(\Omega+\omega^\prime)} (\frac{1}{\omega-\Omega}+\frac{1}{\Omega+\omega^\prime}) R_{ji}(\Omega) \nonumber \\
    &-\frac{1}{\omega^\prime+\omega}\dashint_{-\infty}^{+\infty}d\Omega~ (\frac{1}{\omega-\Omega}+\frac{1}{\Omega+\omega^\prime}) R_{ji}(\Omega) \\
    &=\frac{1-e^{-it(\omega+\omega^\prime)}}{(\omega+\omega^\prime)} \mathcal{S}_{ji}(\omega)
    +\dashint_{-\infty}^{+\infty}d\Omega~ \frac{e^{-it(\Omega+\omega^\prime)}}{(\omega-\Omega)(\Omega+\omega^\prime)} R_{ji}(\Omega) -\frac{1}{\omega^\prime+\omega} \mathcal{S}_{ji}(\omega)+\frac{1}{\omega^\prime+\omega} \mathcal{S}_{ji}(-\omega^\prime) \\
    &=-\frac{e^{-it(\omega+\omega^\prime)}}{(\omega+\omega^\prime)} \mathcal{S}_{ji}(\omega)
    +\dashint_{-\infty}^{+\infty}d\Omega~ \frac{e^{-it(\Omega+\omega^\prime)}}{(\omega-\Omega)(\Omega+\omega^\prime)} R_{ji}(\Omega) +\frac{1}{\omega^\prime+\omega} \mathcal{S}_{ji}(-\omega^\prime),\label{eqn:Hard_int}
\end{align}
where
\begin{align}
    &\gamma_{ji}(\omega,t) =  \int_{-\infty}^{t}ds~ e^{i\omega s} \left<\tilde{R}_j(s)R_i\right>_{\tilde{\rho}_R},~~\lim_{t\to+\infty} \gamma_{ji}(\omega,t)=\gamma_{ji}(\omega) \\
    &\bar{\gamma}_{ji}(\omega,t) =  \int_{t}^{+\infty}ds~ e^{i\omega s} \left<\tilde{R}_j(s)R_i\right>_{\tilde{\rho}_R},\\
    &\gamma_{ji}(\omega,t)+\bar{\gamma}_{ji}(\omega,t)=\gamma_{ji}(\omega).
\end{align}
Let us also notice that
Let us notice that:
\begin{align}
    &\gamma_{ji}(\omega,t)=\Gamma_{ij}^*(\omega)+\Gamma_{ji}^{(t)}(\omega), \\
    &\gamma_{ji}(\omega,-t)=\Gamma_{ij}^*(\omega)-{\Gamma_{ij}^{(t)}}^*(\omega,t).
\end{align}
Now the second term in equation \eqref{eqn:Hard_int} can be computed in the following way.
\begin{align}
    &\dashint_{-\infty}^{+\infty}d\Omega~ \frac{e^{-it(\Omega+\omega^\prime)}}{(\omega-\Omega)(\Omega+\omega^\prime)} R_{ji}(\Omega) = 
    \frac{1}{2\pi} e^{-it \omega^\prime} \int_{-\infty}^{+\infty}ds~ \left<\tilde{R}_j(s)R_i\right>_{\tilde{\rho}_R} \dashint_{-\infty}^{+\infty}d\Omega~ \frac{e^{-i\Omega(t-s)}}{(\omega-\Omega)(\Omega+\omega^\prime)}\\
    &=\frac{1}{2\pi} e^{-it \omega^\prime} \int_{-\infty}^{+\infty}ds~ \left<\tilde{R}_j(s)R_i\right>_{\tilde{\rho}_R}
    \frac{1}{\omega+\omega^\prime}\left(\dashint_{-\infty}^{+\infty}d\Omega~e^{-i\Omega(t-s)}\frac{1}{\omega-\Omega}+\dashint_{-\infty}^{+\infty}d\Omega~e^{-i\Omega(t-s)}\frac{1}{\Omega+\omega^\prime}\right) \\
    &=\frac{1}{2\pi} e^{-it \omega^\prime} \int_{-\infty}^{+\infty}ds~ \left<\tilde{R}_j(s)R_i\right>_{\tilde{\rho}_R}
    \frac{1}{\omega+\omega^\prime}\left(e^{-i\omega(t-s)}\dashint_{-\infty}^{+\infty}d\Omega~e^{i\Omega(t-s)}\frac{1}{\Omega}-e^{i\omega^\prime(t-s)}\dashint_{-\infty}^{+\infty}d\Omega~e^{i\Omega(t-s)}\frac{1}{\Omega}\right) \\
    &=\frac{i}{\pi} e^{-it \omega^\prime} \int_{-\infty}^{+\infty}ds~ \left<\tilde{R}_j(s)R_i\right>_{\tilde{\rho}_R}
    \frac{1}{\omega+\omega^\prime}\left(e^{-i\omega(t-s)}\dashint_{0}^{+\infty}d\Omega~\frac{\sin(\Omega(t-s))}{\Omega}-e^{i\omega^\prime(t-s)}\dashint_{0}^{+\infty}d\Omega~\frac{\sin{\Omega(t-s)}}{\Omega}\right) \\
    &=\frac{i}{\pi} e^{-it \omega^\prime} \int_{-\infty}^{+\infty}ds~ \left<\tilde{R}_j(s)R_i\right>_{\tilde{\rho}_R}
    \frac{1}{\omega+\omega^\prime}\left(e^{-i\omega(t-s)} \frac{\pi}{2} \mathrm{sgn} (t-s)-e^{i\omega^\prime(t-s)}\frac{\pi}{2} \mathrm{sgn} (t-s)\right) \\
    &=\frac{i}{2}  \frac{1}{\omega+\omega^\prime} e^{-it \omega^\prime} \int_{-\infty}^{+\infty}ds~\mathrm{sgn} (t-s) \left<\tilde{R}_j(s)R_i\right>_{\tilde{\rho}_R}
    \left(e^{-i\omega(t-s)} -e^{i\omega^\prime(t-s)}  \right) \\
    &=\frac{i}{2}  \frac{1}{\omega+\omega^\prime} e^{-it (\omega +\omega^\prime) } \int_{-\infty}^{+\infty}ds~\mathrm{sgn} (t-s) e^{i\omega s} \left<\tilde{R}_j(s)R_i\right>_{\tilde{\rho}_R}
    -\frac{i}{2}  \frac{1}{\omega+\omega^\prime}  \int_{-\infty}^{+\infty}ds~\mathrm{sgn} (t-s) e^{-i\omega^\prime s} \left<\tilde{R}_j(s)R_i\right>_{\tilde{\rho}_R}\\
    &=\frac{i}{2}  \frac{1}{\omega+\omega^\prime} e^{-it (\omega +\omega^\prime) } \left(\gamma_{ji}(\omega,t)-\bar{\gamma}_{ji}(\omega,t)\right)
    -\frac{i}{2}  \frac{1}{\omega+\omega^\prime}  \left(\gamma_{ji}(-\omega^\prime,t)-\bar{\gamma}_{ji}(-\omega^\prime,t)\right)\\
    &=\frac{i}{2}  \frac{1}{\omega+\omega^\prime} e^{-it (\omega +\omega^\prime) } \left(2\gamma_{ji}(\omega,t)-{\gamma}_{ji}(\omega)\right)
    -\frac{i}{2}  \frac{1}{\omega+\omega^\prime}  \left(2\gamma_{ji}(-\omega^\prime,t)-{\gamma}_{ji}(-\omega^\prime)\right).
\end{align}
Where we used the following relation
\begin{align}
    \int_0^{+\infty}dt \frac{\sin at}{t}=\mathrm{sgn}(a)\frac{\pi}{2}. 
\end{align}
Therefore we have:
\begin{align}
    A_{ij}(\omega,\omega^\prime,t)&=\frac{i}{2}  \frac{1}{\omega+\omega^\prime}  \left({\gamma}_{ji}(-\omega^\prime)-2\gamma_{ji}(-\omega^\prime,t)\right)-\frac{i}{2}  \frac{1}{\omega+\omega^\prime} e^{-it (\omega +\omega^\prime) } \left({\gamma}_{ji}(\omega)-2\gamma_{ji}(\omega,t)\right)
    \\&+ \frac{1}{\omega^\prime+\omega} \mathcal{S}_{ji}(-\omega^\prime)-\frac{e^{-it(\omega+\omega^\prime)}}{\omega+\omega^\prime} \mathcal{S}_{ji}(\omega).
\end{align}
Using analogical methods one can arrive at
\begin{align}
    B_{ij}(\omega^\prime,\omega,t) &=\frac{i}{2}  \frac{1}{\omega+\omega^\prime} e^{-it (\omega +\omega^\prime) } \left({\gamma}_{ji}(-\omega)-2\gamma_{ji}(-\omega,-t)\right)
    -\frac{i}{2}  \frac{1}{\omega+\omega^\prime}  \left({\gamma}_{ji}(\omega^\prime)-2\gamma_{ji}(\omega^\prime,-t)\right)\\
    &- \frac{1}{\omega^\prime+\omega} \mathcal{S}_{ji}(\omega^\prime)+\frac{e^{-it(\omega+\omega^\prime)}}{\omega+\omega^\prime} \mathcal{S}_{ji}(-\omega).
\end{align}

We obtain explicit expression for $\gamma_{ij}(\omega,\omega^\prime,t)$
\begin{align}
    \gamma_{ij}(\omega,\omega^\prime,t)&=
    -\frac{i}{2}\frac{1}{\omega-\omega^\prime} \left(e^{-it(\omega-\omega^\prime)}-1\right) \left(\gamma_{ji}(\omega)-2\gamma_{ji}(\omega,t)\right) \nonumber\\
    &-\frac{i}{2}\frac{1}{\omega-\omega^\prime} \left[\left(\gamma_{ji}(\omega)-2\gamma_{ji}(\omega,t)\right)-\left(\gamma_{ji}(\omega^\prime)-2\gamma_{ji}(\omega^\prime,t)\right)\right] \nonumber\\
    &-\frac{1}{\omega-\omega^\prime} \left(e^{-it(\omega-\omega^\prime)}-1\right) \mathcal{S}_{ji}(\omega)
    -\frac{1}{\omega-\omega^\prime} \left(\mathcal{S}_{ji}(\omega)-\mathcal{S}_{ji}(\omega^\prime)\right) \nonumber\\
    &+\frac{i}{2}\frac{1}{\omega-\omega^\prime} \left(e^{-it(\omega-\omega^\prime)}-1\right) \left(\gamma_{ji}(\omega^\prime)-2\gamma_{ji}(\omega^\prime,-t)\right) \nonumber\\
    &-\frac{i}{2}\frac{1}{\omega-\omega^\prime} \left[\left(\gamma_{ji}(\omega)-2\gamma_{ji}(\omega,-t)\right)-\left(\gamma_{ji}(\omega^\prime)-2\gamma_{ji}(\omega^\prime,-t)\right)\right] \nonumber\\
    &+\frac{1}{\omega-\omega^\prime} \left(e^{-it(\omega-\omega^\prime)}-1\right) \mathcal{S}_{ji}(\omega^\prime)
    -\frac{1}{\omega-\omega^\prime} \left(\mathcal{S}_{ji}(\omega)-\mathcal{S}_{ji}(\omega^\prime)\right),
\end{align}
and for $\Xi_{ij}(\omega,\omega^\prime,t)$
\begin{align}
    \Xi_{ij}(\omega,\omega^\prime,t)&=-\frac{1}{4}\frac{1}{\omega-\omega^\prime} \left(e^{-it(\omega-\omega^\prime)}-1\right) \left(\gamma_{ji}(\omega)-2\gamma_{ji}(\omega,t)\right) \nonumber\\
    &-\frac{1}{4}\frac{1}{\omega-\omega^\prime} \left[\left(\gamma_{ji}(\omega)-2\gamma_{ji}(\omega,t)\right)-\left(\gamma_{ji}(\omega^\prime)-2\gamma_{ji}(\omega^\prime,t)\right)\right] \nonumber\\
    &-\frac{1}{2i}\frac{1}{\omega-\omega^\prime} \left(e^{-it(\omega-\omega^\prime)}-1\right) \mathcal{S}_{ji}(\omega)\\
    &-\frac{1}{4}\frac{1}{\omega-\omega^\prime} \left(e^{-it(\omega-\omega^\prime)}-1\right) \left(\gamma_{ji}(\omega^\prime)-2\gamma_{ji}(\omega^\prime,-t)\right) \nonumber\\
    &+\frac{1}{4}\frac{1}{\omega-\omega^\prime} \left[\left(\gamma_{ji}(\omega)-2\gamma_{ji}(\omega,-t)\right)-\left(\gamma_{ji}(\omega^\prime)-2\gamma_{ji}(\omega^\prime,-t)\right)\right] \nonumber\\
    &-\frac{1}{2i}\frac{1}{\omega-\omega^\prime} \left(e^{-it(\omega-\omega^\prime)}-1\right) \mathcal{S}_{ji}(\omega^\prime),
\end{align}
where the terms for which $\omega=\omega^\prime$ are treated in the sense of a limit $\lim_{\omega^\prime \to \omega}$: 
\begin{align}
    &\gamma_{ij}(\omega,\omega,t)=\lim_{\omega^\prime \to \omega}\gamma_{ij}(\omega,\omega^\prime,t),\\
    &\Xi_{ij}(\omega,\omega,t)=\lim_{\omega^\prime \to \omega}\Xi_{ij}(\omega,\omega^\prime,t).
\end{align}

As we observe, the formal long time limit ($\lim_{t \to \infty}$) for $\Xi_{ij}(\omega,\omega^\prime,t)$ and $\gamma_{ij}(\omega,\omega^\prime,t)$ does not exist due to the presence of the oscillating terms. But for times, which are large enough $t \approx +\infty$ (with respect to correlation time $\left<\tilde{R}_j(t)R_i\right>_{\tilde{\rho}_R}=0$), and terms $\omega \neq \omega^\prime$ we can write:
\begin{align}
    \gamma_{ij}(\omega,\omega^\prime,t)&\stackrel{t\approx +\infty}{\approx}
    \frac{i}{2}\frac{1}{\omega-\omega^\prime} \left(e^{-it(\omega-\omega^\prime)}-1\right) \left(\gamma_{ji}(\omega)+\gamma_{ji}(\omega^\prime)\right) \nonumber\\
    &-\frac{1}{\omega-\omega^\prime} \left(e^{-it(\omega-\omega^\prime)}-1\right) \left(\mathcal{S}_{ji}(\omega)-\mathcal{S}_{ji}(\omega^\prime)\right)
    -\frac{2}{\omega-\omega^\prime} \left(\mathcal{S}_{ji}(\omega)-\mathcal{S}_{ji}(\omega^\prime)\right) \nonumber\\
    &=\frac{i}{2}\frac{1}{\omega-\omega^\prime} \left(e^{-it(\omega-\omega^\prime)}-1\right) \left(\gamma_{ji}(\omega)+\gamma_{ji}(\omega^\prime)\right)-\frac{1}{\omega-\omega^\prime} \left(e^{-it(\omega-\omega^\prime)}+1\right) \left(\mathcal{S}_{ji}(\omega)-\mathcal{S}_{ji}(\omega^\prime)\right),\\
    \Xi_{ij}(\omega,\omega^\prime,t)&\stackrel{t\approx +\infty}{\approx}\frac{1}{4}\frac{1}{\omega-\omega^\prime} \left(e^{-it(\omega-\omega^\prime)}-1\right) \left(\gamma_{ji}(\omega)-\gamma_{ji}(\omega^\prime)\right)+\frac{1}{2}\frac{1}{\omega-\omega^\prime} \left(\gamma_{ji}(\omega)-\gamma_{ji}(\omega^\prime)\right) \nonumber\\
    &-\frac{1}{2i}\frac{1}{\omega-\omega^\prime} \left(e^{-it(\omega-\omega^\prime)}-1\right) \left(\mathcal{S}_{ji}(\omega)+\mathcal{S}_{ji}(\omega^\prime)\right),
\end{align}
and for terms for which $\omega^\prime=\omega$ we obtain
\begin{align}
    &\gamma_{ij}(\omega,\omega,t)\stackrel{t\approx +\infty}{\approx}
    t \gamma_{ji}(\omega)
    -2 \frac{\partial}{\partial \omega} \mathcal{S}_{ji}(\omega), \\
    &\Xi_{ij}(\omega,\omega,t) \stackrel{t \approx + \infty}{\approx}
    t \mathcal{S}_{ji}(\omega)+\frac{1}{2}\frac{\partial}{\partial \omega} \gamma_{ji} (\omega).
\end{align}
This result allows to construct the long-time limit $\tilde{K}_{\infty}^{(2)}(t)$ of the  superoperator  $\tilde{K}^{(2)}(t)$
\begin{align}
    \tilde{K}_{\infty}^{(2)}(t) = t \tilde{L}+D_t, \label{eqn:LongTimesCumulant}
\end{align}
where $\tilde{L}$ is the (interaction picture) generator of the Davies GKSL equation, and $D_t$ contains all other components of $\gamma_{ij}(\omega,\omega^\prime,t)$ (also $ -2 \frac{\partial}{\partial \omega} \mathcal{S}_{ji}(\omega)$ term). The form of the long-time limit of the cumulant superoperator $\tilde{K}_{\infty}^{(2)}(t)$ in equation \eqref{eqn:LongTimesCumulant}, strongly suggests that the eigenvectors of $\tilde{K}_{\infty}^{(2)}(t)$ are arbitrarily close to the eigenvectors of $\tilde{L}$. Therefore the stationary state of the cumulant equation is the same as for the renormalized Davies-GKSL equation.

It is important to mention that for the Ohmic spectral density (that is a typical case), with exponential cut-off
\begin{align}
    J(\omega) = \frac{\omega^S}{\omega_c^{S-1}} e^{-\frac{\abs{\omega}}{\omega_c}},~~S\ge 0,
\end{align}
where $\omega_c$ is the cut-off frequency, the superoperator $\tilde{K}^{(2)}(t)$ converges to its long-times form $\tilde{K}_{\infty}^{(2)}(t)$ at pace $\mathcal{O}(t^{-2})$, i.e.,
\begin{align}
    \norm{\tilde{K}_{\infty}^{(2)}(t)-\tilde{K}^{(2)}(t)}_1 \stackrel{t \approx +\infty}{\le} \frac{C}{t^2}.
\end{align}
Here, $\norm{\cdot}_1$ is an induced (superoperator) norm on $\mathcal{B}(\mathcal{H_\mathcal{S}})$, and $0<C<+\infty$ is a finite constant (cut-off dependent).

\color{black}
We notice that the time derivative of the above reproduces the corresponding terms of the generator of the Bloch-Redfield equation. Interestingly, we observe the presence of the terms that are constant in time. The above results has been numerically confirmed for the agreement with formula \eqref{eqn:Cumulant_Matrix}.

Additionally, for times that are large enough $\gamma_{ij}(\omega,\omega,t)$ satisfies the detailed-balance condition, i.e, 
\begin{align}
    &\frac{\gamma_{ji}(-\omega,-\omega,t)}{\gamma_{ij}(\omega,\omega,t)}
    =\frac{t \gamma_{ij}(-\omega)
    +2 \frac{\partial}{\partial \omega} \mathcal{S}_{ij}(-\omega)}{t \gamma_{ji}(\omega)
    -2 \frac{\partial}{\partial \omega} \mathcal{S}_{ji}(\omega)}
    =\frac{ \gamma_{ij}(-\omega)
    +\frac{1}{t} 2 \frac{\partial}{\partial \omega} \mathcal{S}_{ij}(-\omega)}{ \gamma_{ji}(\omega)
    -\frac{1}{t}2 \frac{\partial}{\partial \omega} \mathcal{S}_{ji}(\omega)}\stackrel{t\approx + \infty}{\approx} \frac{ \gamma_{ij}(-\omega)
    }{ \gamma_{ji}(\omega)
    }=e^{-\beta \omega}. \label{eqn:DetailedBalance}
\end{align}
What clarifies why the long-times state resulting from the cumulant equation dynamics has the same diagonal elements (in the $H_\mathcal{S}$ basis) as the Gibbs state with respect to $H_\mathcal{S}$.
\color{black}

\end{widetext}
% \bibliographystyle{apsrev4-1}
% \bibliography{references}

\end{document}